\documentclass[journal]{IEEEtran}
\usepackage{caption} 
\usepackage{soul}
\usepackage{lipsum}
\usepackage{indentfirst}
\usepackage{booktabs}
\usepackage{enumerate}
\usepackage{tabu}
\usepackage{tabularx}
\usepackage{environ}         
\usepackage{etoolbox}        
\usepackage[dvipsnames]{xcolor}

\usepackage{cite}

\usepackage{graphicx} 
\usepackage{wrapfig}
\usepackage[labelformat=simple]{subcaption}

\usepackage{float}
\usepackage{eurosym}
\usepackage{textpos}
\usepackage[pscoord]{eso-pic}

\newcommand{\placetextbox}[3]{
  \setbox0=\hbox{#3}
  \AddToShipoutPictureFG*{
    \put(\LenToUnit{#1\paperwidth},\LenToUnit{#2\paperheight}){\vtop{{\null}\makebox[0pt][c]{#3}}}%
  }%
}%

\usepackage{amsfonts, eqnarray, amsmath, amssymb}
\usepackage{amsthm}
\usepackage{mathtools}
\usepackage{xfrac}
\usepackage{breqn}
\usepackage{bm}

\newtheorem{theorem}{Theorem}
\newtheorem{proposition}[theorem]{Proposition}

\usepackage{algorithm}
\usepackage{algpseudocode}
\usepackage{etoolbox}

\makeatletter
\newcommand*{\algrule}[1][\algorithmicindent]{%
  \hspace*{.2em}
  \vrule 
  \hspace*{\dimexpr#1-.2em-.4pt}%
}

\newcommand{\StatePar}[1]{%
  \State\parbox[t]{\dimexpr\linewidth-\ALG@thistlm}{\strut #1\strut}%
}
\renewcommand{\ALG@beginalgorithmic}{\offinterlineskip}

\newcount\ALG@printindent@tempcnta
\def\ALG@printindent{%
  \ifnum \theALG@nested > 0
    \ifx\ALG@text\ALG@x@notext
    \else
      \unskip
      \ALG@printindent@tempcnta=1
      \loop
        \algrule[\csname ALG@ind@\the\ALG@printindent@tempcnta\endcsname]%
        \advance \ALG@printindent@tempcnta 1
        \ifnum \ALG@printindent@tempcnta<\numexpr\theALG@nested+1\relax
      \repeat
        \fi
    \fi
}
\patchcmd{\ALG@doentity}{\noindent\hskip\ALG@tlm}{\ALG@printindent}{}{\errmessage{failed to patch}}
\makeatother

\algrenewcommand\algorithmicend{\strut\textbf{end}}
\algrenewcommand\algorithmicdo{\strut\textbf{do}}
\algrenewcommand\algorithmicwhile{\strut\textbf{while}}
\algrenewcommand\algorithmicfor{\strut\textbf{for}}
\algrenewcommand\algorithmicforall{\strut\textbf{for all}}
\algrenewcommand\algorithmicloop{\strut\textbf{loop}}
\algrenewcommand\algorithmicrepeat{\strut\textbf{repeat}}
\algrenewcommand\algorithmicuntil{\strut\textbf{until}}
\algrenewcommand\algorithmicprocedure{\strut\textbf{procedure}}
\algrenewcommand\algorithmicfunction{\strut\textbf{function}}
\algrenewcommand\algorithmicif{\strut\textbf{if}}
\algrenewcommand\algorithmicthen{\strut\textbf{then}}
\algrenewcommand\algorithmicelse{\strut\textbf{else}}

\algrenewcommand\algorithmicrequire{\strut\textbf{Input:}}
\algrenewcommand\algorithmicensure{\strut\textbf{Output:}}

\let\oldState\State
\renewcommand{\State}{\oldState\strut}

\usepackage{array}

\usepackage{url}

\NewEnviron{resizealign}{\sbox0{
    $\begin{matrix}\displaystyle\BODY\end{matrix}$}%
  \sbox1{$(\theequation)$}%
  \sbox2{\parbox{\dimexpr \wd0 + 2\wd1}%
    {\begin{align}\BODY\end{align}}}
  \noindent\resizebox{\columnwidth}{!}{\usebox2}%
}

\begin{document}
\title{Optical Front/Mid-haul with Open Access-Edge Server~Deployment~Framework~for~Sliced~O-RAN}
%
\author{Sourav~Mondal,~\IEEEmembership{Member,~IEEE}
        and~Marco~Ruffini,~\IEEEmembership{Senior~Member,~IEEE}
\thanks{S. Mondal and M. Ruffini are with CONNECT Centre for Future Networks and Communication, Trinity College Dublin, University of Dublin, Dublin 2, Ireland (e-mail: somondal@tcd.ie, marco.ruffini@tcd.ie).}}


\placetextbox{0.5}{0.04}{This is the final version to be published in IEEE Transactions on Network and Service Management (TNSM). Copyright @ IEEE.}%

\maketitle

\begin{abstract}
The fifth-generation of mobile radio technologies is expected to be agile, flexible, and scalable while provisioning ultra-reliable and low-latency communication (uRLLC), enhanced mobile broadband (eMBB), and massive machine type communication (mMTC) applications. An efficient way of implementing these is by adopting cloudification, network function virtualization, and network slicing techniques with open-radio access network (O-RAN) architecture where the base-band processing functions are disaggregated into virtualized radio unit (RU), distributed unit (DU), and centralized unit (CU) over front/mid-haul interfaces. However, cost-efficient solutions are required for designing front/mid-haul interfaces and time-wavelength division multiplexed (TWDM) passive optical network (PON) appears as a potential candidate. Therefore, in this paper, we propose a framework for the optimal placement of RUs based on long-term network statistics and connecting them to open access-edge servers for hosting the corresponding DUs and CUs over front/mid-haul interfaces while satisfying the diverse QoS requirements of uRLLC, eMBB, and mMTC slices. In turn, we formulate a two-stage integer programming problem and time-efficient heuristics for users to RU association and flexible deployment of the corresponding DUs and CUs. We evaluate the O-RAN deployment cost and latency requirements with our TWDM-PON-based framework against urban, rural, and industrial areas and show its efficiency over the optical transport network (OTN)-based framework.
\end{abstract}

\IEEEpeerreviewmaketitle

\vspace{-0.5\baselineskip}
\section{Introduction} \label{sec1}
\IEEEPARstart{T}{he} fifth-generation (5G) of mobile radio access networks (RANs) envisions to support a diverse set of applications which are classified into three broad categories, i.e., uRLLC, eMBB, and mMTC. Hence, the 5G networks are expected to possess features like flexibility, scalability, manageability, and customizability \cite{5G_vision}. In addition to high throughput (peak uplink: 10 Gbps and peak downlink: 20 Gbps), an extremely high user density ($\sim 10^6$ devices/km$^2$) is expected from 5G RANs that leads to a dense base station deployment. Therefore, for a diverse and dense RAN deployment, initially, cloud-radio access network (C-RAN) was considered as a potential architecture by the 5G infrastructure public private partnership (5GPPP) because it enables efficient network management and resource sharing in real-time through a centralized architecture \cite{c2oran}. In the C-RAN architecture, the traditional base stations are decoupled into two parts, viz., remote radio heads (RRHs) and base-band units (BBUs). The RRHs are distributed as remote antenna elements, whereas a cluster of BBUs (known as BBU pool) is placed at a centralized location. The RRHs are connected to the BBU pool by front-haul links, typically using the common public radio interface (CPRI) that requires continuous data streaming (split-8) at high throughput with low-latency and low-jitter ($\leq 65$ ns) \cite{habibi}. However, to foster more open and smarter RANs, major mobile network operators across the globe are collaborating within the \emph{Open-RAN (O-RAN) Alliance} to standardize the O-RAN architecture by adopting dis-aggregated virtual BBU function processing units like RUs, DUs, and CUs (by exploiting software defined everything (SDx) and network function virtualization (NFV) on commercial off-the-shelf (COTS) hardware), which are interconnected by open front/mid-haul interfaces. This architecture relies on \emph{programmability, openness, resource sharing}, and \emph{edgefication} to handle complex network management issues through artificial intelligence-based techniques \cite{oran}.\par
%
Another critical issue of the C-RAN architecture is that continuous bit streaming is required in the front-haul links at a very high data rate that does not scale with actual user traffic and mainly depends on RRH configurations \cite{itu_5g}. This issue is overcome by using different split options among RUs, DUs, and CUs in O-RAN architecture that enables a flexible and scalable network deployment \cite{mgain}. Nonetheless, designing the RU-DU (front-haul) interface is still very challenging as it demands high capacity (typically working on split-7.2 but considerably less demanding than the RRH-BBU split-8 CPRI), low-latency, and low-jitter ($\leq 40$ $\mu$s) links. The DU-CU (mid-haul) interface, on the other hand, demands less throughput and can tolerate relatively higher latency. To meet the next-generation front-haul interface (NGFI) requirements, OTNs and TWDM-PONs are recommended by ITU-T as their latest standards can support up to $100$ Gbps or more aggregated datarate \cite{itu_5g}. As the O-RAN allows RUs to lease on-demand DU and CU processing resources, a better operational cost reduction and energy efficiency can be achieved when a large number of RUs can be aggregated to edge cloud nodes. Thus, the concept of using shared PONs to improve flexibility across access and metro \cite{8606981} and to provide connectivity to edge cloud nodes \cite{9129653} is gaining momentum. Note that microwave and free-space optics can also be used for implementing the front/mid-haul links \cite{fh_desgn}, but they fail to provide the same level of capacity and reliability as optical fiber-based OTN and TWDM-PON. Our motivations behind choosing TWDM-PON are its cost efficiency, the possibility of ubiquitous deployment as they are normally used to provide highly-dense FTTx services, and the convenience to scale up network throughput by introducing additional wavelengths. To the best of our knowledge, the design and planning of O-RAN front/mid-haul networks with TWDM-PON and its merits over OTN are not examined so far in the existing literature.\par
\begin{table}[!t]
\centering
\caption{Abbreviations Employed}
\label{table0}
\resizebox{\columnwidth}{!}{%
\begin{tabular}{cl}
\toprule
\textbf{Abbreviation}      & \multicolumn{1}{c}{\textbf{Meaning or Description}}                        \\ \toprule
3GPP                 & The 3rd Generation Partnership Project               \\ \hline
5GPPP                & The 5G Infrastructure Public Private Partnership     \\ \hline
BBU                  & Base-band processing unit                            \\ \hline
CNF                  & Containerized network function                       \\ \hline
COTS                 & Commercial off-the-shelf                             \\ \hline
CPRI                 & Common public radio interface                        \\ \hline
C-RAN                & Cloud-radio access network                           \\ \hline
CU                   & Centralized unit                                     \\ \hline
DU                   & Decentralized unit                                   \\ \hline
eMBB                 & Enhanced mobile broadband                            \\ \hline
FTTx                 & Fiber-to-the-home/curb/business/office               \\ \hline
GOPS                 & Giga operations per second                           \\ \hline
ILP                  & Integer linear programming                           \\ \hline
ITU-T                & International Telecommunication Union-Telecommunication           \\ \hline
M-DBA                & Mobile dynamic bandwidth allocation                  \\ \hline
mMTC                 & Massive machine-type communication                   \\ \hline
NFV                  & Network function virtualization                      \\ \hline
NGFI                 & Next-generation front-haul interface                 \\ \hline
OFDM                 & Orthogonal frequency division multiplexing           \\ \hline
OLT                  & Optical line terminal                                \\ \hline
ONU                  & Optical network unit                                 \\ \hline
O-RAN                & Open radio access network                            \\ \hline
OTN                  & Optical transport network                            \\ \hline
PDCP                 & Packet data convergence protocol                     \\ \hline
PON                  & Passive optical network                              \\ \hline
RRC                  & Radio resource controller                            \\ \hline
RRH                  & Remote radio head                                    \\ \hline
RU                   & Radio unit                                           \\ \hline
SDx                  & Software defined everything                          \\ \hline
TTI                  & Transmission time interval                           \\ \hline
TWDM                 & Time and wavelength division multiplexed             \\ \hline
UE                   & User equipment                                       \\ \hline
uRLLC                & Ultra-reliable and low-latency communication         \\ \hline
VNF                  & Virtualized network function                    \\ \bottomrule
\end{tabular}
}
\end{table}
\setlength{\textfloatsep}{0pt}
Alongside front/mid-haul network design, integration of edge/cloud servers or processor pools in O-RAN architecture for efficient DU and CU function processing also presents an important research challenge. Over the past decade, the industry, as well as academic research on cloud/edge computing networks, have taken similar momentum as access networks, which created the premise for a confluence of two technologies, viz., \emph{Telecom cloud} and \emph{IT cloud}. Both Telecom network operators and IT cloud providers are interested in controlling the evolution of the Internet at the edge in their favor, which raises the question: \emph{Where is the edge?} \cite{aceg1}. The network operators are motivated to integrate cloud technologies with access networks for supporting low-latency and high-bandwidth edge applications because they have existing and widely distributed physical network infrastructure. On the contrary, the cloud operators are keen to saturate metro areas with massive edge node clusters with very simple access networks such that everything can be provided as a service on COTS hardware. Hence, for providing an open market opportunity to any vendor, \emph{access-edge clouds} for \emph{democratizing the network edge for sustained innovation} is proposed \cite{aceg1}. {These open access-edge cloud servers can be used for hosting DU and CU function processing units as well as executing jobs for edge computing applications.}\par
 
Recently, \emph{network slicing} was proposed for efficient handling of diverse network resource requirements of uRLLC, eMBB, and mMTC applications \cite{3Gpp_NS}. Through network slicing, different logical networks (i.e., slices) are constructed on a shared physical network via SDx, NFV, and cloud/edge computing technologies. Each slice can be considered as an end-to-end virtualized network instance and is customized in terms of communication, computation, and storage resources to meet the specific service requirements \cite{5g_ns}. For example, the uRLLC applications demand a low throughput but a very stringent one-way latency of 1 ms, the eMBB applications demand a one-way latency of 4 ms but an average user experience throughput of 50 Mbps in uplink and 100 Mbps in the downlink, and the mMTC applications demand a one-way latency of 10 ms but a very high device density. Therefore, optimizing the RAN resources to accommodate different network slices is an essential research challenge.\par
Note that two primary aspects of the front/mid-haul design for O-RAN are: (a) identification of RU locations based on user distribution, long-term average throughput, and QoS latency requirements of uRLLC, eMBB, and mMTC applications, and (b) optimally connecting the RUs to open access-edge servers that host the DUs and CUs. In the C-RAN design problem, after the users are assigned to RRHs, optimal locations for BBU pool placement over front-haul (only split-8) interfaces are required. Nevertheless, the flexibility in choosing RU-DU-CU split options and front/mid-haul configurations in O-RAN increases the problem complexity several times more than C-RAN. In this paper, we propose a framework for optimal deployment of network slicing enabled TWDM-PON-based NGFI with integrated open access-edge servers that encompass a minimum cost of O-RAN deployment. Our primary contributions are summarized as follows:
\begin{enumerate}[(i)]
\item {We propose a framework to identify optimal 5G RU, TWDM-PON optical network unit (ONU), and optical line terminal (OLT) locations with a minimum number of access-edge cloud servers to minimize the RAN deployment cost. Thus, we formulate a two-stage integer linear programming (ILP) problem for user equipment (UE) to RU association and flexible placement of corresponding DUs and CUs for uRLLC, eMBB, and mMTC slices.}
\item We compute the global optimal solution of our ILP formulation by commercially available solvers. In addition, we design efficient heuristics for UE to RU association (Lagrangian relaxation based) and connect the corresponding DUs and CUs (greedy approach based). We show that these heuristics can produce near-optimal solutions while ensuring quick convergence with appropriate convergence and optimality gap analysis.
\item We evaluate the proposed framework against urban, rural, and industrial scenarios and present a comparative study on deployment cost, front/mid-haul communication and BBU processing latencies for eMBB, uRLLC, and mMTC slices. We also show that our proposed TWDM-PON-based framework is cost-efficient over OTN-based NGFI framework by using the same RU locations.
\end{enumerate}
\par The rest of this paper is organized as follows. In Section \ref{sec2}, we review some recent related works. In Section \ref{sec3}, we briefly discuss our proposed open access-edge network architecture and system model. In Section \ref{sec4}, the ILPs are formulated and the corresponding heuristics are designed. In Section \ref{sec6}, framework evaluation methods are described and the numerical results are presented. Finally, in Section \ref{sec7}, our primary observations and achievements are summarized.

\begin{figure*}[t!]
\centering
\includegraphics[width=\textwidth,height=10.0cm]{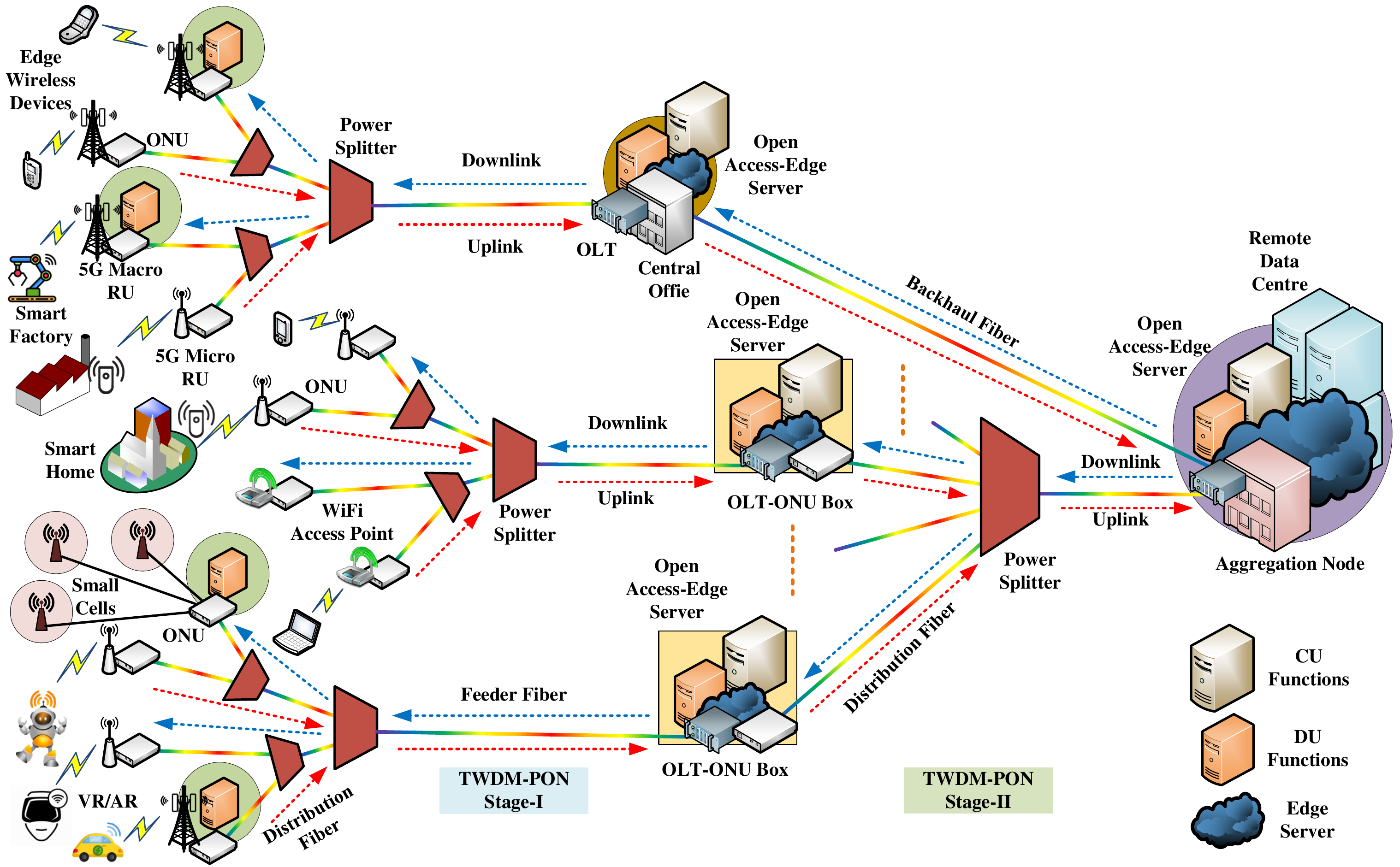}
\caption{The proposed TWDM-PON-based front/mid-haul network architecture for supporting open and intelligent 5G O-RAN (VR/AR: virtual/augmented reality, OLT: optical line terminal, ONU: optical network unit, RU: radio unit, DU: distributed unit, CU: centralized unit).}
\label{architecture}
\end{figure*}\setlength{\textfloatsep}{5pt}
\section{Review of Related Works} \label{sec2}
The vision of the 5G mobile communication network consists of a major leap in terms of bandwidth, latency, device density, and various other QoS requirements from its predecessors. {In the past, the authors of \cite{cell_place} proposed a cell planning method for 4G networks that simultaneously satisfies both cell coverage and capacity constraints. Recently, the authors of \cite{slice_plan} proposed a cell planning framework for dense and heterogeneous 5G networks. The radio resources are orthogonally partitioned among different slices for optimal distribution among cells and allocation to mobile users. Furthermore, the authors of \cite{mm_plan} proposed 5G mobile network planning framework for urban areas in Ibb city, Yemen, where 28 GHz millimeter-wave base stations are considered. However, these frameworks are unsuitable for the 5G O-RAN architecture.} The authors of \cite{gabriel,tdm_place} provided crucial insights on the front-haul network design for 5G RANs. As the SDx and NFV technologies play a significant role in Beyond 5G networks, new planning, and dimensioning models are required to achieve a cost-optimal design that supports a wide range of applications \cite{5g_srv4}. The authors of \cite{5g_core} proposed three optimization models to minimize the network load and data center resources by finding the optimal placement of the data centers, SDx, and NFV mobile network functions. The authors of \cite{o-ran} studied the containerized network function (CNF) placement and resource allocation problem of O-RAN with three-layer hierarchical data centers. Furthermore, the authors of \cite{oran_opt} formulated an optimization problem for RU-DU-CU placement and the allocation of front/mid-haul transmission resources over OTN. The authors of \cite{F_split} formulated an optimization problem for routing, wavelength and bandwidth assignment, and processor pool selection for fine-grained BBU function placement over OTN. However, neither of these works designed TWDM-PON-based front/mid-haul networks for 5G O-RANs that provisions flexible placement of DUs and CUs.\par
An important feature of O-RAN architectures is flexible functional splits and the authors of \cite{5g_fsplt} presented a comprehensive overview of each functional split option with their respective advantages and disadvantages over evolved common public radio interface (eCPRI) and we incorporated this feature in our problem formulation as well. Although the bandwidth demand of front-haul interfaces can be met by TWDM-PONs, the existing dynamic bandwidth allocation algorithms (DBA) fail to satisfy the front-haul latency requirements. Thus, the authors of \cite{5g_fh_bw3} proposed a novel cooperative mobile-DBA (M-DBA) algorithm where each OLT receives future uplink scheduling information from the corresponding BBUs and pre-allocates time-slots by estimating the arrival period to reduce the waiting time of uplink data at the ONUs.\par
%
Recently, for efficient resource management to support a diverse set of eMBB, uRLLC, and mMTC applications, research on RAN slicing, core network slicing, and front/mid-haul slicing has started to gain significant attention \cite{min_max_SM}. The authors of \cite{5g_ns1} formulated the network-slicing process as a weighted throughput maximization problem that involves sharing of computational resources, fronthaul capacity, physical RRHs, and radio resources. The authors of \cite{5g_ns6} presented a slice-based 5G architecture and ``Network Store in a programmable cloud" that efficiently manages network slices through NFV, SDx, and cloud computing. The authors of \cite{5g_ns4} proposed a unified control and network-slicing architecture for a multi-vendor multi-standard PON-based 5G fronthaul network and experimentally demonstrated its flexible resource management and slicing capability. Moreover, they proposed a multi-vendor network-slicing scheme for converged vehicular and fixed access networks in \cite{5g_ns5}. The authors of \cite{5g_fh2} proposed a flexible hierarchical edge cloud architecture to enable 5G optical fronthaul network slicing and a network resource management scheme that jointly allocates bandwidth resources to various network slices. Nonetheless, the performance of network slice enabled 5G O-RAN architecture with TWDM-PON-based front/mid-haul interfaces and integrated open access-edge servers demand a thorough investigation.
\begin{figure}[t!]
\centering
\includegraphics[width=\columnwidth]{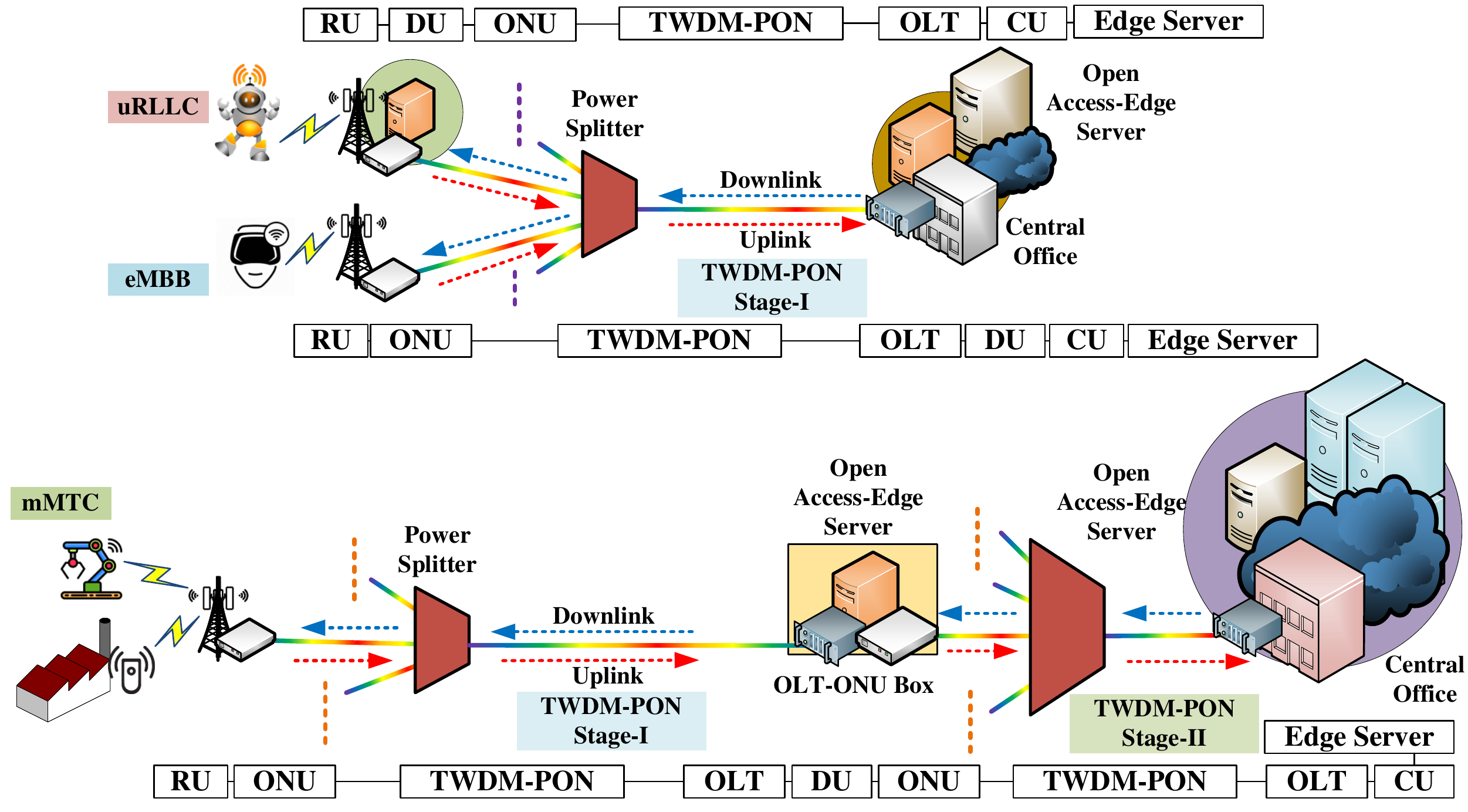}
\caption{Some example RU, DU, and CU inter-connections over TWDM-PON front/mid-haul for uRLLC, eMBB, and mMTC slices.}
\label{use_case}
\end{figure}\setlength{\textfloatsep}{5pt}

\section{Open Access-Edge Cloud Network Design with TWDM-PON-based Front/Mid-haul} \label{sec3}
In this section, we discuss the fundamental aspects of TWDM-PON-based front/mid-haul interfaces for sliced 5G O-RAN with integrated access-edge cloud servers. 
\vspace{-0.5\baselineskip}
\subsection{TWDM-PON-based Front/Mid-haul Network Architecture}
In Fig. \ref{architecture}, we show the proposed TWDM-PON-based front/mid-haul interfaces for 5G O-RAN. Most ONUs are connected to 5G macro, micro, and small cell RUs, and groups of ONUs are connected to OLTs via TWDM-PONs. The TWDM-PONs are being considered for supporting small cell connectivity, with solutions capable of providing slice isolation \cite{9019591} and targeting specific service level agreement targets \cite{9489501}. If we consider the 3GPP recommended RU-DU split-7.2 and DU-CU split-2, then radio antenna and low physical layer circuitry run on RU hardware. This mainly includes physical layer tasks like IQ decompression, precoding, digital beamforming, inverse Fourier transform, cyclic prefix addition, and digital-to-analog conversion. In some cases, very small-capacity open access-edge servers may also be placed along with RUs that can host DUs for processing high physical layer, MAC, and RLC functions. Usually, medium or large-capacity access-edge cloud servers are placed at OLT locations whose resources can be distributed for hosting DU, CU, and edge computing applications. These functions can be implemented either as virtualized network functions (VNFs) or as CNFs \cite{o-ran}. The CUs handle upper layer functions, e.g., RRC, PDCP-C, PDCP-U, and SDAP \cite{o-ran}.\par

Actually, functional splits can be chosen flexibly for the RU-DU and DU-CU interfaces which incur different latency requirements, e.g., max 0.1 ms for split-7.2 and max 10 ms for split-2 \cite{5g_fsplt}. With a split-7.2 RU-DU interface, the front-haul network can span up to 20 km with a 1:64 power-splitter ratio to meet a maximum one-way latency of 100 $\mu$sec. On the other hand, split-2 can be chosen for the DU-CU interface and the mid-haul can span up to 80 km to meet a maximum one-way latency of around 1 msec \cite{pw}. Moreover, the throughput requirement of the mid-haul is nearly 10 times less than that of the front-haul, e.g., with 4x4 MIMO, 100 MHz configuration, around 11.1 Gbps is required for split-7.2 front-haul but only 1.11 Gbps is required for split-2 mid-haul. Therefore, multiple Stage-I TWDM-PONs with DUs located at RU or ONU can be further aggregated through a single Stage-II TWDM-PON.\par
For clarity, please refer to the first example in Fig. \ref{use_case} which shows the RU-DU-CU interfaces for uRLLC (max one-way latency = 1 ms) and eMBB applications (max one-way latency = 4 ms) where Stage-I TWDM-PON is sufficient. A small-capacity server is installed at RU locations for processing the DUs of uRLLC applications and Stage-I TWDM-PON carries its mid-haul traffic. This might be costly but efficient in meeting strict latency requirements. Nonetheless, the DUs and CUs of eMBB applications are processed by servers at Stage-I OLT locations and the Stage-I TWDM-PON carries its front-haul traffic. The second example in Fig. \ref{use_case} shows the RU-DU-CU interfaces for mMTC applications (max one-way latency = 10 ms) where both Stage-I and Stage-II TWDM-PONs are used. An \emph{OLT-ONU box} is used to create a sequential connection among Stage-I OLT, server for DU, and Stage-II ONU. To support a large number of mMTC devices and exploit a higher latency bound, this configuration may prove to be cost-efficient. The edge clouds are always placed after the CUs and play a crucial role in instantiating all the dedicated functions of the uRLLC slice. There are several existing works on edge computing server placement \cite{mec_srvy}, hence this is beyond the scope of this paper. In addition, the cellular core functions run on \emph{remote data centers} where network slice management and orchestration are also performed to create and manage the life cycles of network slices. While latency in PON can be an issue in the upstream direction, mechanisms like the cooperative DBA mentioned previously or other solutions that make use of hardware acceleration \cite{9489802} can substantially reduce this effect.

\begin{figure}[t!]
\centering
\includegraphics[width=\columnwidth]{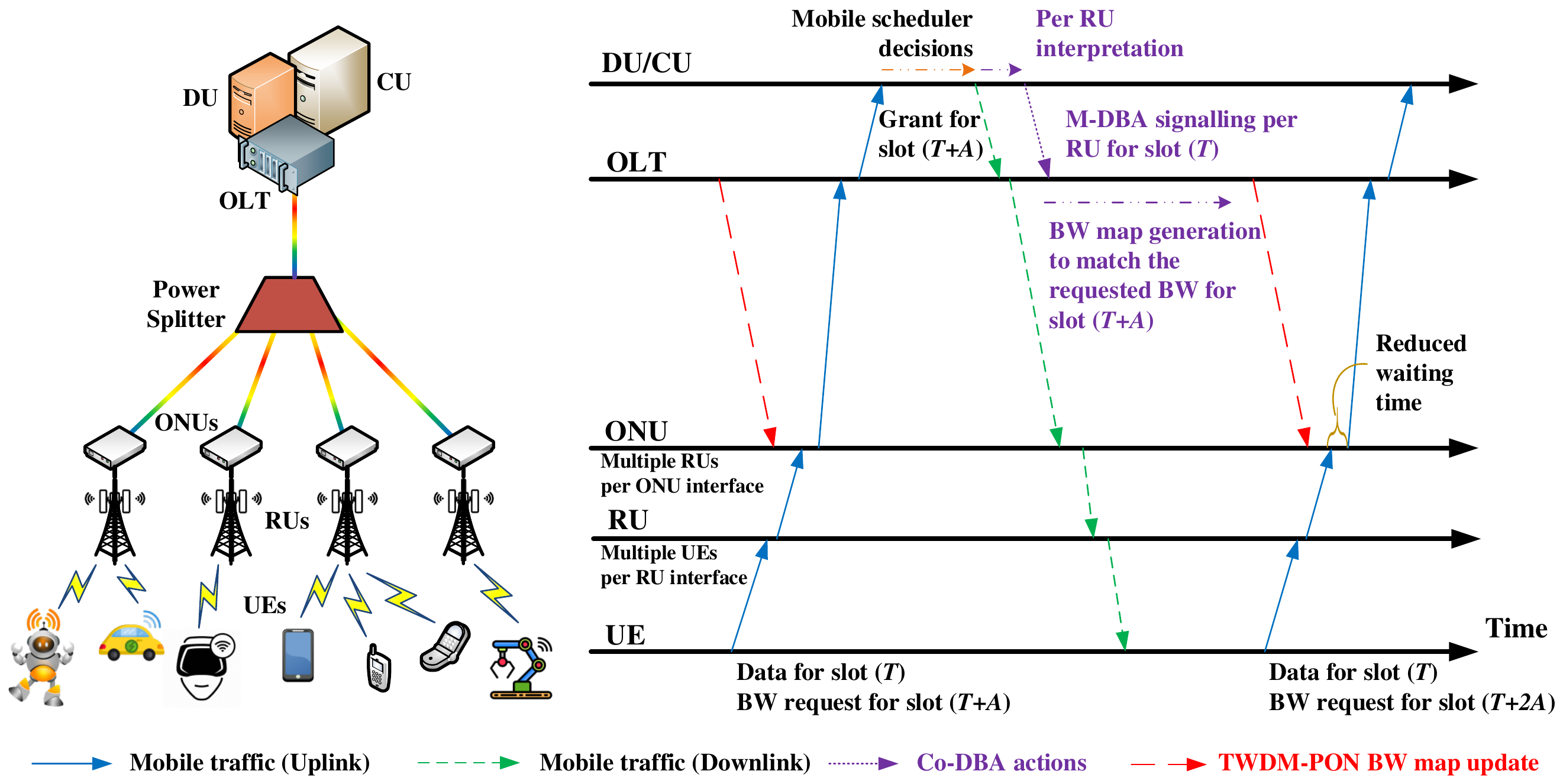}
\caption{The uplink data transmission scheme from UEs to RU, DU, and CU by cooperative M-DBA in TWDM-PON front-haul.}
\label{co_dba}
\end{figure}\setlength{\textfloatsep}{5pt}
\vspace{-\baselineskip}
\subsection{Novel Scheduling Techniques for Ensuring Low-latency}
Ensuring low-latency for uRLLC applications appears to be a critical challenge, but 5G new radio (NR) attempts to solve this by adopting some key techniques, such as frequent data transmission opportunities to minimize waiting time, flexible transmission duration, short UE processing time, short base-band processing time, and grant-free uplink transmission \cite{5g_kpi}. A \emph{numerology-based frame structure} is introduced in 5G NR that allows a flexible set of sub-carrier spacing, i.e., 15, 30, 60, 120, 240, 480 kHz. To achieve a shorter transmission time, \emph{mini-slots} can also be used that contain 2, 4, or 7 OFDM symbols instead of the traditional slot of 14 OFDM symbols \cite{5g_kpi}. For scheduling high-priority uRLLC traffic, especially in uplink, \emph{instant scheduling} and \emph{grant-free scheduling} techniques are used \cite{5g_ul}. In instant scheduling, whenever uRLLC traffic arrives, the ongoing eMBB or mMTC traffic is preempted and uRLLC traffic is scheduled. The grant-free scheduling is adopted where uplink resources are reserved for a single UE (periodic traffic) or a group of contending UEs (non-periodic traffic). To avoid collisions, UEs are allowed to transmit multiple replicas of each packet.\par 
The aggregated traffic from a group of UEs accesses the network through RUs, which are connected with ONUs that use TWDM-PON as front-haul, as shown in Fig. \ref{co_dba}. Usually, a set of dedicated wavelengths in a TWDM-PON are used for catering to the stringent front-haul traffic requirements. A cooperative M-DBA is used to schedule upstream ONU traffic according to the transmit time interval (TTI) cycles so that as soon as the data from UEs reach RU, it goes to ONU, and then, is directly transmitted to OLT and DU/CU with minimal waiting time \cite{5g_fh_bw3}. In this protocol, while sending data in a given slot $(T)$, UEs request wireless capacity for a future slot, say $(T+A)$. Accordingly, the DU takes scheduling decisions and notifies each UE about the allocated wireless resources for the future slot $(T+A)$. In parallel, the DU calculates the corresponding fronthaul traffic load per RU, based on the scheduling allocations for the corresponding UEs. The DU notifies the traffic load per RU for the future slot to the OLT and the OLT generates a bandwidth map for the future slot $(T+A)$ corresponding to the traffic identifier.
\vspace{-0.5\baselineskip}
\subsection{Communication and RU-DU-CU Processing Models}
To propose a TWDM-PON-based O-RAN planning framework, it is important to consider appropriate models for the communication and computation aspects of the network. In this O-RAN front/mid-haul design framework, we formulate an offline optimization problem for the network deployment stage, where we need to find the maximum possible throughput of a 5G NR RU by the following formula \cite{5g_kpi}:\par
\begin{resizealign}
    W_b = \sum_{p=1}^{N_c} \left(a^{(p)}\nu^{(p)}Q_m^{(p)}f^{(p)}R_{max} \frac{12\times N_{PRB}^{BW(p),\mu}}{T_s^\mu} (1-OH^{(p)}) \right), \label{eq01}
\end{resizealign}
\par \hspace{-1em}where, the number of aggregated carriers in a frequency band is denoted by $N_c$, the percentage of resources used for uplink/downlink transmissions in the carrier $p$ is denoted by $a^{(p)}$, the number of MIMO layers is denoted by $\nu^{(p)}$, the modulation order is denoted by $Q_m^{(p)}$, the capability mismatch between base-band and RF for UEs is denoted by $f^{(p)}$, the coding rate is denoted by $R_{max}$, the chosen numerology is denoted by $\mu$, the average symbol duration in seconds is denoted by $T_s^\mu$, and the maximum physical resource block (PRB) in the available bandwidth of a single carrier is denoted by $N_{PRB}^{BW(p),\mu}$, and $OH^{(p)}$ denotes the overhead. However, the downlink throughput of a UE is dependent on the power transmitted by the connected RU and the path loss. We can calculate the throughput of UE-$u$ as follows \cite{on_off}:
\begin{align}
    W_u = B\times \log_2 \left(1+\frac{P_b \mathcal{L}(d_{ub})}{\sigma^2 + \sum_{b'\neq b} P_{b'} \mathcal{L}(d_{ub'})} \right), \label{eq02}
\end{align}
where, $B$ denotes the bandwidth, $P_b$ denotes the power transmitted by RU-$b$, $\sigma^2$ denotes the noise power, and $\mathcal{L}(d_{ub})$ denotes the path loss in dB. The considered models for path loss of macro and small-cell RUs against a distance $d$ (km) are respectively given by (derived from \cite{3Gpp_PL}):
\begin{align}
    \mathcal{L}_M(d) &= 128.1 + 37.6\times \log_{10}(d), \label{eq03} \\
    \mathcal{L}_S(d) &= 37 + 30\times \log_{10}(d). \label{eq04}    
\end{align}
\par A similar model can be used for the uplink throughput of the UEs. If we know the maximum throughput requirements of UEs from any slice, then we can find the maximum allowable UE to RU distance of that slice by using (\ref{eq02})-(\ref{eq04}). In a network planning problem, we need to consider the long term network statistics, but mobile user throughput requirement shows \emph{spatio-temporal} and \emph{tidal wave characteristics}, i.e., UE density varies over geography and peak throughput demand arises only during certain hours of a day \cite{data_mi}. Therefore, if we consider the maximum throughput of all UEs, then there will be severe under-utilization of resources at low-throughput hours. Thus, we use the average UE throughput demand by considering a \emph{semi-static service request model}, i.e., each UE requests a constant amount of resources for its desired service over an interval of time \cite{o-ran}.\par 
Recently, the IEEE 802.1CM standard was developed for front/mid-haul interfaces \cite{IEEE_802.CM} and is compatible with Ethernet-based TWDM-PON standards. When UEs are attached to RUs, the corresponding DU and CU processing data in the front/mid-haul interfaces are transmitted as periodic bursts of Ethernet frames. Data is transmitted as undivided blocks in the network \cite{oran_opt}. For a burst interval of $\delta_t$ (sec), the number of frames in a burst, denoted by $\mathcal{B}$, can be calculated as $\mathcal{B} = \lceil R_D\times \delta_t/\mathcal{P}\rceil$, where $R_D$ denotes required data rate and $\mathcal{P}$ denotes the payload size of an Ethernet frame (up to 1500 Bytes) \cite{gabriel}. Hence, the actual throughput of a flow can be calculated by $(\mathcal{B}\times F/\delta_t)$, where $F$ is the Ethernet frame size (1542 Bytes, assuming the best case of maximum packet size). The total computational effort for the BBU functions (here BBU indicates the total digital processing across RU, DU, and CU) per TTI in giga operations per second (GOPS) is given by the following expression \cite{F_split}:
\begin{align}
    \mathcal{C}_{BBU} = \left(3\mathcal{A} + \mathcal{A}^2 + \frac{1}{3} \times \mathcal{M}\times \Psi\times \mathcal{S}\right)\times \frac{\Phi}{10}, \label{eq05}
\end{align}
where, $\mathcal{A}$ denotes the number of MIMO antennas, $\mathcal{M}$ denotes the number of modulation bits, $\Psi$ denotes the coding rate, $\mathcal{S}$ denotes the number of MIMO layers, and $\Phi$ denotes the number of PRB. We distribute this total computational effort $\mathcal{C}_{BBU}$ among RU, DU, and CU based on the ``PHY split" and ``RLC-PDCP split" in \cite{mgain}. With the considered split-7.2 (RU-DU) and split-2 (DU-CU), 40\% processing is done by RU, 50\% processing is done by DU, and 10\% processing is done by CU. Note that the RU functions are typically implemented on dedicated hardware rather than on open access-edge servers, and this will be considered in our optimization problem formulation. The total RU-DU-CU function processing time depends on the number of PRBs, the MCS index, and the CPU frequency and can be computed by the polynomial expression provided in \cite{bbu_lat}. Accordingly, we can choose a maximum limit for the sum of RU, DU, and CU processing latencies.

\vspace{-\baselineskip}
\subsection{Optimal TWDM-PON-based Front/Mid-haul Design}
In a practical RAN deployment scenario, UEs are randomly distributed across certain geographic areas and can be classified as either of uRLLC, eMBB, and mMTC application users depending on their throughput and latency demands. These UEs require a set of end-to-end resources from RUs, DUs, and CUs. This chain of resources is known as \emph{service function request (SFR)} \cite{o-ran}. Therefore, to accommodate all SFRs generated from all UEs in the network design and planning stage, firstly we need to identify a set of possible 5G RU locations belonging to uRLLC, eMBB, and mMTC slices over the considered area such that each UE can be connected to one RU only. The RU locations are also the possible locations for the ONUs of Stage-I TWDM-PONs. In addition, we also need to identify a set of locations for OLTs for Stage-I and Stage-II TWDM-PONs and placement locations of open access-edge servers for hosting DUs and CUs. Our objective is to install a minimum number of 5G RUs, OLTs, and open access-edge servers because this minimizes the overall cost of 5G O-RAN deployment. Nonetheless, we must ensure that the front/mid-haul communication and total RU, DU, and CU processing latencies of the uRLLC, eMBB, and mMTC slices are satisfied.\par 
Several UE to base station association mechanisms based on dynamic resource requirements exist in literature for heterogeneous wireless networks \cite{asc1,user_game} as well as C-RAN \cite{rrh-bbu3,yao}. However, we are formulating a UE to RU association problem for O-RAN with long-term network statistics. Firstly, we attempt to find an optimal set of RU locations based on the long-term average SFR requirements of UEs, and secondly, these optimal RU locations are used to find a minimum number of TWDM-PON interconnections between ONUs and OLTs. Besides, we want to install a minimum number of open access-edge servers at RU or Stage-I ONU, Stage-I OLT, and Stage-II OLT locations for hosting DUs and CUs. We incorporate flexible DU and CU deployment in our formulation, but the DUs and CUs corresponding to each SFR must be placed at one location only. Although the front/mid-haul deployment for the O-RAN problem can be formulated in several possible ways, the primary benefit of this two-stage ILP formulation is that the optimal solution can be evaluated by commercial solvers with a reasonably large dataset. After the ILP formulations, we employ this framework to design TWDM-PON-based front/mid-haul against urban, rural, and industrial scenarios. 

\begin{table}[!t]
\centering
\caption{Access-Edge Cloud Network Parameters and Sets}
\label{table1}
\resizebox{\columnwidth}{!}{%
\begin{tabular}{cl}
\toprule
\textbf{Symbol}      & \multicolumn{1}{c}{\textbf{Definition}}                                    \\ \toprule
$\mathcal{S}$        & Set of all possible network slices                                         \\ \hline
$\mathcal{U}_{5G_s}$ & Set of 5G NR UEs in slice $s$                                              \\ \hline
$\mathcal{B}_s$      & Set of possible 5G RU locations in slice $s$                              \\ \hline
$\mathcal{O}$        & Set of possible OLT locations of Stage-I TWDM-PONs                         \\ \hline
$\mathcal{Q}$        & Set of possible OLT locations of Stage-II TWDM-PONs                        \\ \hline
$D_{u_s b_s}$        & Distance between UE-$u_s\in\mathcal{U}_{5G_s}$ and RU-$b_s\in\mathcal{B}_s$ \\ \hline
$D_{b_s r_I}$          & Distance between RU-$b_s$ and RN-$r_I$ of Stage-I TWDM-PON                  \\ \hline
$D_{r_I o}$             & Distance between RN-$r_I$ and OLT-$o$ of Stage-I TWDM-PON                    \\ \hline
$D_{or_{II}}$             & Distance between Stage-I OLT-$o$ and RN-$r_{II}$ of Stage-II TWDM-PON           \\ \hline
$D_{r_{II}q}$             & Distance between RN-$r_{II}$ and OLT-$q$ of Stage-II TWDM-PON                   \\ \hline
$D_{max}^e$          & Maximum allowable distance of $e\in\{b_s,o,q\}$                             \\ \hline
$W_{b_s}$            & Maximum supported throughput of the RU-$b_s$                              \\ \hline
$W_{u_s}$            & Throughput requirement of the UE-$u_s$                                          \\ \hline
$R_o$                & Maximum supported throughput of the Stage-I TWDM-PON                       \\ \hline
$R_q$                & Maximum supported throughput of the Stage-II TWDM-PON                      \\ \hline
$\eta_{b_s,RU}$          & Required computation resources for RU from RU-$b_s$                       \\ \hline
$H_{b_s}$            & Maximum available computation resources for RU at RU-$b_s$                \\ \hline
$\Gamma_{b_s,DU}$    & Required computation resources for DU of RU-$b_s$                       \\ \hline
$GD_{b_s}$           & Maximum available computation resources for DU at RU-$b_s$                \\ \hline
$GD_{o}$             & Maximum available computation resources for DU at Stage-I ONU-$o$          \\ \hline
$\Gamma_{b_s,CU}$    & Required computation resources for CU of RU-$b_s$                       \\ \hline
$GC_{o}$             & Maximum available computation resources for CU at Stage-I OLT-$o$          \\ \hline
$GC_{q}$             & Maximum available computation resources for CU at Stage-II OLT-$q$         \\ \hline
$\Delta_s^{OTA}$     & Over-the-air latency requirement of slice-$s$         \\ \hline
$\Delta_{b_s}^{BBU}$ & Total BBU processing latency requirement of RU-$b_s$  \\ \hline
$\delta^{TTI}$       & Transmit time interval time of the UEs  \\ \hline
$\delta_{b_s o}/\delta_{oq}$ & The reduced waiting time for data ONUs at locations $b_s$ and $o$ \\ \hline
$v_c$                  & Speed of light in optical fiber ($2\times 10^8$ m/s)                       \\ \hline
$c$                  & Speed of electromagnetic wave in the air ($3\times 10^8$ m/s)              \\ \bottomrule
\end{tabular}
}
\end{table}
\setlength{\textfloatsep}{5pt}

\section{Optimization Problem Formulations} \label{sec4}
In this section, firstly, we formulate an ILP for associating mobile users and RUs based on the long-term network statistics of uRLLC, eMBB, and mMTC applications. Once all the optimal RU locations for different slices are known, we formulate a second ILP to find the optimal TWDM-PON-based front/mid-haul links and placement locations of open access-edge servers to host the respective DUs and CUs.
\vspace{-0.5\baselineskip}
\subsection{Mobile User Equipment and RU Association Problem}
We consider that the mobile UEs can be sliced into $S$ sets according to their QoS throughput and latency requirements. Let $\mathcal{U}_{5G_s} = \{1,2,\dots,U_s\}$ denote the set of 5G NR UEs in slice $s\in \mathcal{S}=\{1,2,\dots,S\}$, where $U_s$ denotes the maximum number of UEs in slice $s$. The average uplink and downlink throughput demands of each UE-$u_s \in \mathcal{U}_{5G_s}$ over its daily active hours are denoted by $W^{UL}_{u_s}$ and $W^{DL}_{u_s}$, respectively. Furthermore, $\mathcal{B}_s = \{1, 2, \dots, B_s\}$ denotes the set of possible 5G RU locations corresponding to slice $s$. The maximum uplink and downlink throughput supported by each RU-$b_s\in \mathcal{B}_s$ are denoted by $W^{UL}_{b_s}$ and $W^{DL}_{b_s}$, respectively. Nonetheless, RUs from all slices will cover the total area so that UEs from all slices can be randomly distributed all over the area. The distance between UE-$u_s$ and 5G RU-$b_s$ are denoted by $D_{u_s b_s}$ and the maximum allowable UE-RU distance for slice $s$ is denoted by $D_{b_s}^{max}$. Note that all the considered network parameters are described in Table \ref{table1} and the decision variables are defined in Table \ref{table1x}.\par
\textbf{Objective:} We consider binary variables $\theta_{b_s}$ to indicate if every 5G RU-$b_s$ from slice $s$ are installed. We also consider binary variables $x_{u_sb_s}$ to indicate if a UE-$u_s$ is connected to a RU-$b_s\in\mathcal{B}_s$ from slice $s$. As we want to install minimum number of RUs, the objective is given as:\par
\begin{resizealign}
    \mathcal{P}_1: \min_{\bm{\theta}_{b},\bm{x}} \quad \alpha\sum_{s}\sum_{b_s} \theta_{b_s} + \beta \sum_{s}\sum_{b_s}\sum_{u_s} \left(\mathcal{T}_{u_s b_s}^{UL} + \mathcal{T}_{u_s b_s}^{DL}\right), \label{eq06}
\end{resizealign} \vspace{-0.5\baselineskip}
\par \hspace{-1em}where, $\mathcal{T}_{u_s b_s}^{UL}$ and $\mathcal{T}_{u_s b_s}^{DL}$ represent the uplink and downlink over-the-air (OTA) latencies, $\alpha = 1/(\sum_s B_s)$ unit and $\beta = 1/(\sum_s U_s)$ unit/sec are weight factors to impose a very low importance on latency than the number of RUs as $\alpha \gg \beta$. Although the OTA latency upper bound is guaranteed in (\ref{eq10})-(\ref{eq11}), we are adding the sum of latencies in (\ref{eq06}) for efficiently designing a Lagrangian relaxation-based heuristic. Nonetheless, a problem formulation without the second term in the objective should also produce the same optimal solution.\par
\textbf{Connectivity constraints:} The binary variables $x_{u_sb_s}$ variables can take value 1 only if UE-$u_s$ is within the coverage distance of RU-$b_s$ as follows:
\begin{align}
    x_{u_sb_s} &\leq \left\lfloor \frac{D^{max}_{b_s}}{D_{u_sb_s}} \right\rfloor, \forall s\in \mathcal{S}, u_s\in \mathcal{U}_{5G_s}, b_s \in \mathcal{B}_s. \label{eq07}
\end{align}
\par We can determine the maximum allowed distance $D^{max}_{b_s}$ for each slice $s\in\mathcal{S}$ by using the maximum throughput, transmit power, and noise power (-174 dBm/Hz) in (\ref{eq02})-(\ref{eq04}). Each UE-$u_s$ can be associated with an RU-$b_s$ only if it is installed, which is ensured as follows:
\begin{align}
    \theta_{b_s} \geq x_{u_sb_s}, \forall s\in \mathcal{S}, u_s \in \mathcal{U}_{5G_s}, b_s \in \mathcal{B}_s. \label{eq08}
\end{align}
\par Moreover, each UE-$u_s$ accesses the RAN through some RU-$b_s$ and we consider that each UE is associated with only one RU. This is guaranteed by the following constraint:
\begin{align}
    \sum_{b_s\in\mathcal{B}_s} x_{u_sb_s} = 1, \forall s\in \mathcal{S}, u_s \in \mathcal{U}_{5G_s}. \label{eq09}
\end{align}
\par \textbf{Latency constraints:} In this work, we consider both the uplink and downlink throughput demands while UE to RU association. Thus, a group of UEs can be associated with a RU only when the uplink and downlink OTA latency bound ($\Delta_s^{OTA}$) of the corresponding slice $s$ can be supported by that RU-$b_s$ while transmitting the data generated in each TTI duration ($\delta^{TTI}$). To ensure this, we introduce the following constraints $\forall s\in \mathcal{S}, u_s \in \mathcal{U}_{5G_s}, b_s\in\mathcal{B}_s$:\par
\begin{resizealign}
    \mathcal{T}_{u_s b_s}^{UL} &= \left\{\frac{x_{u_s b_s} D_{u_s b_s}}{c} \right\} + \left\{\frac{\sum_{u_s} x_{u_s b_s} W_{u_s}^{UL}\delta^{TTI}}{W_{b_s}^{UL}}\right\}  \leq \Delta_s^{OTA}, \label{eq10}\\
    \mathcal{T}_{u_s b_s}^{DL} &= \left\{\frac{x_{u_s b_s} D_{u_s b_s}}{c}\right\} + \left\{\frac{\sum_{u_s} x_{u_s b_s} W_{u_s}^{DL}\delta^{TTI}}{W_{b_s}^{DL}}\right\} \leq \Delta_s^{OTA}, \label{eq11}
\end{resizealign}
%
\par \hspace{-1em}where, the first terms of (\ref{eq10})-(\ref{eq11}) denote the \emph{wireless signal propagation latency} and the second terms denote the \emph{data transmission latency} in each TTI. The maximum available uplink and downlink throughput of each RU-$b_s$ can be determined from (\ref{eq01}) and these constraints implicitly ensure that the throughput requirements of all the UEs are not more than the total available wireless throughput. Clearly, the above-formulated optimization problem is an ILP with a linear objective function and linear constraints.
\begin{table}[!t]
\centering
\caption{Optimization Decision Variables}
\label{table1x}
\resizebox{\columnwidth}{!}{%
\begin{tabular}{cl}
\toprule
\textbf{Variable}      & \multicolumn{1}{c}{\textbf{Definition}}                        \\ \toprule
$\theta_{b_s}$         & indicates if RU-$b_s$ from slice $s$ are installed (binary)              \\ \hline
$x_{u_sb_s}$           & indicates if a UE-$u_s$ is connected to a RU-$b_s$ (binary)     \\ \hline
$\mathcal{T}_{u_s b_s}^{UL}$    & indicates the uplink over-the-air latency (non-negative)                            \\ \hline
$\mathcal{T}_{u_s b_s}^{DL}$    & indicates the downlink over-the-air latency (non-negative)                      \\ \hline
$\theta_o$                 & indicates if a Stage-I OLT is installed (binary)                        \\ \hline
$\theta_q$                 & indicates if a Stage-II OLT is installed (binary)                   \\ \hline
$y_{b_s o}$                & indicates if RU-$b_s$ is connected to Stage-I OLT-$o$ (binary)            \\ \hline
$z_{oq}$                   & indicates if Stage-I OLT-$o$ is connected to Stage-II OLT-$q$ (binary)    \\ \hline
$\omega_{b_s}^d$           & indicates if DU of slice $s$ is installed at RU-$b_s$ (binary)                                   \\ \hline
$\omega_{os}^d$                 & indicates if DU of slice $s$ is installed at Stage-I OLT-$o$ (binary)                   \\ \hline
$\omega_{os}^c$                 & indicates if CU of slice $s$ is installed at Stage-I OLT-$o$ (binary)                          \\ \hline
$\omega_{qs}^c$                  & indicates if CU of slice $s$ is installed at Stage-II OLT-$q$ (binary)                       \\ \hline
$\beta_{os}$                 & indicates the number of RUs connected to each OLT-$o$ (integer)                           \\ \hline
$A_{oqs}^c$                  & linearizes the integer and binary product term $(\beta_{os} \times K_{oqs}^c)$          \\ \hline
$Z_{oqs}^c$                  & linearizes the integer and binary product term $(\beta_{os} \times L_{oqs}^c)$          \\ \hline
$\chi_{b_so}^d$           & linearizes binary variable product term $(\omega_{b_s}^d \times y_{b_s o})$                           \\ \hline
$\Upsilon_{b_so}^d$       & linearizes binary variable product term $(\omega_{os}^d \times y_{b_s o})$                           \\ \hline
$K_{oqs}^c$               & linearizes binary variable product term $(\omega_{os}^c \times z_{oq})$                           \\ \hline
$L_{oqs}^c$               & linearizes binary variable product term $(\omega_{oqs}^c \times z_{oq})$                           \\ \hline
$\mathcal{T}_{b_s o}^{UL}$      & indicates the uplink Stage-I TWDM-PON latency (non-negative)                           \\ \hline
$\mathcal{T}_{b_s o}^{DL}$      & indicates the downlink Stage-I TWDM-PON latency (non-negative)                           \\ \hline
$\mathcal{T}_{oq}^{UL}$         & indicates the uplink Stage-II TWDM-PON latency (non-negative)                           \\ \hline
$\mathcal{T}_{oq}^{DL}$         & indicates the downlink Stage-II TWDM-PON latency (non-negative)                           \\ \hline
$\mathcal{T}_{rdc}^{UL}$      & indicates the uplink BBU processing latency (non-negative)                           \\ \hline
$\mathcal{T}_{rdc}^{DL}$      & indicates the downlink BBU processing latency (non-negative)             \\ \bottomrule
\end{tabular}
}
\end{table}
\setlength{\textfloatsep}{5pt}
\subsection{Lagrangian Relaxation Heuristic for UE-RU Association}
As the locations of the RUs are unknown, the problem of associating UEs to RUs is an \emph{NP-hard problem} \cite{mobility_garcia}. A general observation was made around the 1970s that many hard integer programming problems can be considered as easy problems but are complicated by a small set of side constraints. Since then, \emph{Lagrangian relaxation} has gained popularity as the best existing algorithm for problems of this type \cite{LagRel2}. Thus, to solve the UE to RU association problem, we design a Lagrangian heuristic algorithm by relaxing constraints (\ref{eq09}). This Lagrangian relaxation problem is summarized as:\par
\begin{resizealign}
    \mathcal{R}_1: \quad \max_{\bm{\nu}} \min_{\bm{\theta}_{b},\bm{x}} \quad\alpha\sum_{s}\sum_{b_s} \theta_{b_s} + \beta \sum_{s}\sum_{b_s}\sum_{u_s} \left(\mathcal{T}_{u_s b_s}^{UL} + \mathcal{T}_{u_s b_s}^{DL}\right) \nonumber\\
    \quad\quad\quad\quad\quad\quad +\sum_{s}\sum_{u_s} \nu_{u_s} \left(1-\sum_{b_s} x_{u_s b_s} \right)  \label{eq35}
\end{resizealign}
\par \vspace{-\baselineskip}
\begin{align}
    \text{subject to } & (7), (8), (10), (11), \nonumber\\
    & \theta_{b_s} \in \{0,1\}, \forall s\in \mathcal{S}, b_s\in\mathcal{B}_s, \label{eq36} \\
    & x_{u_s b_s} \in \{0,1\}, \forall s\in \mathcal{S}, u_s \in \mathcal{U}_{5G_s}, b_s\in\mathcal{B}_s, \label{eq37}
\end{align}
\par \hspace{-1em}where, $\nu_{u_s}$ are Lagrange multipliers used for relaxing constraints (\ref{eq09}). For fixed values of these multipliers, the relaxed problem $\mathcal{R}_1$ will yield an optimal value of the objective that serves as a \emph{lower bound (LB)} of the original problem $\mathcal{P}_1$.\par
\begin{proposition} \label{prp1}
    The solutions of the Lagrangian relaxation problem $\mathcal{R}_1$ with given $\nu_{u_s}$ are given by:
    \begin{align*}
        x_{u_s b_s} &= \begin{cases}
              1, & \text{if } \left\{\frac{2\beta D_{u_s b_s}}{c} + \frac{\beta U_s W_{u_s}^{UL}\delta^{TTI}}{W_{b_s}^{UL}} \right.\\
              & \left. \quad + \frac{\beta U_s W_{u_s}^{DL}\delta^{TTI}}{W_{b_s}^{DL}} - \nu_{u_s} \right\} <0 \text{ and } \theta_{b_s} = 1,\\
              0, & \text{otherwise},
              \end{cases}
    \end{align*}
    \begin{align*}
        \theta_{b_s} &= \begin{cases}
              1, & \text{if } \alpha+\sum_{u_s}\min\left\{0,\frac{2\beta D_{u_s b_s}}{c} + \frac{\beta U_s W_{u_s}^{UL}\delta^{TTI}}{W_{b_s}^{UL}} \right.\\
                 &\left. \quad\quad\quad\quad\quad + \frac{\beta U_s W_{u_s}^{DL}\delta^{TTI}}{W_{b_s}^{DL}} - \nu_{u_s} \right\} < 0, \\
              0, & \text{otherwise}.
              \end{cases}
    \end{align*}
\end{proposition}
\begin{IEEEproof}
    For given values of Lagrange multipliers $\nu_{u_s} \forall s$, we can minimize the objective function by preferably choosing $x_{u_s b_s} = 1$, if the corresponding coefficient is $\left\{\frac{2\beta D_{u_s b_s}}{c} + \frac{\beta U_s W_{u_s}^{UL}\delta^{TTI}}{W_{b_s}^{UL}} + \frac{\beta U_s W_{u_s}^{DL}\delta^{TTI}}{W_{b_s}^{DL}} - \nu_{u_s} \right\} <0$. Otherwise, it is best to set $x_{u_s b_s} = 0$ \cite{LagRel2}. Note that the weight factor $\beta = 1/(\sum_s U_s)$. However, setting $x_{u_s b_s} = 1$ implies that we must also set $\theta_{b_s} = 1$ according to constraint (\ref{eq08}), i.e., $\theta_{b_s} \geq x_{u_sb_s}, \forall s\in \mathcal{S}, u_s \in \mathcal{U}_{5G_s}, b_s \in \mathcal{B}_s$. Now, if we set $\theta_{b_s} = 1$, then we add an excess value of $\Delta f = \alpha + \sum_{u_s}\min\left\{0,\frac{2\beta D_{u_s b_s}}{c} + \frac{\beta U_s W_{u_s}^{UL}\delta^{TTI}}{W_{b_s}^{UL}} + \frac{\beta U_s W_{u_s}^{DL}\delta^{TTI}}{W_{b_s}^{DL}}- \nu_{u_s} \right\}$ to the objective value. If only $\Delta f$ is negative, the objective value is reduced. Hence, we set $\theta_{b_s} = 1$ when $\Delta f <0$.
\end{IEEEproof}
\par An optimal solution for the problem $\mathcal{R}_1$ can be obtained by exploiting Proposition \ref{prp1}. However, it acts only as an LB of the problem $\mathcal{P}_1$, denoted by $\mathcal{P}_{lb}$, and may not be feasible because (\ref{eq09}) is relaxed. Hence, we want to find a feasible solution that acts as the \emph{upper bound (UB)} for $\mathcal{P}_1$. Nonetheless, a feasible solution may not always guarantee optimality \cite{LagRel2}. Therefore, we need to explore other feasible solutions whose performance are not worse than the best-known UB. To obtain a UB, denoted as $\mathcal{P}_{ub}$, we formulate the following problem:\par
\begin{resizealign}
    \mathcal{R}_2: \quad \min_{\bm{\theta}_{b},\bm{x}} \sum_{s}\sum_{b_s} \theta_{b_s} + \sum_{s}\sum_{b_s}\sum_{u_s} \left(\mathcal{T}_{u_s b_s}^{UL} + \mathcal{T}_{u_s b_s}^{DL} -2\Delta_s^{OTA}\right)  \label{eq38}
\end{resizealign} \par \vspace{-\baselineskip}
\begin{align}
    \text{\hspace{-4em}subject to}\quad  (7), (8), (9), (13), (14). \nonumber
\end{align}
\par In $\mathcal{R}_2$, the coupling constraint (\ref{eq08}) and the integrality constraints (\ref{eq36})-(\ref{eq37}) are not relaxed. Note that the latency term is modified to $\beta \sum_{s}\sum_{b_s}\sum_{u_s} (\mathcal{T}_{u_s b_s}^{UL} + \mathcal{T}_{u_s b_s}^{DL} - 2\Delta_s^{OTA})$ in the objective, which is equivalent to relaxing (\ref{eq10})-(\ref{eq11}). Now, we design a heuristic for $\mathcal{R}_2$ to check the feasibility of solution from $\mathcal{R}_1$. We define a term $\Delta w_{u_s} = (\mathcal{T}_{u_s b'_s}^{UL} + \mathcal{T}_{u_s b'_s}^{DL})-(\mathcal{T}_{u_s b_s}^{UL} + \mathcal{T}_{u_s b_s}^{DL})$, where $b'_s$ is the initially assigned RU of UE-$u_s$ and $b_s$ is the re-assigned RU of UE-$u_s$. We want to associated each UE-$u_s$ to only one RU-$b_s$ such that $\Delta w_{u_s}$ is maximized subject to (\ref{eq07}), because this minimizes the objective (\ref{eq38}). We use $\theta_{b_s}^{lb}$ from $\mathcal{R}_1$ but if a UE could not be attached to any RU, then we declare the problem as infeasible. The Lagrangian relaxation method for UE-RU association is summarized in Algorithm \ref{alg1}.\par
The original problem $\mathcal{P}_1$ always considers the UB as its best objective value because the UB can guarantee a feasible solution. Nonetheless, different values of Lagrange multipliers can produce different UBs and LBs. Thus, we iteratively adjust the values of $\nu_{u_s}$ using the \emph{sub-gradient method} \cite{Bartsekas} and terminate when the UB and LB are close to each other or the maximum number of iterations ($n_{max}$) is reached. The values of $\nu_{u_s}$ are updated as follows:\par
\begin{resizealign}
    \nu_{u_s}^{(n+1)} = \nu_{u_s}^{(n)} + \sigma_s^{(n)}\left(1-\sum\nolimits_{b_s} x_{u_s b_s}^{(n)} \right), \forall s\in \mathcal{S}, u_s \in \mathcal{U}_{5G_s}, \label{eq39}
\end{resizealign}
\par \hspace{-1em}where, $\sigma_s^{(n)}$ is the step size at the $n$-th iteration, given by:
\begin{align}
    \sigma_s^{(n)} = \frac{\lambda^{(n)} (\mathcal{P}_{opt}-\mathcal{P}_{lb}^{(n)})}{\sum_{u_s}\left(1-\sum_{b_s} x_{u_s b_s}\right)^2}, \forall s\in \mathcal{S},\label{eq40}
\end{align}

\begin{algorithm}[t!]
\caption{Heuristic Algorithm for UE-RU Assignment} \label{alg1}
\hspace*{\algorithmicindent} \textbf{Input:} $\mathcal{S}, \mathcal{U}_{5G_s}, \mathcal{B}_s, D_{b_s}^{max}, D_{u_s b_s}, W_{u_s}^{DL}, W_{u_s}^{UL}, W_{b_s}^{DL}, W_{b_s}^{UL}$.\\
\hspace*{\algorithmicindent} \textbf{Output:} Near-optimal solution: $x_{u_s b_s}^*$, $\theta_{b_s}^*$, and $\mathcal{P}_{opt}^*$.\\
\hspace*{\algorithmicindent} \textbf{Initialize:} $\nu_{u_s}= 0$, $x_{u_s b_s} = 0$, $\theta_{b_s} = 0$, $\mathcal{P}_{lb} = 0$, \\
\hspace*{\algorithmicindent} $\mathcal{P}_{ub} = +\infty$, $\mathcal{P}_{opt} = +\infty$, $n = 1$, $feasible = 1$.
\begin{algorithmic}[1]
\While{$\mathcal{P}_{lb}^{(n)} \neq \mathcal{P}_{ub}^{(n)}$ and $n < n_{max}$}
    \StatePar{Find a LB solution $x_{u_s b_s}^{lb}$ and $\theta_{b_s}^{lb}$ by solving $\mathcal{R}_1$ using Proposition \ref{prp1};} \label{line2}
    \State Calculate the LB objective value $\mathcal{P}_{lb}^{(n)}$; \label{line3}
    \State $\theta_{b_s}^{ub} \gets \theta_{b_s}^{lb}, \forall s\in \mathcal{S}, b_s\in\mathcal{B}_s$;
    \For{Each UE-$u_s, \forall s\in \mathcal{S}$} \Comment{\textit{heuristic for} $\mathcal{R}_2$} \label{line5}
        \StatePar{Find a set $\{b_s|\theta_{b_s}^{ub}=1, D_{u_s b_s} \leq D_{b_s}^{max}$, $\mathcal{T}_{u_s b_s}^{UL} \leq \Delta_s^{OTA}$, $\mathcal{T}_{u_s b_s}^{DL} \leq \Delta_s^{OTA}\}$;}
        \If{$|\{b_s\}| = 0$}
            \State $feasible \gets 0$ and $\mathcal{P}_{ub}^{(n)} \gets +\infty$;
            \State break;
        \ElsIf{$|\{b_s\}| = 1$}
            \State Assign UE-$u_s$ to RU-$b_s$;
            \State $x_{u_s b_s}^{ub} \gets 1$;
        \ElsIf{$|\{b_s\}| > 1$}
            \State Find RU-$b_s= \arg\max_{b_s} \{\Delta w_{u_s}\}$ for UE-$u_s$;
            \State $x_{u_s b_s}^{ub} \gets 1$;
        \EndIf
    \EndFor
    \If{$feasible = 0$}
        \State \textbf{return} Infeasible;
    \EndIf
    \State Calculate the UB objective value $\mathcal{P}_{ub}^{(n)}$; \label{line22}
    \If{$\mathcal{P}_{ub}^{(n)} < \mathcal{P}_{opt}$} \Comment{\textit{Update if UB is better}} \label{line23}
        \State $x_{u_s b_s} \gets x_{u_s b_s}^{ub}, \forall s\in \mathcal{S}, u_s \in \mathcal{U}_{5G_s}, b_s\in\mathcal{B}_s$;
        \State $\theta_{b_s} \gets \theta_{b_s}^{ub}, \forall s\in \mathcal{S}, b_s\in\mathcal{B}_s$;
        \State $\mathcal{P}_{opt} \gets \mathcal{P}_{ub}^{(n)}$;
    \EndIf
    \State Update step size $\sigma_s$ according to (\ref{eq40});
    \State Update Lagrangian multipliers $\nu_{u_s}$ by (\ref{eq39});
    \State $n \gets n+1$    \label{line30}
\EndWhile
\State \textbf{return} $x_{u_s b_s}$, $\theta_{b_s}$, and $\mathcal{P}_{opt}$;
\end{algorithmic}
\end{algorithm}

\hspace{-1em}with $\lambda^{(n)}$ as a decreasing scalar parameter and $\mathcal{P}_{opt}$ denotes the best-known solution or UB from previous iterations. Commonly $\lambda^{(n)}$ is chosen as a constant satisfying $0<\lambda^{(n)}\leq 2$ and then halved if $\mathcal{P}_{lb}^{(n)}$ remains constant for several iterations. However, as there is no well-known stopping criteria of this method, the Algorithm \ref{alg1} needs to be stopped after a finite number iterations $(n_{max})$ and there may be an optimality gap.
\vspace{-\baselineskip}
\setcounter{theorem}{0}
\begin{theorem}
    The optimality gap in the solution for $\mathcal{P}_1$ produced by the Algorithm \ref{alg1} tends to zero when the number of iterations $n\to\infty$.
\end{theorem}
\begin{IEEEproof}
    From the properties of Lagrangian relaxation \cite{netflow}, we can state that if for some Lagrangian multiplier vector $\bm{\nu}$, the solutions $\theta_{b_s}^*, x_{u_s b_s}^*$ from $\mathcal{R}_1$ is feasible in the optimization problem $\mathcal{P}_1$, then $\theta_{b_s}^*, x_{u_s b_s}^*$ is an optimal solution for $\mathcal{P}_1$. A feasibility certificate can be obtained from the solution of $\mathcal{R}_2$. However, as sub-gradient method is used to evaluate $\mathcal{R}_1$ and is terminated after a finite number of iterations, there may be an optimality gap between the true optimal solution $\mathcal{P}_{opt}^*$ and $\mathcal{P}_{lb}^{(n)}$. Note that, with our chosen step-size (\ref{eq40}), we ensure that $[\lambda^{(n)}(\mathcal{P}_{opt}-\mathcal{P}_{lb}^{(n)})] \geq 0$, $\lim_{n\to\infty} [\lambda^{(n)}(\mathcal{P}_{opt}-\mathcal{P}_{lb}^{(n)})] = 0$, and $\sum_{n=1}^{\infty}[\lambda^{(n)}(\mathcal{P}_{opt}-\mathcal{P}_{lb}^{(n)})] = \infty$. Now, the optimality gap in the solution produced by the sub-gradient method after $n=n_{max}$ (finite) iterations is given by \cite{Bartsekas}:
    \begin{align}
        \left|\mathcal{P}_{opt}^*-\mathcal{P}_{lb}^{(n)}\right| \leq \frac{R^2+\sum_{n=1}^{n_{max}}[\lambda^{(n)}(\mathcal{P}_{opt}-\mathcal{P}_{lb}^{(n)})]^2}{(2/G)\sum_{n=1}^{n_{max}}[\lambda^{(n)}(\mathcal{P}_{opt}-\mathcal{P}_{lb}^{(n)})]}, \label{n01}
    \end{align}
    where, $\|x_{u_s b_s}^{(1)}-x_{u_s b_s}^*\| \leq R$ and $\sum_{u_s}\left(1-\sum_{b_s} x_{u_s b_s}\right)^2 \leq G$. Thus, we can choose $R=1$ and $G=U_s$ for $\mathcal{P}_1$ to show the optimality gap in the solution produced by Algorithm \ref{alg1} as:
    \begin{align}
        \left|\mathcal{P}_{opt}^*-\mathcal{P}_{lb}^{(n)}\right| \leq \frac{1+\sum_{n=1}^{n_{max}}[\lambda^{(n)}(\mathcal{P}_{opt}-\mathcal{P}_{lb}^{(n)})]^2}{(2/U_s)\sum_{n=1}^{n_{max}}[\lambda^{(n)}(\mathcal{P}_{opt}-\mathcal{P}_{lb}^{(n)})]}. \label{n02}
    \end{align}
    \par Clearly, under the aforementioned conditions, the optimality gap $|\mathcal{P}_{opt}^*-\mathcal{P}_{lb}^{(n)}| \to 0$ as $n\to\infty$.
\end{IEEEproof}
After initialization of the Lagrangian multipliers, LB and the optimal values, lines \ref{line2}-\ref{line30} iteratively change the Lagrangian multipliers as well as update the LB and UB to find the optimal value $\mathcal{P}_{opt}$. In each iteration, the LB is evaluated by lines \ref{line2}-\ref{line3} with a complexity of $\texttt{O}(\sum_s U_s B_s)$ and the UB is evaluated by lines \ref{line5}-\ref{line22} with a complexity of $\texttt{O}(\sum_s U_s B_s)$. Updating Lagrangian multipliers in lines \ref{line23}-\ref{line30} involves a complexity of $\texttt{O}(\sum_s U_s)$. Therefore, the convergence rate of Algorithm 1 with an optimality gap of $|\mathcal{P}_{opt}^*-\mathcal{P}_{lb}^{(n)}|$ is $\texttt{O}(n_{max}(\sum_s U_s B_s))$ as $n_{max}$ is the maximum number of iterations. Note that, in general Algorithm \ref{alg1} can achieve a desired accuracy $|\mathcal{P}_{opt}^*-\mathcal{P}_{lb}^{(n)}| \leq \epsilon$ with a convergence rate of $\texttt{O}(1/\epsilon^2)$ \cite{Bartsekas}.

\vspace{-0.5\baselineskip}
\subsection{DU-CU Placement and Front/Mid-haul Design Problem}
We denote the set of Stage-I OLTs by $\mathcal{O} = \{1,2,\dots,O\}$ and the set of Stage-II OLTs by $\mathcal{Q} = \{1,2,\dots,q\}$. The RUs are connected to the Stage-I OLTs through a remote node (RN) where the power splitter is located. Similarly, the Stage-II OLTs are also connected to Stage-I OLTs through an RN. We denote the distances between RUs and Stage-I RNs by $D_{b_s r_I}$, the distances between Stage-I RNs and OLTs by $D_{r_I o}$, the distances between Stage-I OLTs and Stage-II RNs by $D_{or_{II}}$, and the distances between Stage-II RNs and ONUs by $D_{r_{II} q}$.\par
\textbf{Objective:} We use binary variables $\theta_o$ and $\theta_q$ to indicate if a Stage-I OLT and a Stage-II OLT are installed, respectively. The cost of installing a Stage-I ONU and a Stage-II ONU are denoted by $C_o$ and $C_q$. Moreover, we use the binary variables $y_{b_s o}$ to denote if RU-$b_s$ is connected to Stage-I OLT-$o$ and the binary variables $z_{oq}$ to denote if Stage-I OLT-$o$ is connected to Stage-II OLT-$q$. The cost of optical fiber and installation per km is denoted by $C_f$. Furthermore, we use the binary variables $\omega_{b_s}^d$ and $\omega_{os}^d$ to indicate if DU of slice $s$ is installed at RU-$b_s$ and Stage-I OLT-$o$ locations, and the binary variables $\omega_{os}^c$ and $\omega_{qs}^c$ to indicate if CU of slice $s$ is installed at Stage-I OLT-$o$ and Stage-II OLT-$q$ locations. The DUs and CUs can be placed only if the corresponding node is installed, i.e., $\omega_{os}^d \leq \theta_o$, $\omega_{os}^c \leq \theta_o, \forall s\in \mathcal{S}, o\in\mathcal{O}$ and $\omega_{qs}^c \leq \theta_q, \forall s\in \mathcal{S}, q\in\mathcal{Q}$. The cost of installation of RU processors, open access-edge servers at Stage-I OLT and Stage-II OLT locations are proportional to the maximum available GOPS/TTI, i.e., $G_{b_s}$, $G_o$, $G_q$, and $C_g$ denotes cost/GOPS/TTI. The objective to minimize the network installation cost is given as follows:
\vspace{-0.5\baselineskip}
\begin{align}
    &\mathcal{P}_2: \quad\min_{\bm{\theta}_o,\bm{\theta}_q,\bm{y},\bm{z}} \left\{\sum_{o\in\mathcal{O}} (C_o\theta_o + C_f\rho_o) + \sum_{q\in\mathcal{Q}} (C_q\theta_q + C_f\rho_q) \quad\quad \right.\nonumber\\
    &\left.+ \sum_{s\in\mathcal{S}}\sum_{b_s\in\mathcal{B}_s} C_g G_{b_s}\omega^d_{b_s} + \sum_{o\in\mathcal{O}} C_g G_o \theta_o + \sum_{q\in\mathcal{Q}} C_g G_q \theta_q \right\}, \label{eq12} 
\end{align}
where, the length of optical fibers associated with Stage-I and Stage-II TWDM-PONs are given by:
\begin{align}
    \rho_o &= D_{r_I o} + \sum_{s\in\mathcal{S}}\sum_{b_s\in\mathcal{B}_s} y_{b_s o} D_{b_s r_I}, \forall o\in\mathcal{O}, \label{eq13}\\
    \rho_q &= D_{r_{II} q} + \sum_{o\in\mathcal{O}} z_{oq} D_{or_{II}}, \forall q\in\mathcal{Q}. \label{eq14}
\end{align}

\vspace{-0.2\baselineskip}
%
\par \textbf{Connectivity constraints:} The binary variable $y_{b_s o}$ indicate if RU-$b_s$ is attached to Stage-I OLT-$o$ and the binary variable $z_{oq}$ indicate if Stage-I OLT-$o$ is attached to Stage-II OLT-$q$. However, their intermediate distances must not be greater than the maximum length of the TWDM-PON ($\sim 20$ km):
\begin{align}
    y_{b_s o} &\leq \left\lfloor \frac{D^{max}_o}{D_{b_s r_I}+D_{r_I o}} \right\rfloor, \forall s\in \mathcal{S}, b_s \in \mathcal{B}_s, o\in \mathcal{O}, \label{eq15}\\
    z_{oq} &\leq \left\lfloor \frac{D^{max}_q}{D_{or_{II}}+D_{r_{II}q}} \right\rfloor, \forall o\in \mathcal{O}, q\in \mathcal{Q}. \label{eq16}
\end{align}
\par Moreover, we can connect an RU-$b_s$ to a Stage-I OLT-$o$ only if it is installed. Similarly, we can connect a Stage-I OLT-$o$ to a Stage-II OLT-$q$ only if it is installed. These conditions are ensured by the following constraints:
\begin{align}
    \theta_o &\geq y_{b_s o}, \forall s\in \mathcal{S}, b_s \in \mathcal{B}_s, o\in \mathcal{O}, \label{eq17}\\
    \theta_q &\geq z_{oq}, \forall o\in \mathcal{O}, q\in \mathcal{Q}. \label{eq18}
\end{align}
\par In addition, we must ensure that each RU-$b_s$ is connected to only one Stage-I OLT-$o$ and each installed Stage-I OLT-$o$ is connected to only one Stage-II OLT-$q$ if required.
\begin{gather}
    \sum_{o\in \mathcal{O}} y_{b_s o} = 1, \forall s\in \mathcal{S}, b_s \in \mathcal{B}_s, \label{eq19}\\
    \theta_q \leq \sum_{q\in \mathcal{Q}} z_{oq} \leq \theta_o, \forall o\in \mathcal{O}, q\in \mathcal{Q}. \label{eq20}
\end{gather}
\par Furthermore, we ensure that DUs and CUs corresponding to each RU-$b_s$ are placed at one location only.
\begin{align}
    \omega_{b_s}^d + \sum_{o \in \mathcal{O}} \omega_{b_s o}^d &= 1, \forall s\in \mathcal{S}, b_s \in \mathcal{B}_s, \label{eq21}\\
    \omega_{os}^c + \sum_{q\in \mathcal{Q}} \omega_{oqs}^c &= \theta_o, \forall s\in \mathcal{S}, o\in \mathcal{O}. \label{eq22}
\end{align}
\par \textbf{Linearization constraints:} As we attempt to establish a chain of connections among RUs, DUs, and CUs, a set of binary variable product terms arise. Thus, we introduce a set of binary variables $\chi_{b_so}^d$, $\Upsilon_{b_so}^d$, $K_{oqs}^c$, and $L_{oqs}^c$ for linearizing the binary variable product terms $(\omega_{b_s}^d y_{b_s o})$, $(\omega_{os}^d y_{b_s o})$, $(\omega_{os}^c z_{oq})$, and $(\omega_{oqs}^c z_{oq})$, and the corresponding constraints are:\par
\begin{resizealign}
    (1-\chi_{b_so}^d) &\leq (1-\omega_{b_s}^d) + (1-y_{b_s o}), \forall s\in \mathcal{S}, b_s \in \mathcal{B}_s, o\in \mathcal{O}, \label{eq23}\\
    (1-\Upsilon_{b_so}^d) &\leq (1-\omega_{b_s o}^d) + (1-y_{b_s o}), \forall s\in \mathcal{S}, b_s \in \mathcal{B}_s, o\in \mathcal{O}, \label{eq24}\\
    (1-K_{oqs}^c) &\leq (1-\omega_{os}^c) + (1-z_{oq}), \forall s\in \mathcal{S}, o\in \mathcal{O}, q\in \mathcal{Q}, \label{eq25}\\
    (1-L_{oqs}^c) &\leq (1-\omega_{oqs}^c) + (1-z_{oq}), \forall s\in \mathcal{S}, o\in \mathcal{O}, q\in \mathcal{Q}. \label{eq26}
\end{resizealign}
\par In addition, we introduce integer variables $\beta_{os} = \sum_{b_s} y_{b_s o}$ that denote the number of RUs connected to each OLT-$o$. To linearize the products $\beta_{os} K_{oqs}^c$ and $\beta_{os} L_{oqs}^c$, we introduce the binary variables $A_{oqs}^c$ and $Z_{oqs}^c$, and the following constraints:\par
\begin{resizealign}
    (B_s - A_{oqs}^c) \leq B_s(1 - K_{oqs}^c) + (B_s - \beta_{os}), \forall s\in \mathcal{S}, o\in \mathcal{O}, q\in \mathcal{Q}, \label{eq27}\\
    (B_s - Z_{oqs}^c) \leq B_s(1 - L_{oqs}^c) + (B_s - \beta_{os}), \forall s\in \mathcal{S}, o\in \mathcal{O}, q\in \mathcal{Q}. \label{eq28}
\end{resizealign}
\par \textbf{Latency constraints:} If the DU is placed at RU-$b_s$ location, then its corresponding Stage-I TWDM-PON with OLT-$o$ will act as a mid-haul link, but if the DU is placed at the OLT-$o$ location, then the Stage-I TWDM-PON will act as a front-haul link. This is ensured for the uplink and downlink by the following constraints $\forall s\in \mathcal{S}, b_s \in \mathcal{B}_s, o\in \mathcal{O}$:
\begin{align}
    &\mathcal{T}_{b_s o}^{UL} = y_{b_s o}\left\{\delta_{b_so} + \frac{D_{b_sr_I}+D_{r_Io}}{v_c}\right\} + \left\{\frac{\sum_{s}\sum_{b_s} \chi_{b_so}^d U_{b_s}^{UL}}{R_o^{UL}} \right.\nonumber\\
    &\left.+ \frac{\sum_{s}\sum_{b_s} \Upsilon_{b_so}^d V_{b_s}^{UL}}{R_o^{UL}}\right\} \delta^{TTI} \leq \omega_{b_s}^d \Delta^{FH} + \omega_{os}^d \Delta^{MH}_s, \label{eq29}\\
    &\mathcal{T}_{b_s o}^{DL} = y_{b_s o}\left\{\frac{D_{b_sr_I}+D_{r_Io}}{v_c}\right\} + \left\{\frac{\sum_{s}\sum_{b_s} \chi_{b_so}^d U_{b_s}^{DL}}{R_o^{DL}} \right.\nonumber\\
    &\left.+ \frac{\sum_{s}\sum_{b_s} \Upsilon_{b_so}^d V_{b_s}^{DL}}{R_o^{DL}} \right\} \delta^{TTI} \leq \omega_{b_s}^d \Delta^{FH} + \omega_{os}^d \Delta^{MH}_s. \label{eq30}
\end{align}
\par In (\ref{eq29}), the parameters $\delta_{b_so}$ denote the \emph{reduced waiting time for uplink data at ONUs associated to RUs} within each TTI interval. The second term denotes the signal propagation latency and the third term denotes the uplink data transmission latency per TTI. The parameters $V_{b_s}^{UL}$ and $V_{b_s}^{DL}$ denote the required uplink and downlink front-haul throughput and $U_{b_s}^{UL}$ and $U_{b_s}^{DL}$ denote the required uplink and downlink mid-haul throughput of RU-$b_s$. The parameters $R_o^{UL}$ and $R_o^{DL}$ denote the maximum supported throughput of the Stage-I TWDM-PON. At the right side of the inequality, the parameter $\Delta^{FH}$ denotes the maximum front-haul latency and the parameter $\Delta^{MH}_s$ denotes the maximum mid-haul latency of slice $s$. As there is no waiting time in the downlink of TWDM-PON, we consider only signal propagation and data transmission latencies in (\ref{eq30}). Note that the DUs are placed either at RU/ONU or at OLT locations of Stage-I TWDM-PON. Therefore, the Stage-II TWDM-PON is required as mid-haul only if the CU is placed at the OLT of the Stage-II TWDM-PON and the uplink and downlink constraints are given below:
\begin{align}
    \mathcal{T}_{oq}^{UL} &= z_{oq} \left\{\delta_{oq} + \frac{D_{or_{II}}+D_{r_{II}q}}{v_c}\right\} + \left\{\frac{\sum_{s}\sum_{o} Z_{oqs}^c U_{o}^{UL}}{R_q^{UL}}\right\} \delta^{TTI}\nonumber \\
    &\quad\quad\quad\quad\quad \leq \omega_{qs}^c \Delta^{MH}_s, \forall s\in \mathcal{S}, o\in \mathcal{O}, q\in \mathcal{Q}, \label{eq31}\\
    \mathcal{T}_{oq}^{DL} &= z_{oq} \left\{\frac{D_{r_{II}q}+D_{r_{II}q}}{v_c}\right\} + \left\{\frac{\sum_{s}\sum_{o} Z_{oqs}^c U_{o}^{DL}}{R_q^{DL}}\right\} \delta^{TTI} \nonumber \\
    &\quad\quad\quad\quad\quad \leq \omega_{qs}^c \Delta^{MH}_s, \forall s\in \mathcal{S}, o\in \mathcal{O}, q\in \mathcal{Q}. \label{eq32}
\end{align}
\par Alongside the communication latency constraints, we also consider the processing latency constraints for the RU-DU-CU functions as follows $\forall s\in\mathcal{S}, b_s\in\mathcal{B}_s, o\in \mathcal{O}, q\in \mathcal{Q}$:\par
\begin{resizealign}
    \mathcal{T}_{rdc}^{UL} = \left\{\frac{\eta_{b_s,RU}^{UL}}{H_{b_s}^{UL}} + \frac{\omega_{b_s}^d \Gamma_{b_s,DU}^{UL}}{GD_{b_s}^{UL}} + \frac{\sum_{s}\sum_{b_s} \Upsilon_{b_so}^d \Gamma_{b_s,DU}^{UL}}{GD_o^{UL}}  \quad\quad\quad\quad\quad\quad\quad\quad \right. \nonumber\\
    \left.+ \frac{\sum_{s}\sum_{o} A_{oqs}^c \Gamma_{b_s,CU}^{UL}}{GC_o^{UL}} + \frac{\sum_{s}\sum_{o} Z_{oqs}^c \Gamma_{b_s,CU}^{UL}}{GC_q^{UL}} \right\} \leq \frac{\Delta_{b_s}^{BBU}}{\delta^{TTI}}, \label{eq33}\\
    \mathcal{T}_{rdc}^{DL} = \left\{\frac{\eta_{b_s,RU}^{DL}}{H_{b_s}^{DL}} + \frac{\omega_{b_s}^d \Gamma_{b_s,DU}^{DL}}{GD_{b_s}^{DL}} + \frac{\sum_{s}\sum_{b_s} \Upsilon_{b_so}^d \Gamma_{b_s,DU}^{DL}}{GD_o^{DL}} \quad\quad\quad\quad\quad\quad\quad\quad \right. \nonumber\\
    \left.+ \frac{\sum_{s}\sum_{o} A_{oqs}^c \Gamma_{b_s,CU}^{DL}}{GC_o^{DL}} + \frac{\sum_{s}\sum_{o} Z_{oqs}^c \Gamma_{b_s,CU}^{UL}}{GC_q^{DL}} \right\} \leq \frac{\Delta_{b_s}^{BBU}}{\delta^{TTI}}. \label{eq34}
\end{resizealign}
\par In both of these constraints, the first term denotes the RU processing latency at RU location $b_s$, the second term denotes the DU processing latency at RU location $b_s$, the third term denotes the DU processing latency at Stage-I OLT location $o$, the fourth term denotes the CU processing latency at Stage-I OLT location $o$, and the fifth term denotes the CU processing latency at Stage-II OLT location $q$. The parameters $\eta_{b_s,RU}^{UL}$ and $\eta_{b_s,RU}^{DL}$ denote the required GOPS/TTI for uplink and downlink RU functions, parameters $\Gamma_{b_s,DU}^{UL}$ and $\Gamma_{b_s,DU}^{DL}$ denote the required GOPS/TTI for uplink and downlink DU functions, and parameters $\Gamma_{b_s,CU}^{UL}$ and $\Gamma_{b_S,CU}^{DL}$ denote the required GOPS/TTI for uplink and downlink CU functions. Moreover, the parameters $H_{b_s}^{UL}$ and $H_{b_s}^{DL}$ denote the maximum available GOPS/TTI for uplink and downlink RU function processing, parameters $GD_{b_s}^{UL}$ and $GD_{b_s}^{DL}$ denote the maximum available GOPS/TTI for uplink and downlink DU function processing at RU-$b_s$ such that $G_{b_s} = (GD_{b_s}^{UL} + GD_{b_s}^{DL})$, parameters $GD_o^{UL}$ and $GD_o^{DL}$ denote the maximum available GOPS/TTI for uplink and downlink DU function processing at OLT-$o$, and parameters $GC_o^{UL}$ and $GC_o^{DL}$ denote the maximum available GOPS/TTI for uplink and downlink CU function processing at OLT-$o$ such that $G_o = (GD_o^{UL} + GD_o^{DL} + GC_o^{UL} + GC_o^{DL})$, and parameters $GC_q^{UL}$ and $GC_q^{DL}$ denote the maximum available GOPS/TTI for uplink and downlink CU function processing at OLT-$q$ such that $G_q = (GC_q^{UL} + GC_q^{DL})$. 



\vspace{-0.5\baselineskip}
\subsection{Heuristic for Front/Mid-haul and DU-CU Deployment}
After evaluating the UE-RU assignment problem $\mathcal{P}_1$, we know the locations of installed RUs corresponding to each slice and their respective throughput demands. Now, we proceed to solve the problem $\mathcal{P}_2$ and design a heuristic algorithm for connecting the installed RUs with TWDM-PON-based front/mid-haul links and placing open access-edge servers for hosting DUs and CUs. The main focus of this algorithm is to connect all the active RUs with a minimum number of open access-edge servers for hosting DUs and CUs while satisfying the communication and BBU processing latency constraints. This, in turn, reduces the cost of TWDM-PON and server deployment according to the objective (\ref{eq12}) of $\mathcal{P}_2$. The throughput requirement between the RU-DU interface and the DU-CU interface, mainly depends on the RU configurations, because we consider 100\% PRB usage for the network planning purpose. Let us denote $\mathfrak{B}_s$ as the set of installed but unassigned RUs, $\mathfrak{O}$ as the final set of installed Stage-I OLTs, and $\mathfrak{Q}$ as the final set of installed Stage-II OLTs. Initially, all RUs from all slices are unassigned, i.e, $\mathfrak{B}_s = \mathcal{B}_s$ and neither of Stage-I and Stage-II OLTs are installed, i.e., $\mathfrak{O} = \emptyset$, $\mathfrak{Q} = \emptyset$. In the first iteration, we activate only one Stage-I OLT-$o$ and add it to $\mathfrak{O}$. Firstly, we consider placing DUs at RU locations, and hence, the Stage-I TWDM-PON acts like a mid-haul interface. Then we sequentially choose one RU-$b_s$ from each slice and associate with an OLT-$o\in\mathfrak{O}$ with minimum distance, i.e., $o = \min_{o}\{D_{b_s o} = (D_{b_s r_I} + D_{r_I o}) |o\in \mathfrak{O}, D_{b_s o} \leq D_o^{max}\}$, and check if the following front-haul communication and RU+DU processing latencies are satisfied:\par
\begin{algorithm}[t!]
\caption{Heuristic for Front/Mid-haul \& DU-CU Deploy}\label{alg2}
\hspace*{\algorithmicindent}\textbf{Input:} $\mathcal{S},\mathcal{B}_s, \mathcal{O}, \mathcal{Q}, D_{b_s o}, D_{oq}, D_{o}^{max}, D_{q}^{max}, U_{b_s}^{U/DL},$\\
\hspace*{\algorithmicindent}$V_{b_s}^{U/DL}, R_o^{U/DL}, R_q^{U/DL}, \Gamma_{b_s,CU}^{U/DL}, \Gamma_{b_s,DU}^{U/DL}, H_{b_s}^{U/DL}, GD_{b_s}^{U/DL},$\\ \hspace*{\algorithmicindent}$GD_o^{U/DL}, GC_o^{U/DL}, GC_q^{U/DL}, \Delta^{FH}, \Delta^{MH}_s, \Delta_{b_s}^{BBU}$.\\
\hspace*{\algorithmicindent}\textbf{Output:} $y_{b_s o}^*, z_{oq}^*, \omega_{b_s}^{d*}, \omega_{b_s o}^{d*}, \omega_{os}^{c*}, \omega_{oqs}^{c*}, \theta_o^*, \theta_q^*$.\\
\hspace*{\algorithmicindent}\textbf{Initialize:} $\mathfrak{B}_s = \mathcal{B}_s, \mathfrak{O} = \{o\}, \mathfrak{Q} = \{q\}, sucs = 0$.
\begin{algorithmic}[1]
\While{$sucs \neq 1$ \text{\textbf{and}} $|\mathfrak{O}| \neq |\mathcal{O}|$} \Comment{Stage-I design} \label{ln1}
    \For{$b_s \gets 1$ \textbf{to} $\max\{B_s\}$}   \label{ln2}
        \For{$s \gets 1$ \textbf{to} $S$}
            \If{$b_s \leq B_s$}
                \State Find $o = \min_{o}\{D_{b_s o}|D_{b_s o}\leq D_o^{max}\}$;
                \If{(\ref{eq41})-(\ref{eq44}) are satisfied} \Comment{Case-1}
                    \State Calculate the total cost;
                \EndIf
                \If{(\ref{eq45})-(\ref{eq48}) are satisfied} \Comment{Case-2}
                    \State Calculate the total cost;
                \EndIf
                \If{Case-1 costs minimum}
                    \State $y_{b_s o} \gets 1$, $\theta_o \gets 1$, $\omega_{b_s}^{d} \gets 1$;
                \ElsIf{Case-2 costs minimum}
                    \State $y_{b_s o} \gets 1$, $\theta_o \gets 1$, $\omega_{b_s o}^{d} \gets 1$;
                \Else           \Comment{Some RU is unassigned}
                    \State break;  \Comment{infeasible if $|\mathfrak{O}| = |\mathcal{O}|$}
                \EndIf
            \EndIf
            \State $\mathfrak{B}_s \gets \mathfrak{B}_s \setminus b_s $;
        \EndFor     
    \EndFor     \label{ln22}
    \If{$\mathfrak{B}_s = \emptyset, \forall s$} \Comment{All RUs are assigned}
        \State $sucs \gets 1$;
    \Else           \Comment{some RU is unassigned}     \label{ln25}
        \State $\mathfrak{B}_s \gets \mathcal{B}_s, \forall s$; \Comment{RU set reinitialized}
        \State $\mathfrak{O} \gets \mathfrak{O}\cup \{o'\}$; \Comment{A new Stage-I OLT added}
    \EndIf  \label{ln28}
\EndWhile   \label{ln29}
\State $sucs \gets 0$;    \Comment{Reset the success flag}    \label{ln30}
\While{$sucs \neq 1$ \text{\textbf{and}} $|\mathfrak{Q}| \neq |\mathcal{Q}|$} \Comment{Stage-II design}   \label{ln31}
    \For{$o \gets 1$ \textbf{to} $|\mathfrak{O}|$}  \label{ln33}
        \For{$s \gets 1$ \textbf{to} $S$}
            \If{CU can be processed at $o$}
                \State $\omega_{os}^{c} \gets 1$;
            \Else
                \State Find $q = \min_{q}\{D_{oq}|D_{oq} \leq D_q^{max}\}$;
                \State Subject to constraints (\ref{eq31})-(\ref{eq34});
                \State $\omega_{oqs}^{c} \gets 1$, $z_{oq} \gets 1$;
            \EndIf
        \EndFor 
    \EndFor     \label{ln42}
    \If{All CUs are placed}
        \State $sucs \gets 1$;
    \Else
        \State $\mathfrak{Q} \gets \mathfrak{Q}\cup \{q'\}$; \Comment{New Stage-II OLT added}
    \EndIf
\EndWhile   \label{ln48}
\State \textbf{return} $y_{b_s o}, z_{oq}, \omega_{b_s}^{d}, \omega_{b_s o}^{d}, \omega_{os}^{c}, \omega_{oqs}^{c}, \theta_o, \theta_q$;
\end{algorithmic}
\end{algorithm}
\begin{resizealign}
    \left\{\delta_{b_so} + \frac{D_{b_sr_I}+D_{r_Io}}{v_c}\right\} + \left\{\frac{U_{b_s}^{UL}+\sum_{s}\sum_{b'_s \neq b_s} \chi_{b_so}^d U_{b'_s}^{UL}}{R_o^{UL}} \right\} \delta^{TTI} \leq \Delta^{MH}_s, 
    \label{eq41}\\
    \left\{\frac{D_{b_sr_I}+D_{r_Io}}{v_c}\right\} + \left\{\frac{U_{b_s}^{DL}+\sum_{s}\sum_{b'_s \neq b_s} \chi_{b_so}^d U_{b'_s}^{DL}}{R_o^{DL}} \right\} \delta^{TTI} \leq \Delta^{MH}_s, \label{eq42}\\
    \left\{\frac{\eta_{RU}^{UL}}{H_{b_s}^{UL}} + \frac{\omega_{b_s}^d \Gamma_{b_s,DU}^{UL}}{GD_{b_s}^{UL}}\right\} \leq \frac{\Delta_{b_s}^{BBU}}{\delta^{TTI}}, \quad\quad\quad\quad\quad\quad\quad\quad \label{eq43}\\
    \left\{\frac{\eta_{RU}^{DL}}{H_{b_s}^{DL}} + \frac{\omega_{b_s}^d \Gamma_{DU,b_s}^{DL}}{GD_{b_s}^{DL}}\right\} \leq \frac{\Delta_{b_s}^{BBU}}{\delta^{TTI}}. \quad\quad\quad\quad\quad\quad\quad\quad \label{eq44}
\end{resizealign}
\par Secondly, we consider placing DUs at Stage-I OLT locations, and hence, the Stage-I TWDM-PON acts like a front-haul interface and sequentially associate RUs to OLTs. Again, we sequentially choose one RU-$b_s$ from each slice with minimum distance while checking the following front-haul communication and RU+DU processing latencies are satisfied:\par
\begin{resizealign}
    \left\{\delta_{b_so} + \frac{D_{b_sr_I}+D_{r_Io}}{v_c}\right\} + \left\{\frac{V_{b_s}^{UL}+\sum_{s}\sum_{b'_s \neq b_s} \Upsilon_{b'_s o}^d V_{b'_s}^{UL}}{R_o^{UL}} \right\} \delta^{TTI} \leq \Delta^{FH}, 
    \label{eq45}\\
    \left\{\frac{D_{b_sr_I}+D_{r_Io}}{v_c}\right\} + \left\{\frac{V_{b_s}^{DL}+\sum_{s}\sum_{b'_s \neq b_s} \Upsilon_{b'_s o}^d V_{b'_s}^{DL}}{R_o^{DL}} \right\} \delta^{TTI} \leq \Delta^{FH}, \label{eq46}\\
    \left\{\frac{\eta_{RU}^{UL}}{H_{b_s}^{UL}} + \frac{\Gamma_{b_s,DU}^{UL}+\sum_{s}\sum_{b'_s \neq b_s} \Upsilon_{b'_s o}^d \Gamma_{b'_s,DU}^{UL}}{GD_o^{UL}} \right\} \leq \frac{\Delta_{b_s}^{BBU}}{\delta^{TTI}}, \quad\quad\quad \label{eq47}\\
    \left\{\frac{\eta_{RU}^{DL}}{H_{b_s}^{DL}} + \frac{\Gamma_{b_s,DU}^{DL}+\sum_{s}\sum_{b'_s \neq b_s} \Upsilon_{b'_s o}^d \Gamma_{b'_s,DU}^{DL}}{GD_o^{DL}} \right\} \leq \frac{\Delta_{b_s}^{BBU}}{\delta^{TTI}}. \quad\quad\quad \label{eq48}
\end{resizealign}
\par After this step, we calculate the total cost of OLT installation, fiber deployment, and open access-edge server placement for both the cases i.e., when the DU is located at the RU and at a Stage-I OLT location. Then we choose the option with minimum cost and accordingly set $y_{b_s o} = 1$, $\theta_o = 1$, $\omega_{b_s}^{d} = 1$ (Case-1), and $\omega_{b_s o}^{d} = 1$ (Case-2). Then we remove this RU from $\mathfrak{B}_s$ and continue the same process with the remaining RUs. If a DU server for any RU could not be installed, then the for loop at lines \ref{ln2}-\ref{ln22} breaks. Clearly, in this case $\mathfrak{B}_s \neq \emptyset, \forall s$ and lines (\ref{ln25})-(\ref{ln28}) are executed. This means a new OLT-$o'$ is inserted to $\mathfrak{O}$ and the RU assignment loop at lines \ref{ln2}-\ref{ln22} is restarted by the while loop at lines \ref{ln1}-\ref{ln29}. When supported throughput of the OLTs are different, then prioritize the OLTs with higher throughput support. If all the RUs are attached to the Stage-I OLTs, the \emph{success flag} is set, i.e., $sucs = 1$ and the while loop stops. However, if all the Stage-I OLTs are explored, i.e., $|\mathfrak{O}| \neq |\mathcal{O}|$ and a DU server could not be installed for some RU, then the while loop as well as the algorithm stops due to infeasibility. Next, we reset the success flag at line \ref{ln30} and check if we can also deploy the CUs in open access-edge servers at OLT-$o$ locations by adding the respective processing latency terms $\frac{\Gamma_{b_s,CU}^{UL} + \sum_{s}\sum_{o} A_{oqs}^c \Gamma_{b'_s,CU}^{UL}}{GC_o^{UL}}$ and $\frac{\Gamma_{b_s,CU}^{DL} + \sum_{s}\sum_{o} A_{oqs}^c \Gamma_{b'_s,CU}^{DL}}{GC_o^{DL}}$ to (\ref{eq43})-(\ref{eq44}) or (\ref{eq47})-(\ref{eq48}). If this step is successful, then we set $\omega_{os}^{c} = 1$, otherwise, we find a Stage-II OLT-$q$ with minimum distance from Stage-I OLT-$o$ for installing open access-edge servers to host CUs. Additionally, we need to check if (\ref{eq31})-(\ref{eq34}) are satisfied. We can do this by following a similar process as Stage-I. Here also we start with just one Stage-II OLT in $\mathfrak{Q}$ and iteratively add new Stage_II OLTs to find CU servers for all the RUs. Once a suitable Stage-II OLT-$q$ is found, we set the corresponding $\omega_{oqs}^{c} = 1$, $z_{oq} = 1$. If all the CUs could be successfully placed by the for loop in lines \ref{ln33}-\ref{ln42}, then we set $sucs=1$ again and the algorithm stops successfully. Otherwise, if a CU server could not be installed against some Stag-I OLT, then the while loop in lines \ref{ln31}-\ref{ln48} stops due to infeasibility. The TWDM-PON-based front/mid-haul design and open access-edge server placement algorithm is given in Algorithm \ref{alg2}.\par
\begin{theorem}
    The Algorithm \ref{alg2} provides a $\texttt{O}(\log_e (O\times\sum_s B_s))$ approximation to the optimal solution for $\mathcal{P}_2$.
\end{theorem}
\begin{IEEEproof}
    Let $O^*$ denote the optimum number of Stage-I OLTs. Initially, there are $B_0 = (\sum_s B_s)$ unconnected RUs and $B_t$ denotes the number of RUs remaining to be connected after $t$ greedy iterations. Therefore, after $(t-1)$ iterations, $B_{t-1}$ RUs are still remaining to be connected and there exists some Stage-I OLT that connects at least $(B_{t-1}/O^*)$ RUs (by the pigeonhole principle). As our greedy method attempts to connect the maximum number of the remaining RUs to Stage-I OLTs (with properly placed DUs) at every step, it must select a Stage-I OLT connecting at least $(B_{t-1}/O^*)$ RUs. Hence, the number of remaining RUs to be connected is at most
    \begin{align}
        B_t \leq \left(B_{t-1} - \frac{B_{t-1}}{O^*}\right) = B_{t-1}\left(1-\frac{1}{O^*}\right). \label{n03}
    \end{align}
    \vspace{-\baselineskip}
    \par Thus, the number of remaining RUs decreases by a factor of at least $(1-1/O^*)$ with every iteration and after $t$ iterations, we have $B_t \leq B_0(1-1/O^*)^t$. If the greedy method runs for $(O_G+1)$ iterations, we must have at least one remaining unconnected RU at the $O_G$-th iteration such that\par
    \begin{resizealign}
        1 \leq B_{O_G} \leq B_0\left(1-\frac{1}{O^*}\right)^{O_G} = B_0\left(1-\frac{1}{O^*}\right)^{{O^*}\times\frac{O_G}{O^*}}. \label{n04}
    \end{resizealign}
    \par Now, using the well-known Taylor series approximation $(1-x)^{\frac{1}{x}}\gtrsim (1/e)$ in (\ref{n04}), we have,\par
    \begin{resizealign}
        1 \leq B_0\left(1/e\right)^{\frac{O_G}{O^*}} \Rightarrow  \frac{O_G}{O^*} \leq \log_e (B_0) \Rightarrow O_G \leq O^* \log_e (B_0). \label{n05}
    \end{resizealign}
    \par Similarly, we can show that if the greedy method runs for $(Q_G+1)$ iterations to connect $O$ Stage-I OLTs, then $Q_G \leq Q^* \log_e (O)$, where $Q^*$ denotes the optimal number of Stage-II OLTs. Therefore, the solution for $\mathcal{P}_2$ produced by Algorithm \ref{alg2} can be approximated by a factor of $\texttt{O}(\log_e (B_0) + \log_e O)$ or $\texttt{O}(\log_e (O\times\sum_s B_s))$ to the optimal solution.
\end{IEEEproof}
\par When the problem $\mathcal{P}_2$ is feasible, every iteration of the first while loop of Algorithm \ref{alg2} connects at least one RU to a Stage-I OLT while appropriately placing a DU server. It generates all possible RU and Stage-I OLT combinations with complexity $\texttt{O}((\sum_s B_s) O^2)$ and their feasibility is verified with $\texttt{O}((\sum_s B_s) O)$. Thus, we can analyze that the worst-case complexity of this while loop is $((\sum_s B_s)^2 O^3)$. Similarly, the complexity of designing Stage-II TWDM-PON is $\texttt{O}(O^2 Q^3)$. Therefore, the Algorithm \ref{alg2} converges with a worst-case complexity of $\texttt{O}((\sum_s B_s)^2 O^3)$ (as $((\sum_s B_s)^2 O^3) \gg (O^2 Q^3)$) to a solution upper-bounded by a factor of $(\log_e (O\times\sum_s B_s))$ to the optimal solution.
\begin{figure}[t!]
\centering
\includegraphics[width=0.95\columnwidth,height=5.5cm]{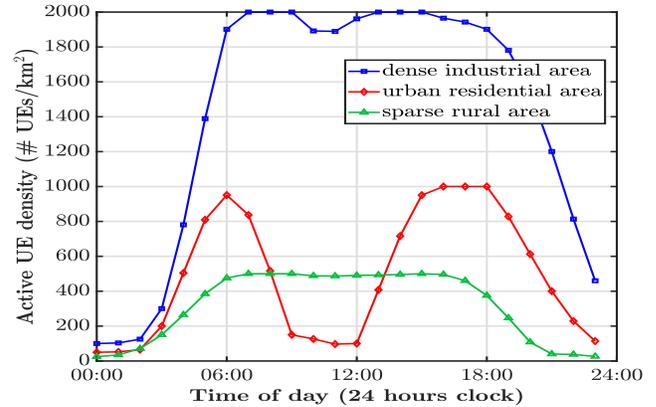}
\caption{Active UE density variation in industrial, urban, and rural areas across different hours of a day.}
\label{ue_density}
\end{figure}
\setlength{\textfloatsep}{5pt}
%
\begin{table}[!b]
\centering
\caption{UE Distribution and Throughput Requirements}
\label{table2}
\resizebox{\columnwidth}{!}{%
\begin{tabular}{lcccccc}
\toprule
               & \textbf{\begin{tabular}[c]{@{}c@{}}Industrial\\ (\%)\end{tabular}} & \textbf{\begin{tabular}[c]{@{}c@{}}Urban\\ (\%)\end{tabular}} & \textbf{\begin{tabular}[c]{@{}c@{}}Rural\\ (\%)\end{tabular}} & \textbf{\begin{tabular}[c]{@{}c@{}}Uplink\\ (Mbps)\end{tabular}} & \textbf{\begin{tabular}[c]{@{}c@{}}Downlink\\ (Mbps)\end{tabular}} \\ \toprule
\textbf{uRLLC} & 25\%                                                             & 30\%                                                          & 20\%                                                          & 10-20                                                            & 30-50                                                              \\ \hline
\textbf{eMBB}  & 25\%                                                             & 50\%                                                          & 60\%                                                          & 50-80                                                            & 100-150                                                            \\ \hline
\textbf{mMTC}  & 50\%                                                             & 20\%                                                          & 20\%                                                          & 10-20                                                            & 10-20                                                    \\ \bottomrule
\end{tabular}
}
\end{table}
\setlength{\textfloatsep}{5pt}
%
\section{Results and Discussions}\label{sec6}
For evaluating the proposed framework, we consider O-RAN deployment in dense industrial, urban residential, and sparsely populated rural areas. The dimensions of these areas vary from $1\times 1$ km$^2$ to $4\times 4$ km$^2$ and the maximum UE density in the industrial, urban, and rural areas are 2000, 1000, and 500 $\#$/km$^2$, respectively. However, to capture the \emph{spatio-temporal} and \emph{tidal wave} characteristics of mobile traffic \cite{data_mi}, we consider the hourly UE density variation at different areas as shown in Fig. \ref{ue_density}. We generate random UE locations according to the maximum UE densities of different areas but only a fraction of them actively transmit data at peak throughput (from Table \ref{table2}) at different time instants. The total number of active UEs at different time instants remains consistent with Fig. \ref{ue_density} and their respective throughput requirements are derived as the average of peak throughput across their active data transmission times. This reduces the over-provisioning of resources at the RAN installation stage. After the RAN deployment, UEs can be dynamically associated with RUs in real-time based on their actual QoS requirements.\par
%
\begin{figure*}[!t]
  \centering
  \subfloat[Industrial Area (km $\times$ km)]{%
    \includegraphics[width=0.333\textwidth,height=5.1cm]{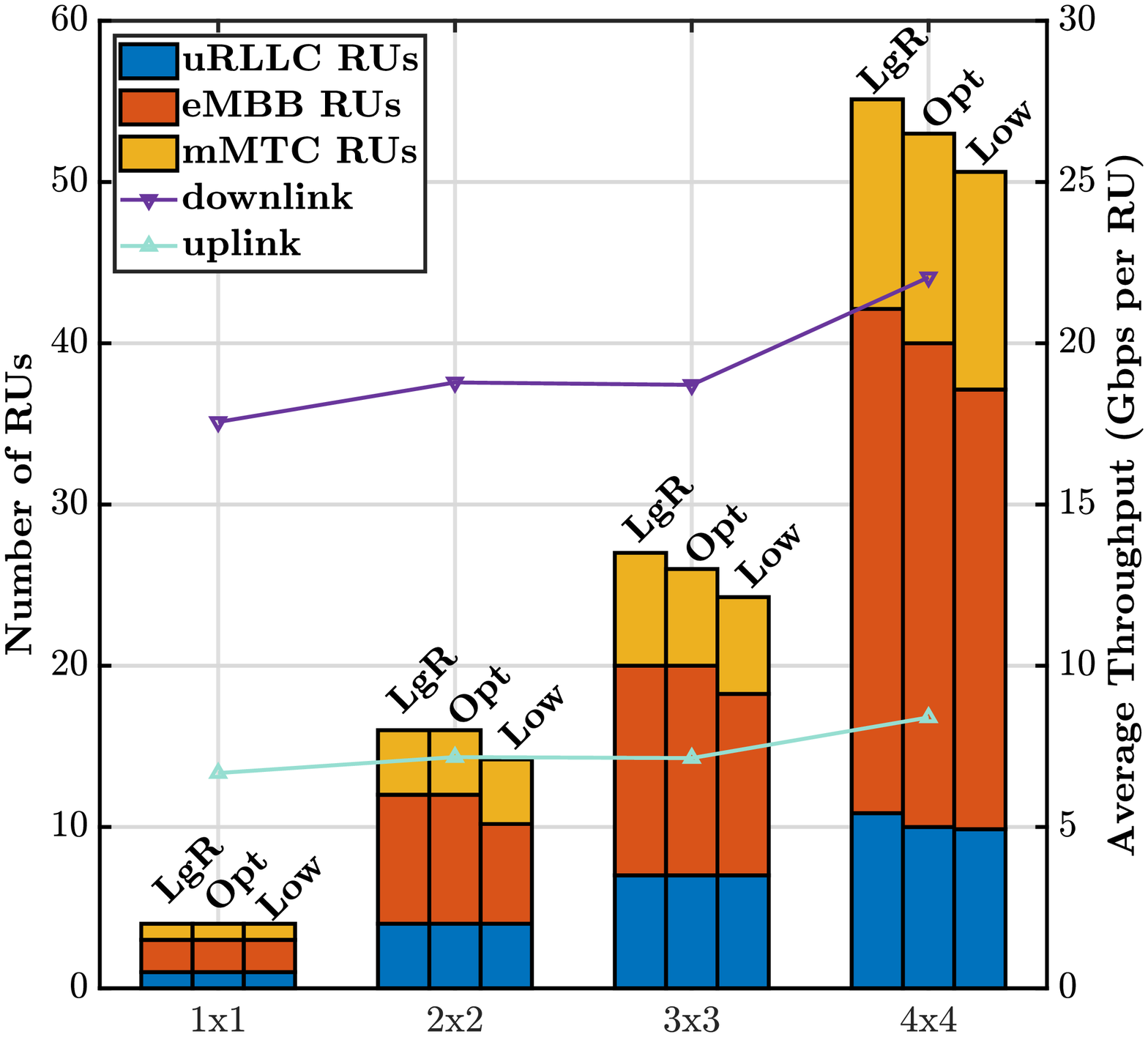}\label{ue_bs1}%
  }
  \subfloat[Urban Area (km $\times$ km)]{%
    \includegraphics[width=0.333\textwidth,height=5.1cm]{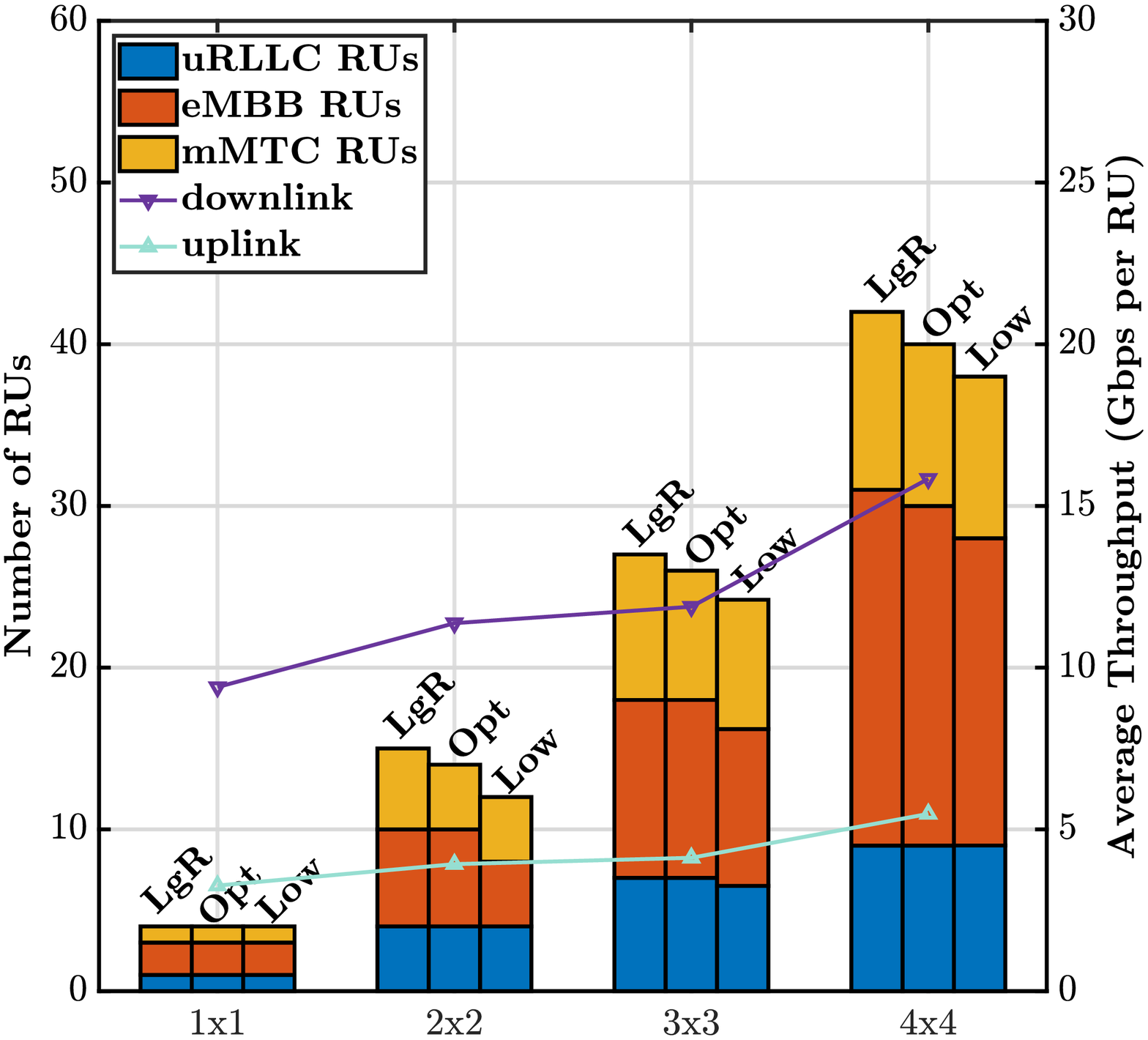}\label{ue_bs2}%
  }
  \subfloat[Rural Area (km $\times$ km)]{%
    \includegraphics[width=0.333\textwidth,height=5.1cm]{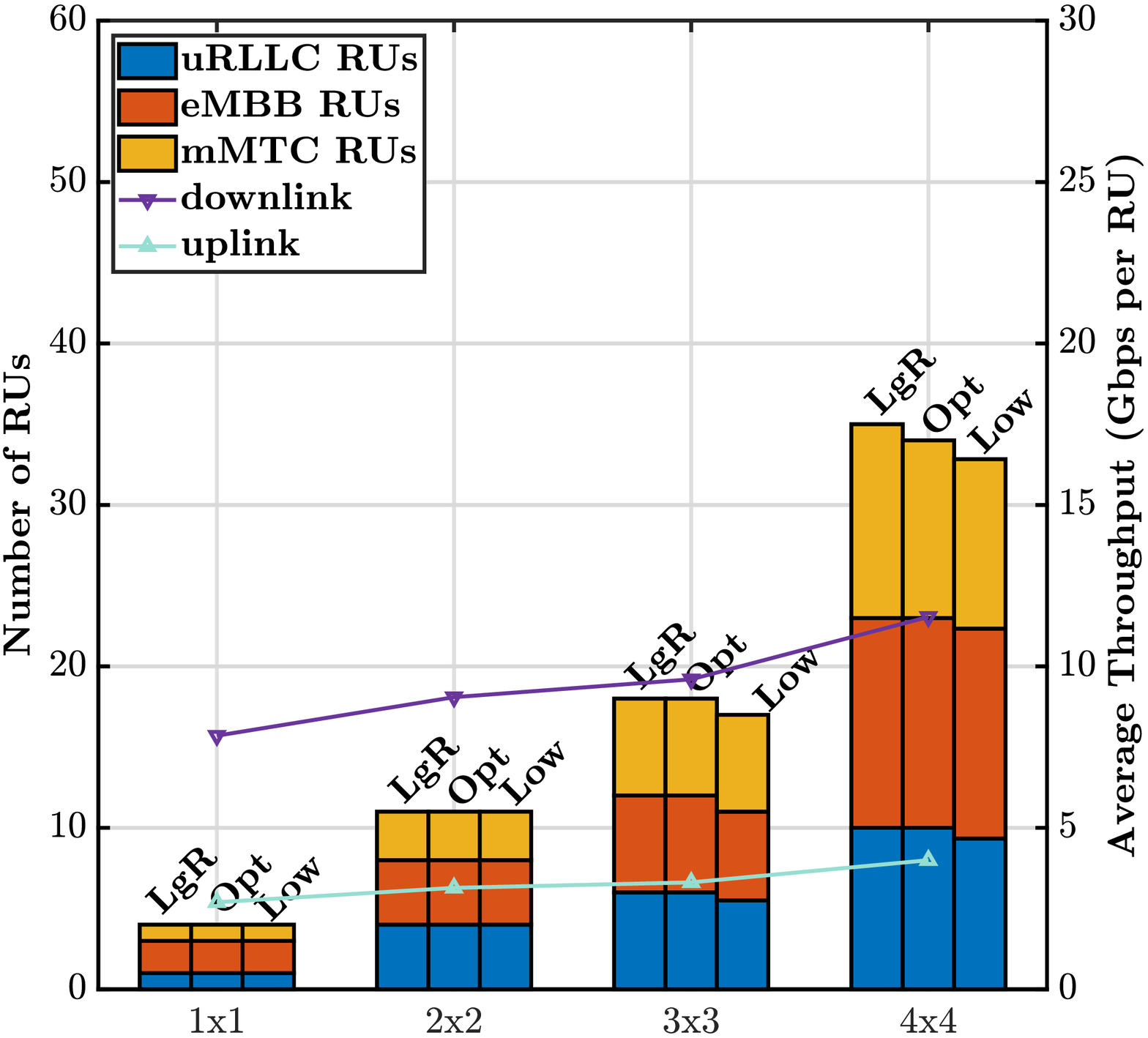}\label{ue_bs3}%
  }

  \caption{Comparison of number of RUs by Lagrangian relaxation heuristic (LgR), the optimal solution (Opt), and the lower bound by integrality relaxation (Low), and the average throughput against the considered (a) industrial, (b) urban, and (c) rural deployment scenarios.}
  \label{ue_bs_cmp}
\end{figure*}
\setlength{\textfloatsep}{1pt}

\begin{figure*}[!t]
  \centering
  \subfloat[Industrial Area (km $\times$ km)]{%
    \includegraphics[width=0.333\textwidth,height=5.1cm]{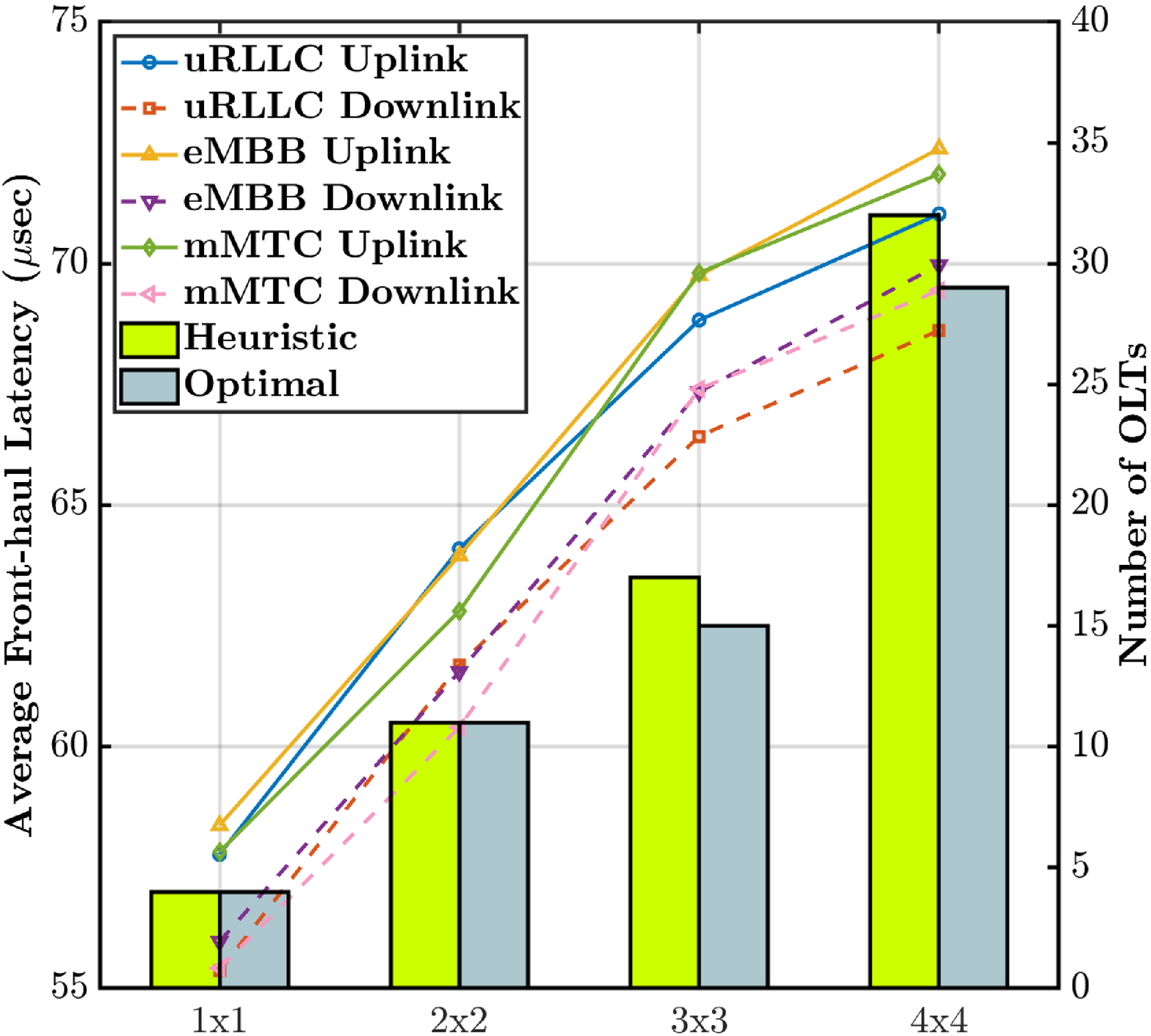}\label{fh_lat1}%
  }
  \subfloat[Urban Area (km $\times$ km)]{%
    \includegraphics[width=0.333\textwidth,height=5.1cm]{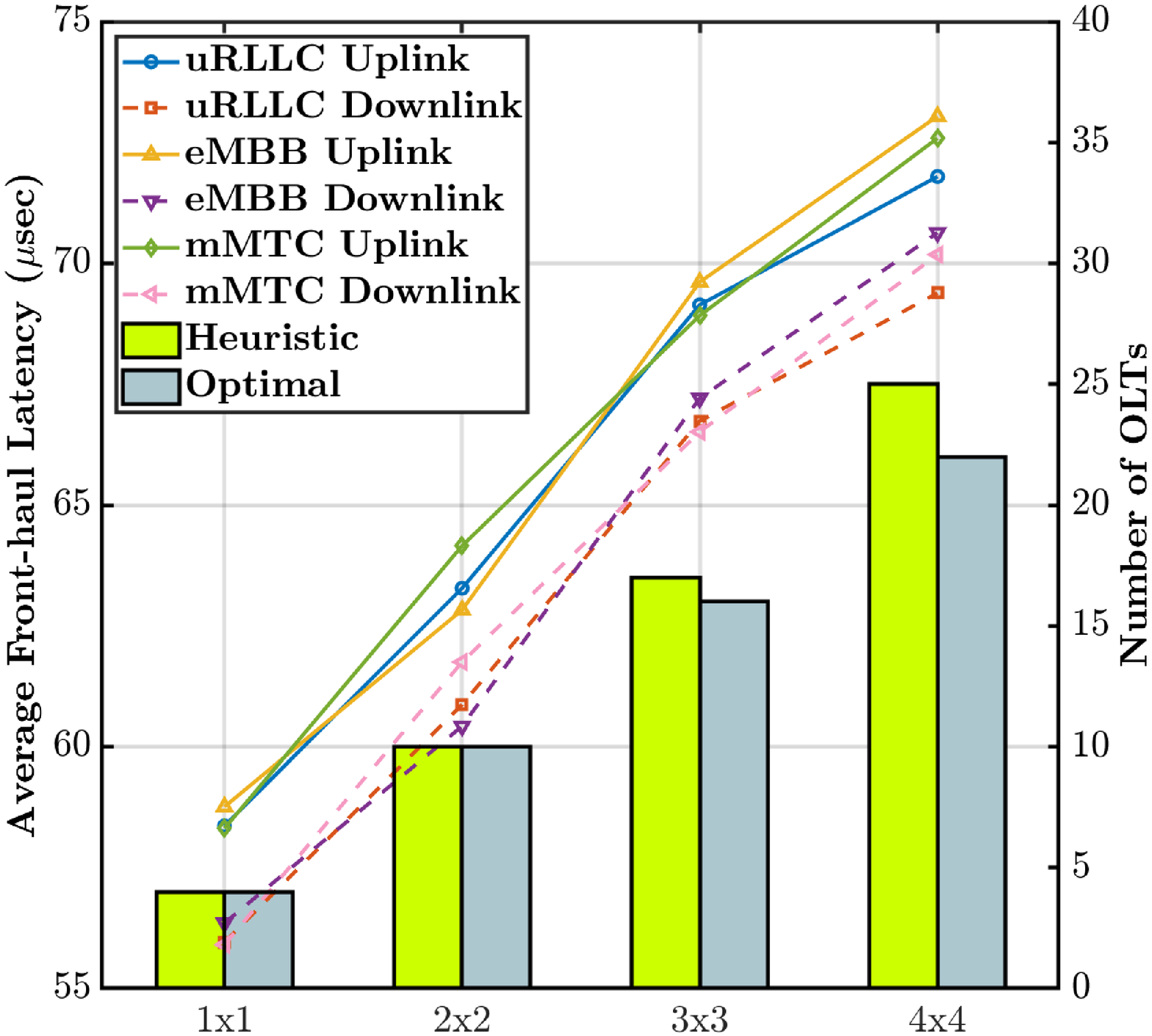}\label{fh_lat2}%
  }
  \subfloat[Rural Area (km $\times$ km)]{%
    \includegraphics[width=0.333\textwidth,height=5.1cm]{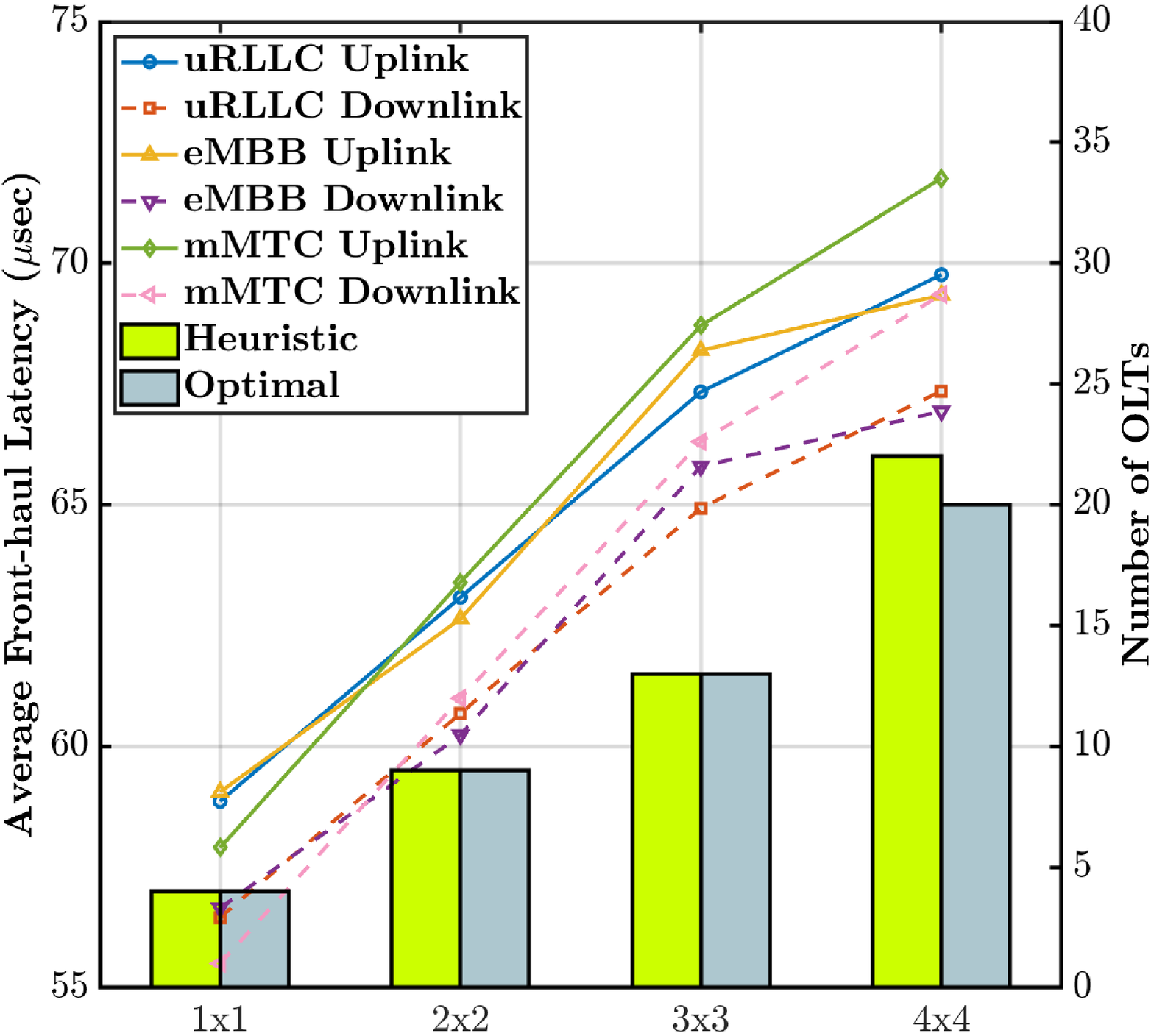}\label{fh_lat3}%
  }

  \caption{Average front-haul uplink and downlink latencies of uRLLC, eMBB, and mMTC slices and the total number of installed OLTs of TWDM-PONs against (a) industrial, (b) urban, and (c) rural areas.}
  \label{fh_latency}
\end{figure*}
\setlength{\textfloatsep}{1pt}
The maximum number of RUs per km$^2$ area is 15 and they are classified in uRLLC, eMBB, and mMTC slices according to the UE distribution shown in Table \ref{table2}. The coverage distance of the RUs varies within 0.5-1 km and the transmission powers of the macro and small-cell RUs are 46 dBm and 30 dBm, respectively. The RU locations are also the locations for Stage-I OLTs. We consider the highest RU configuration as 4x4 MIMO, 2 layers, 100 MHz bandwidth, TTI duration 0.5 msec, and 30 kHz sub-carrier spacing. Thus, the peak wireless throughput supports are uplink: 28 Gbps and downlink: 30 Gbps, respectively \cite{5g_kpi}. Accordingly, their throughput demands for front-haul (split-7.2) are uplink: 9.632 Gbps, downlink: 11.113 Gbps and for mid-haul (split-2) are uplink: 1.111 Gbps, downlink: 1.111 Gbps \cite{IEEE_1914.1}. Both Stage-I and Stage-II TWDM-PONs can support a maximum throughput of 100 Gbps. The average reduced waiting time of uplink data at the ONUs is 5 $\mu$sec \cite{5g_fh_bw3}. We consider the OTA latency bound as 200-400 $\mu$sec and the front-haul latency bound as 100 $\mu$sec for all slices. However, the mid-haul latency bounds for uRLLC, eMBB, and mMTC slices considered are 100 $\mu$sec, 500 $\mu$sec, and 1000 $\mu$sec, respectively. In addition, the computational requirement of each RU is 1800 GOPS/TTI and is divided among RUs, DUs, and CUs. The processing capacities of the open access-edge servers are $10^5$ GOPS/TTI and we choose the BBU processing latency bounds 50, 80, and 100 $\mu$sec for uRLLC, eMBB, and mMTC slices, respectively \cite{mgain}. The cost of open access-edge server installation is \euro 3800, the cost of computational resources is 1.5 \euro/GOPS \cite{F_split}, the cost of optical fiber is 100 \euro/km, the cost of fiber installation is 2500 \euro/km, the cost of a splitter is \euro 200, the cost of ONU is \euro 2000, and the cost of OLT is \euro 16000 \cite{pon_cost}.\par
\begin{figure*}[!t]
  \centering
  \subfloat[Industrial Area (km $\times$ km)]{%
    \includegraphics[width=0.333\textwidth,height=5.1cm]{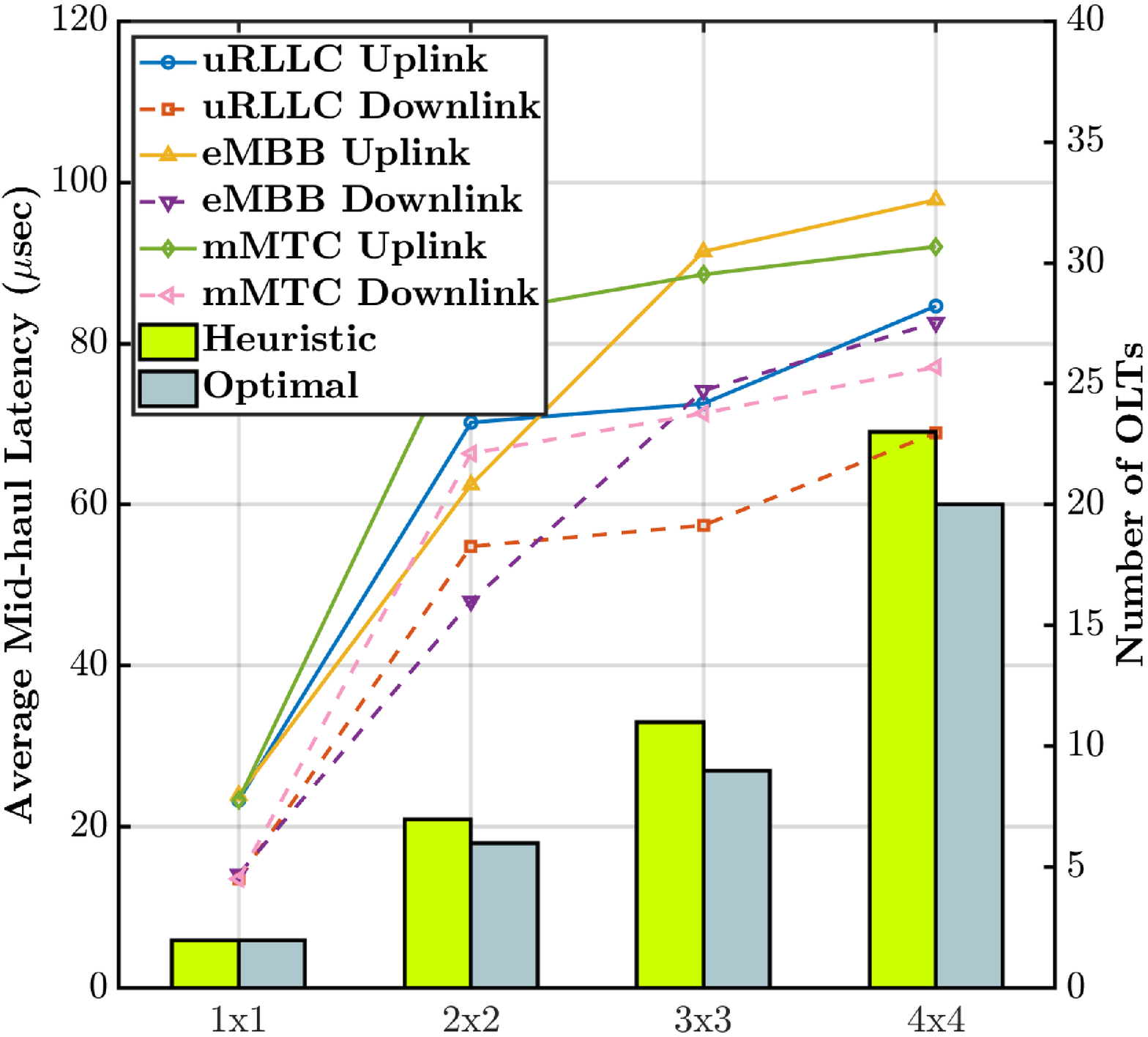}\label{mh_lat1}%
  }
  \subfloat[Urban Area (km $\times$ km)]{%
    \includegraphics[width=0.333\textwidth,height=5.1cm]{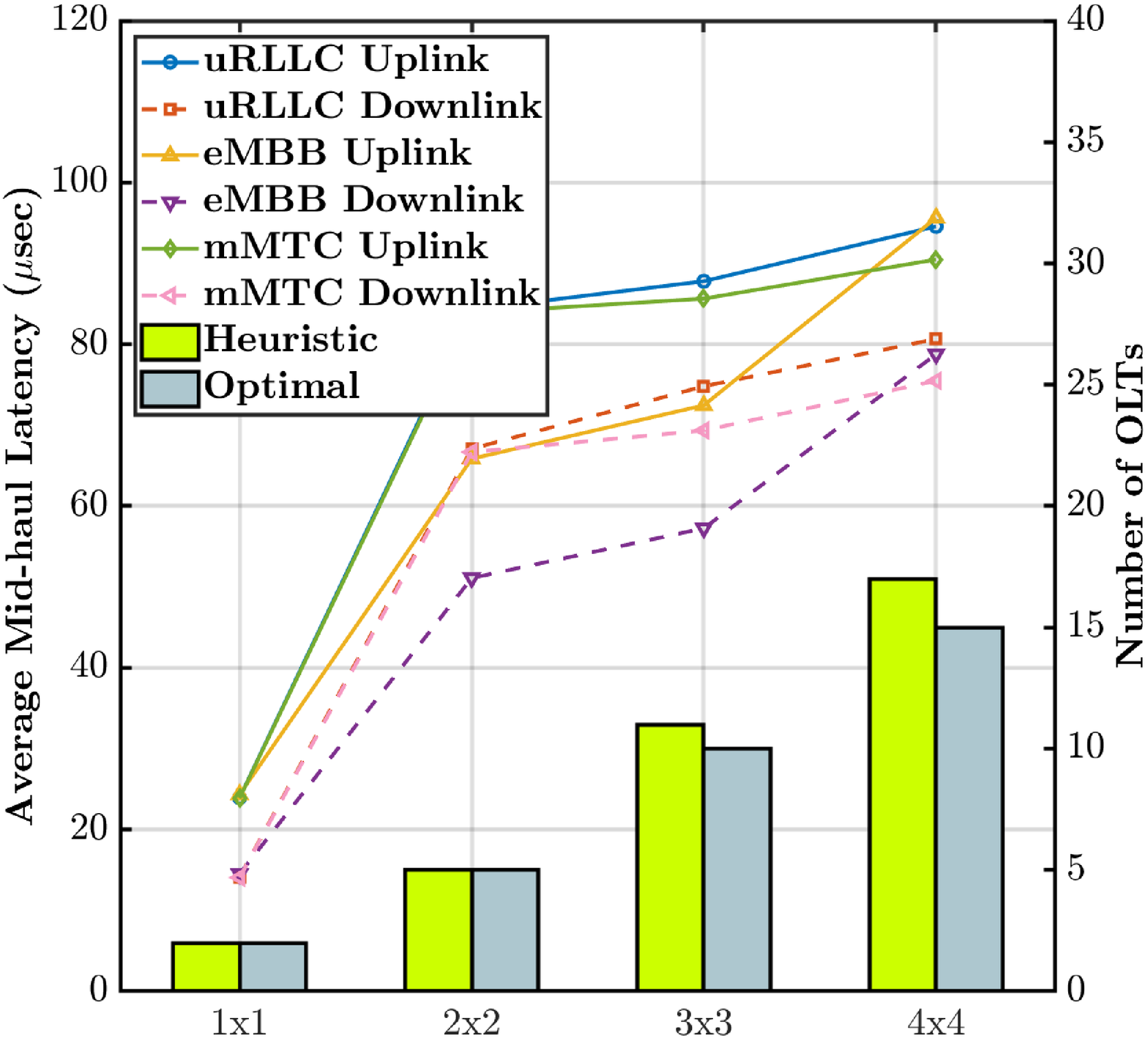}\label{mh_lat2}%
  }
  \subfloat[Rural Area (km $\times$ km)]{%
    \includegraphics[width=0.333\textwidth,height=5.1cm]{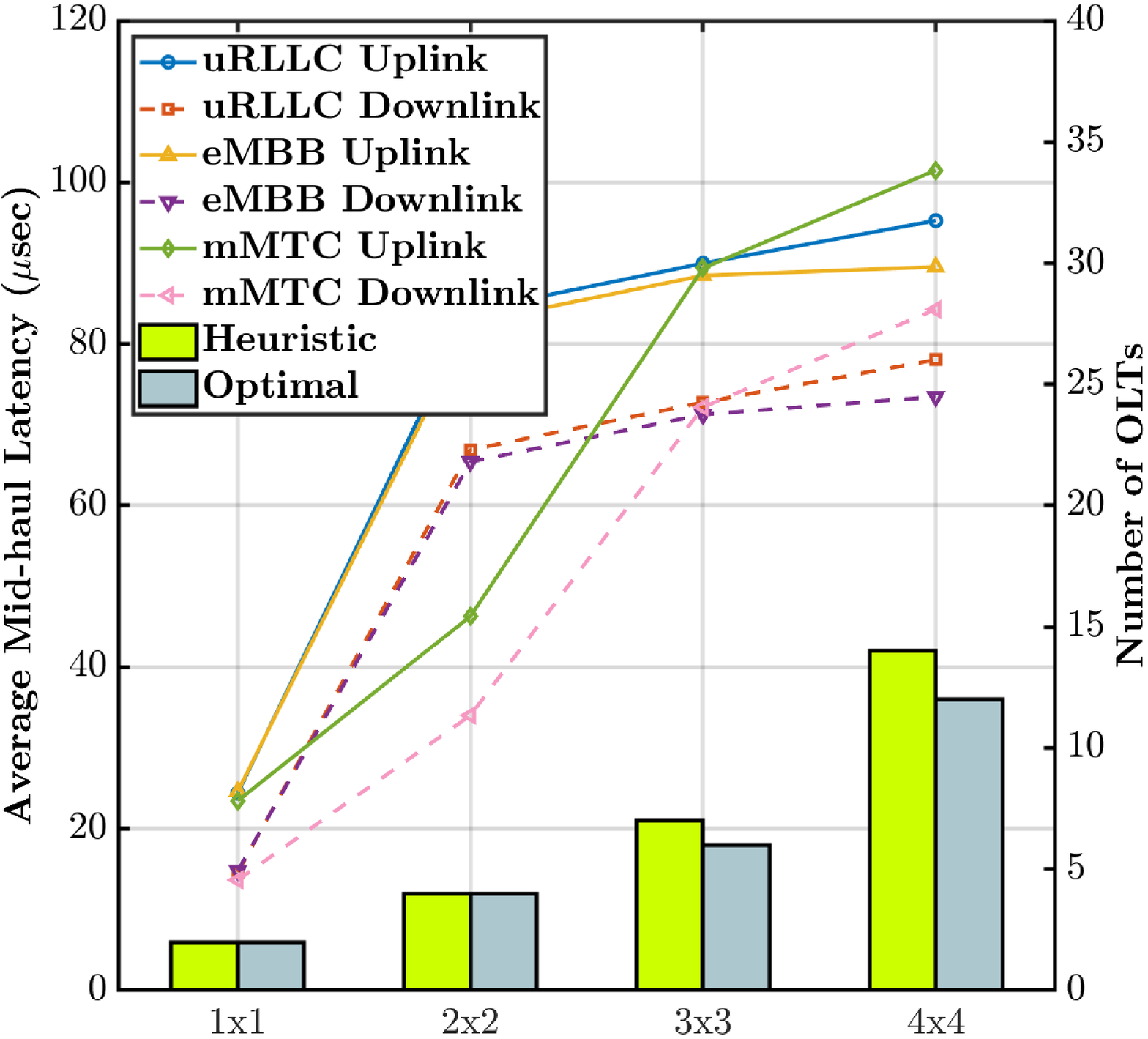}\label{mh_lat3}%
  }

  \caption{Average mid-haul uplink and downlink latencies of uRLLC, eMBB, and mMTC slices and the total number of installed OLTs of TWDM-PONs against (a) industrial, (b) urban, and (c) rural areas.}
  \label{mh_latency}
\end{figure*}
\setlength{\textfloatsep}{1pt}

\begin{figure*}[!t]
  \centering
  \subfloat[Industrial Area (km $\times$ km)]{%
    \includegraphics[width=0.333\textwidth,height=5.1cm]{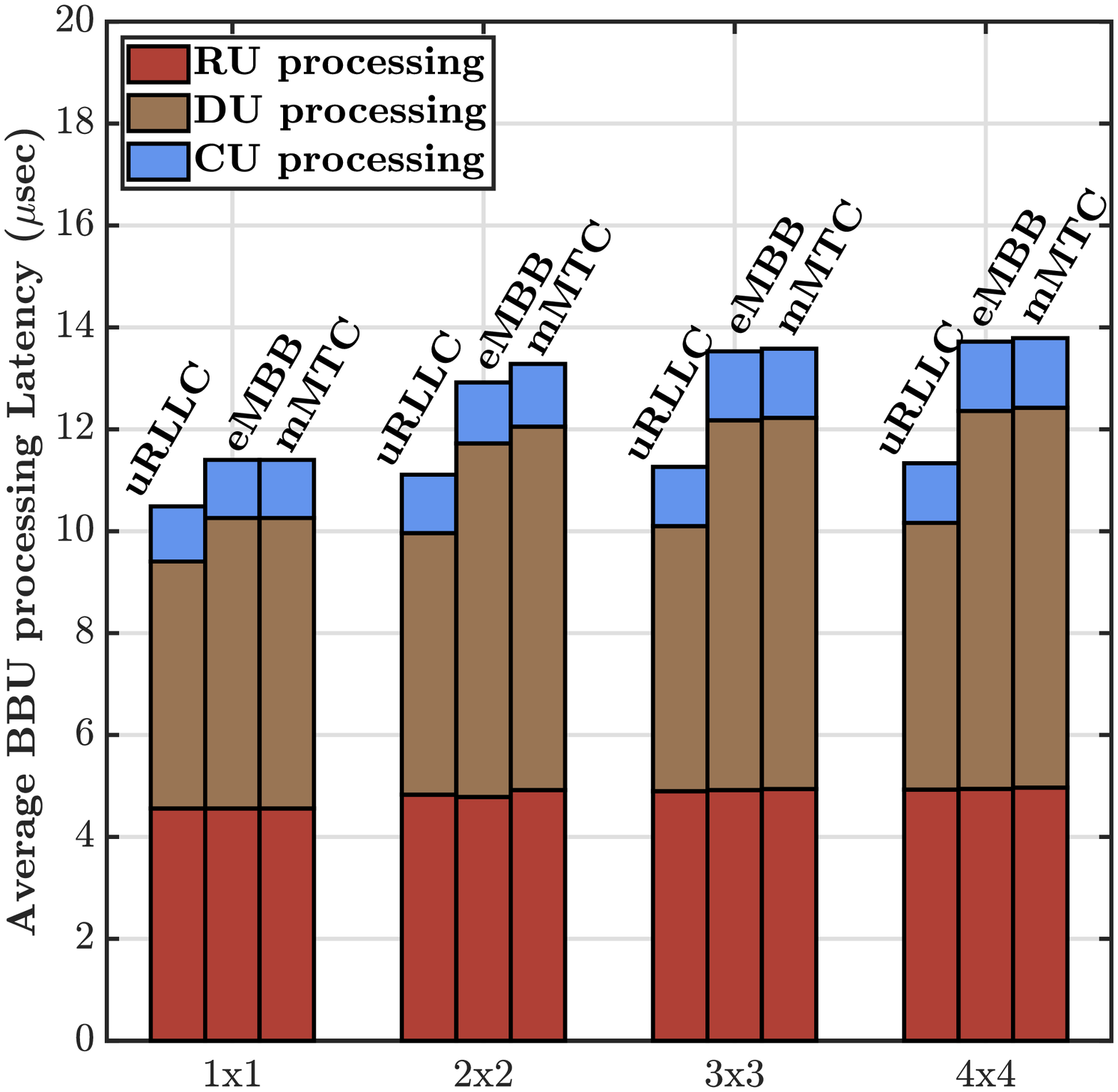}\label{bbu_lat1}%
  }
  \subfloat[Urban Area (km $\times$ km)]{%
    \includegraphics[width=0.333\textwidth,height=5.1cm]{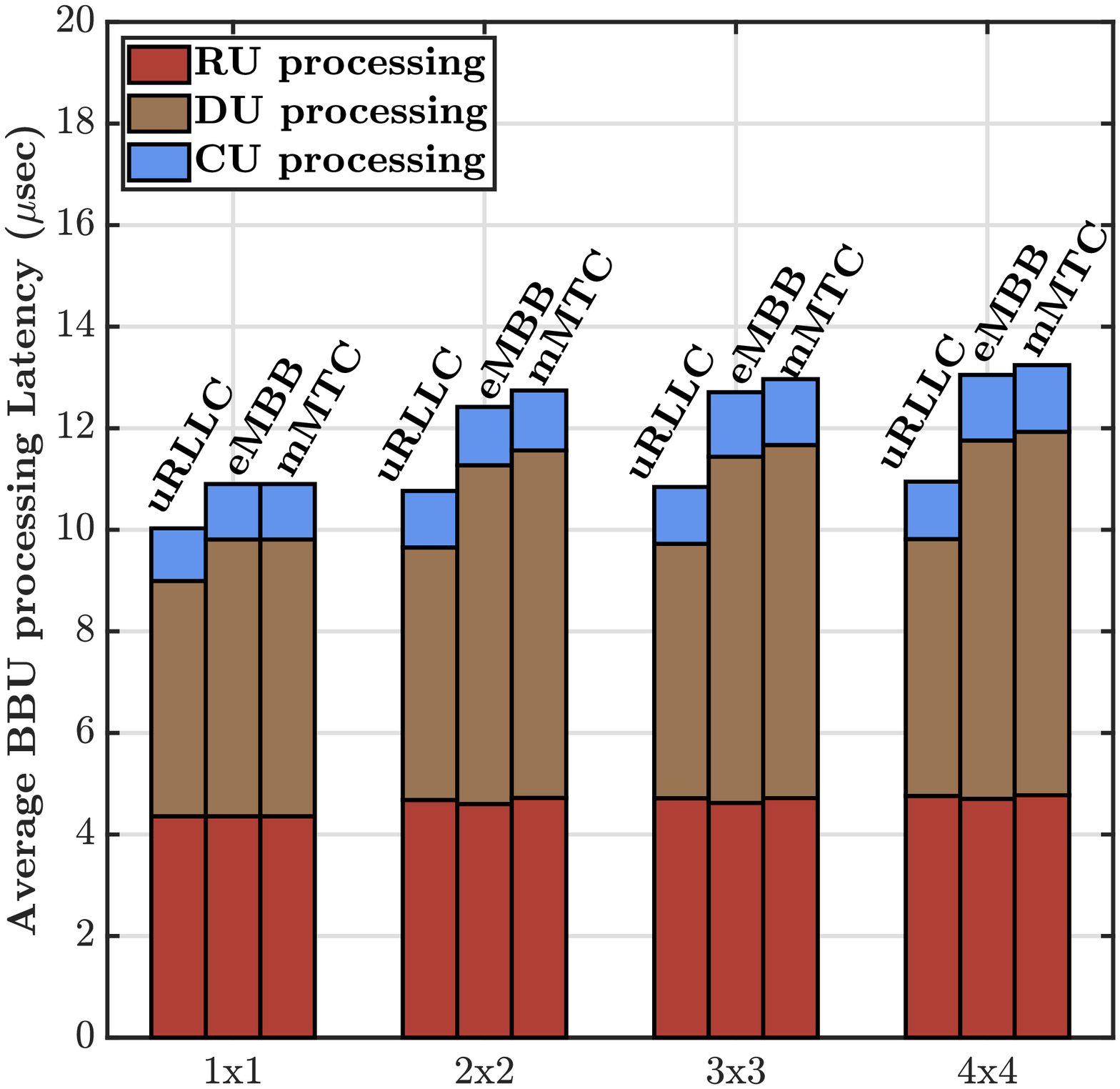}\label{bbu_lat2}%
  }
  \subfloat[Rural Area (km $\times$ km)]{%
    \includegraphics[width=0.333\textwidth,height=5.1cm]{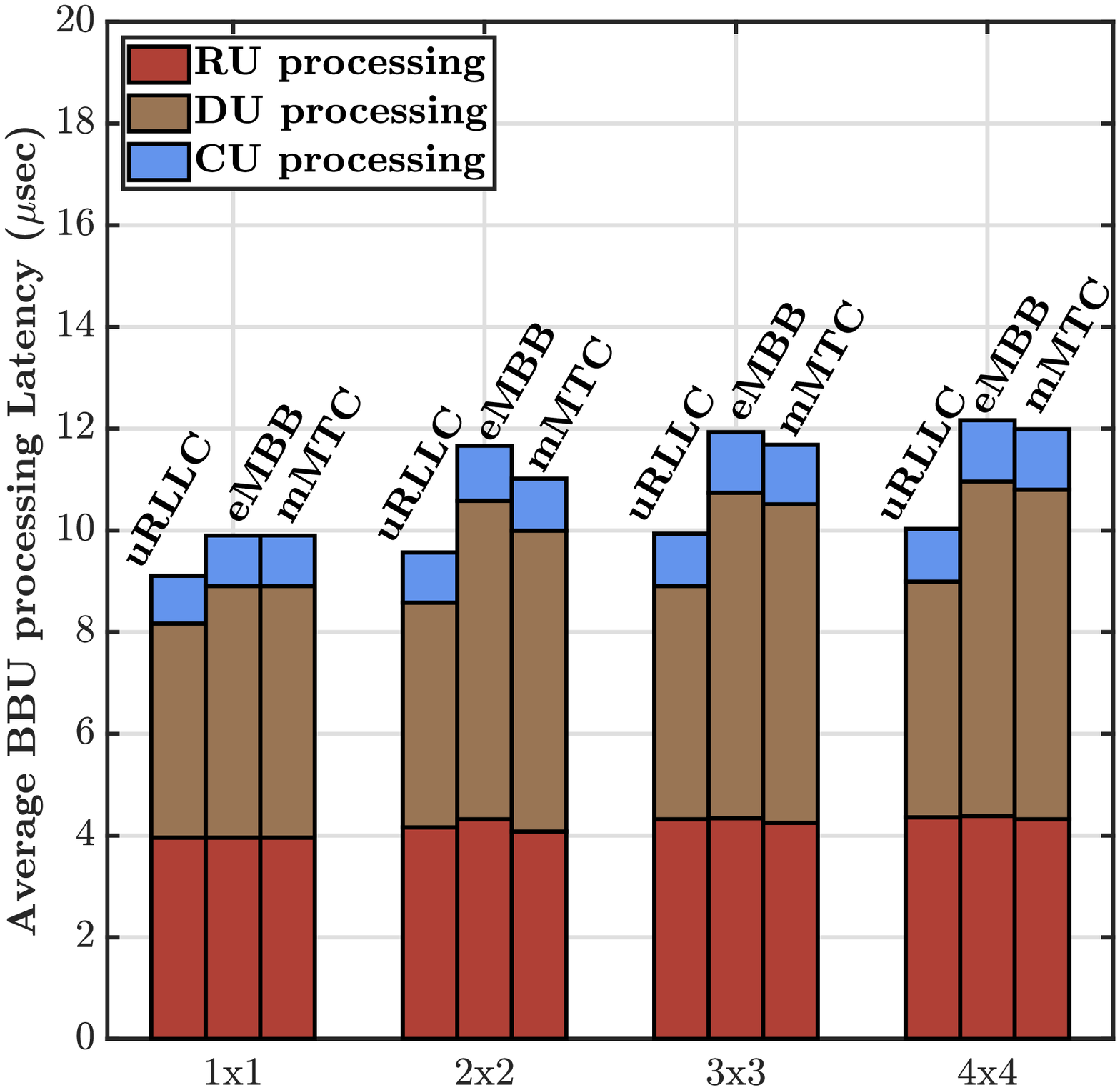}\label{bbu_lat3}%
  }

  \caption{Average BBU function processing latencies of uRLLC, eMBB, and mMTC slices against (a) industrial, (b) urban, and (c) rural areas.}
  \label{bbu_latency}
\end{figure*}
\setlength{\textfloatsep}{1pt}
In Fig. \ref{ue_bs_cmp}, we present the number of required RUs against industrial, urban, and rural areas by evaluating the UE-RU association problem $\mathcal{P}_1$. {As the area size grows from $1\times 1$ km$^2$ to $4\times 4$ km$^2$, the number of UEs increases, which leads to a higher number of RUs against all scenarios. Interestingly, we also observe that the average wireless throughput per RU (total throughput of UEs/number of RUs) also increases. This implies that resources per RU are shared by a higher number of UEs.} The RUs in all the areas are classified into uRLLC, eMBB, and mMTC slices and a major share of the RUs belong to the eMBB slice due to its high throughput demand. Both uRLLC and mMTC slices have similar throughput demands, and hence, a similar number of RUs are required. Each group of the vertical bars in the subplots consists of (left to right) the solution obtained by using the proposed Lagrangian relaxation heuristic (LgR), the optimal solution by IBM Gurobi 9.1 solver (Opt), and the lower bound with integrality relaxation (Low). As the integrality constraints are relaxed, the solutions indicated by Low are not a feasible solution always but provide a minimum benchmark only. {Although IBM Gurobi is much faster than IBM CPLEX, the evaluation of problem $\mathcal{P}_1$ with a dataset of size $5\times 5$ km$^2$ or more on our computer (Intel Core i7 processor, 32 GB RAM) takes more than two days.} Nonetheless, a solution can be obtained very quickly with our proposed heuristic Algorithm \ref{alg1}. On some occasions, it could yield a solution exactly the same as the optimal solution, but at other times, it yields a slightly higher number of RUs.\par
After the UEs are associated with RUs for uRLLC, eMBB, and mMTC slices, we use these RU locations to solve the subsequent DU-CU placement and front/mid-haul design problem $\mathcal{P}_2$. Again, we use the commercially available solver IBM Gurobi 9.1 to evaluate the optimal solutions against the industrial, urban, and rural scenarios. We also employ our proposed heuristic Algorithm \ref{alg2} to evaluate a near-optimal solution in a time-efficient manner. Firstly, we present the average uplink and downlink communication latencies of the front-haul interface in Fig. \ref{fh_latency} and the mid-haul interface in Fig. \ref{mh_latency}. To generate the Fig. \ref{fh_latency}, we consider that the DUs for all slices are deployed at the Stage-I OLT locations, whereas for Fig. \ref{mh_latency}, we consider that the DUs for all slices are deployed at the RU locations. Note that this is done just for comparison purposes as deploying all DUs at RU locations is costlier than OLT locations where several DUs can be aggregated in a single server. Although the peak throughput requirements of both uplink and downlink are similar, the average latencies of the uplink traffic in both front-haul and mid-haul interfaces are slightly higher than the respective downlink traffic due to the waiting time of data at ONUs. We also show the total number of installed OLTs of TWDM-PON obtained from the optimal solution as well as the proposed heuristic Algorithm \ref{alg2} in Figs. \ref{fh_latency}-\ref{mh_latency} against the industrial, urban, and rural scenarios. We can observe that a higher number of OLTs are required for front-haul interfaces than for mid-haul interfaces.\par
\begin{figure*}[!t]
  \centering
  \subfloat[Area (km $\times$ km)]{%
    \includegraphics[width=0.333\textwidth,height=5.1cm]{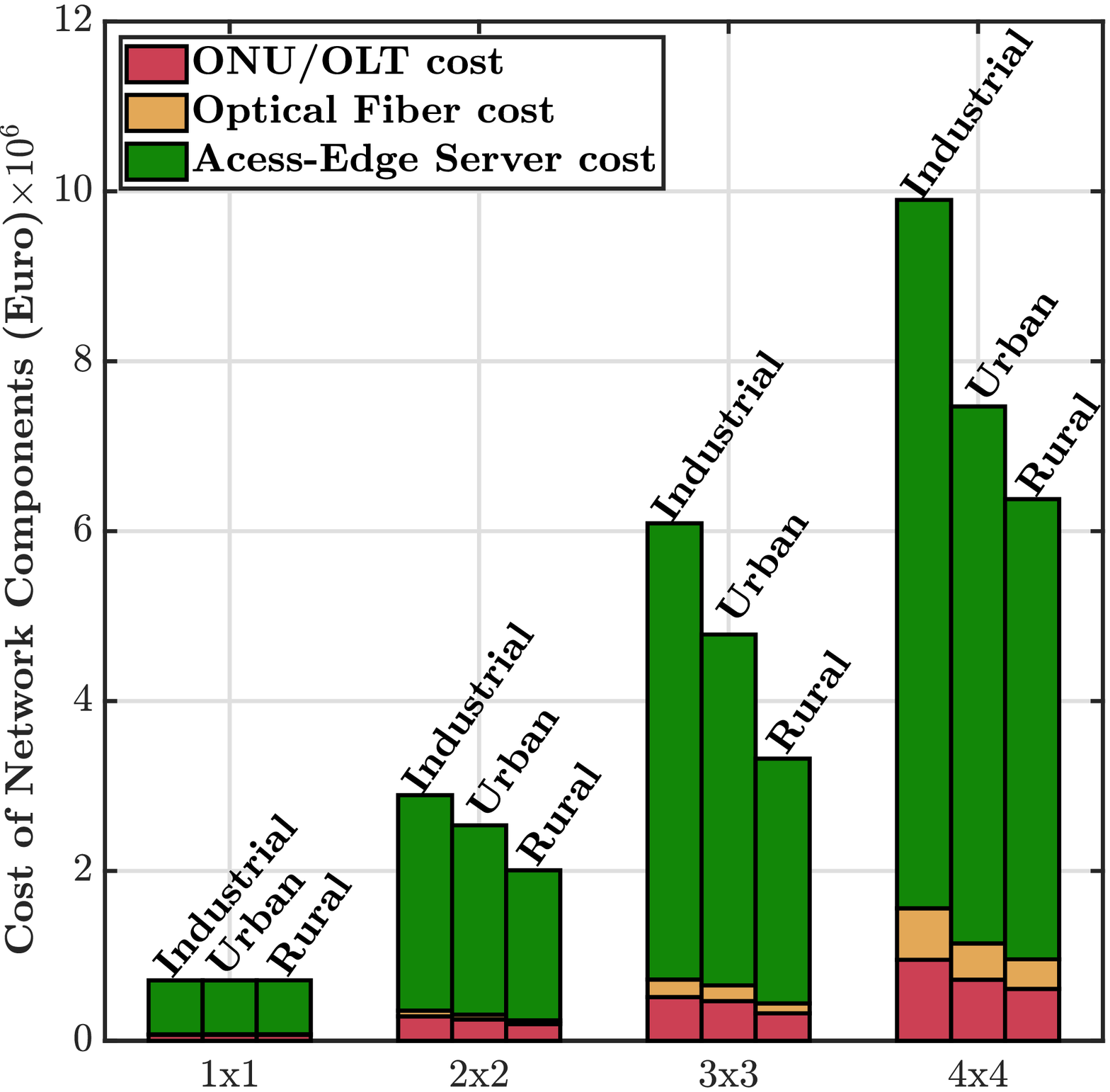}\label{cost_stageI}%
  }
  \subfloat[Area (km $\times$ km)]{%
    \includegraphics[width=0.333\textwidth,height=5.1cm]{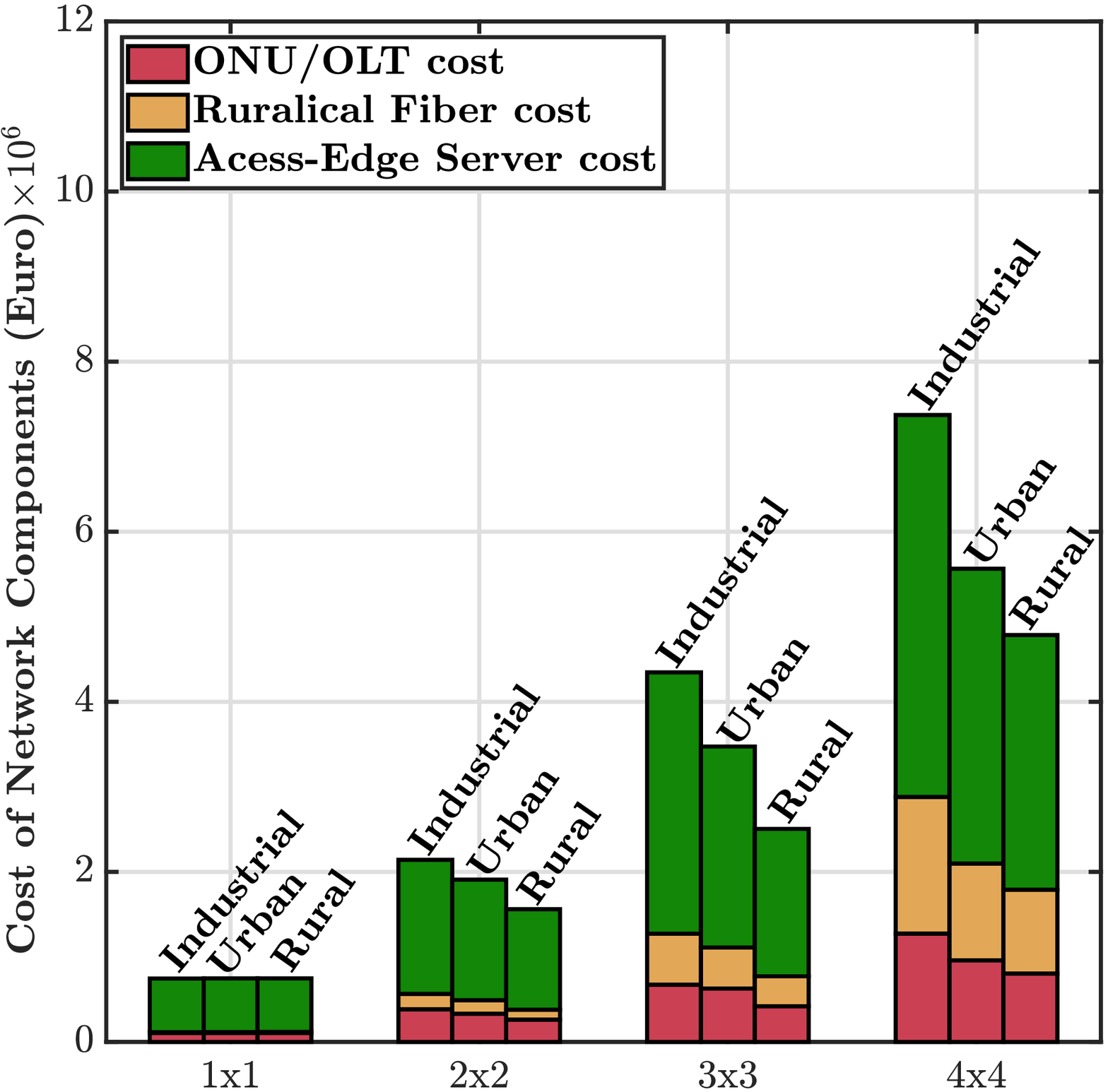}\label{cost_stageII}%
  }
  \subfloat[$N = \lceil \text{(Stage-II datarate)/(Stage-I datarate)} \rceil$]{%
    \includegraphics[width=0.333\textwidth,height=5.1cm]{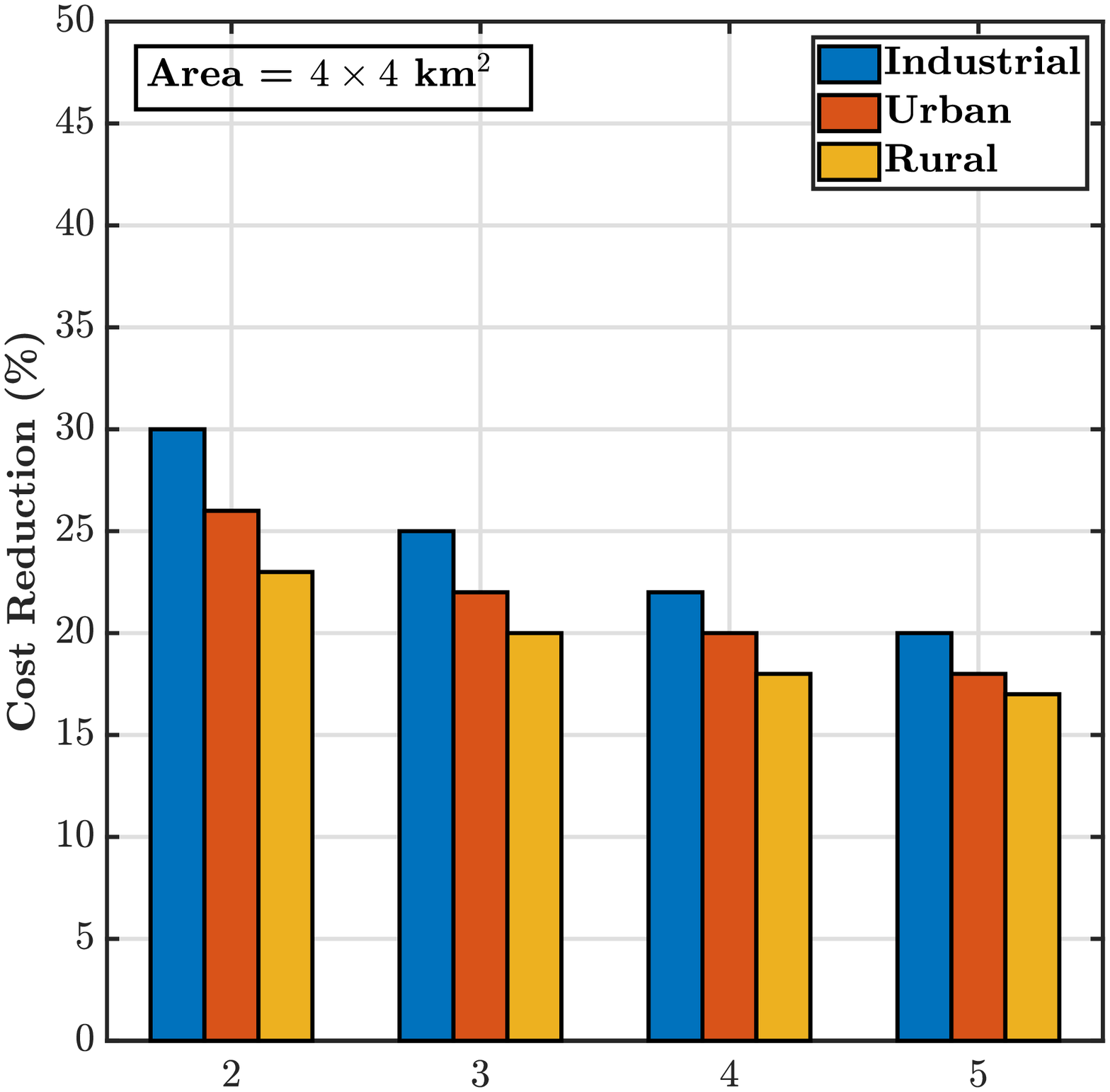}\label{cost_comp1}%
  }

  \caption{Comparison of front/mid-haul interfaces and open access-edge server deployment cost with (a) only Stage-I TWDM-PON and (b) both Stage-I and Stage-II TWDM-PONs. (c) Comparison of cost reduction achieved with a higher datarate Stage-II than Stage-I TWDM-PON.}
\end{figure*}
\setlength{\textfloatsep}{1pt}
Secondly, we present the average virtual BBU, i.e., RU, DU, and CU processing latencies of the uRLLC, eMBB, and mMTC slices in Fig. \ref{bbu_latency}. This result shows the optimal values where the DUs of the uRLLC slice are placed at RU locations, but the DUs of eMBB and mMTC slices are placed at Stage-I OLT locations. We observe that the average BBU (RU-DU-CU) processing latency of the industrial area in Fig. \ref{bbu_lat1} is slightly higher than the urban area in Fig. \ref{bbu_lat2}, which is again higher than the rural area in Fig. \ref{bbu_lat3}. This primarily happens due to the increase in the number of RUs and the required computational efforts. Note that the average RU processing latencies of all three slices are very close to each other because we considered the same RU configuration and 100\% PRB usage for the network planning problem. However, the DU processing latency is lowest for the uRLLC slice against all scenarios because dedicated servers are deployed at RU locations for DU function processing. The DU and CU processing latencies of the eMBB and mMTC slices are almost the same or slightly different from each other because a lesser amount of computational resources are allocated due to a relatively relaxed latency bound of the eMBB and mMTC slices. Interestingly, we can observe that the considered latency bounds for front-haul (100 $\mu$sec), mid-haul (100, 500, and 1000 $\mu$sec), and BBU (RU-DU-CU) function processing (50, 80, and 100 $\mu$sec) are satisfied against all scenarios.\par
\begin{figure}[b!]
\centering
\includegraphics[width=\columnwidth,height=5.5cm]{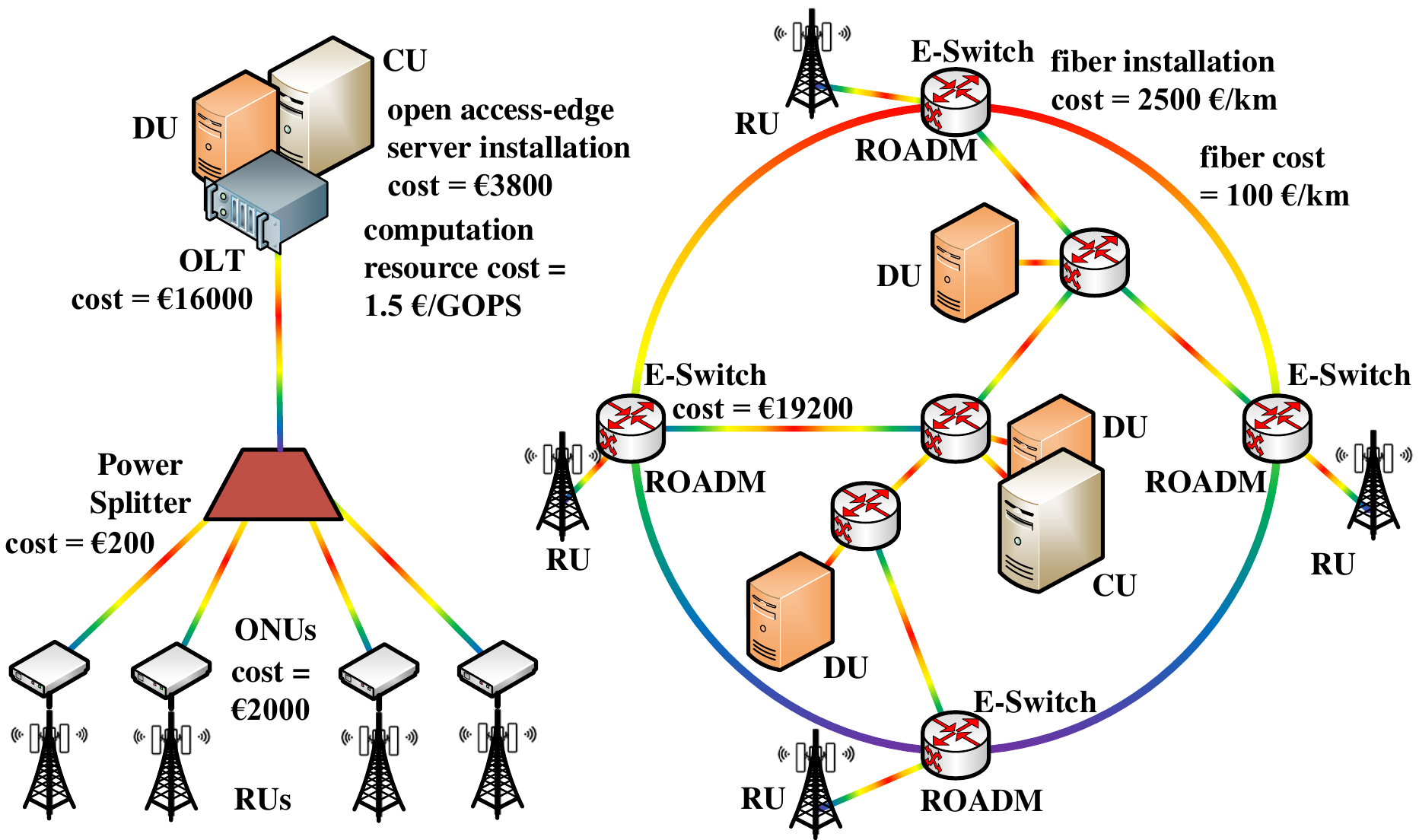}
\caption{Pictorial description of various cost components in TWDM-PON and OTN-based front/mid-haul interfaces.}
\label{cost_params}
\end{figure}
\setlength{\textfloatsep}{5pt}
%
\begin{figure*}[!t]
  \centering
  \subfloat[Industrial Area (km $\times$ km)]{%
    \includegraphics[width=0.333\textwidth,height=5.1cm]{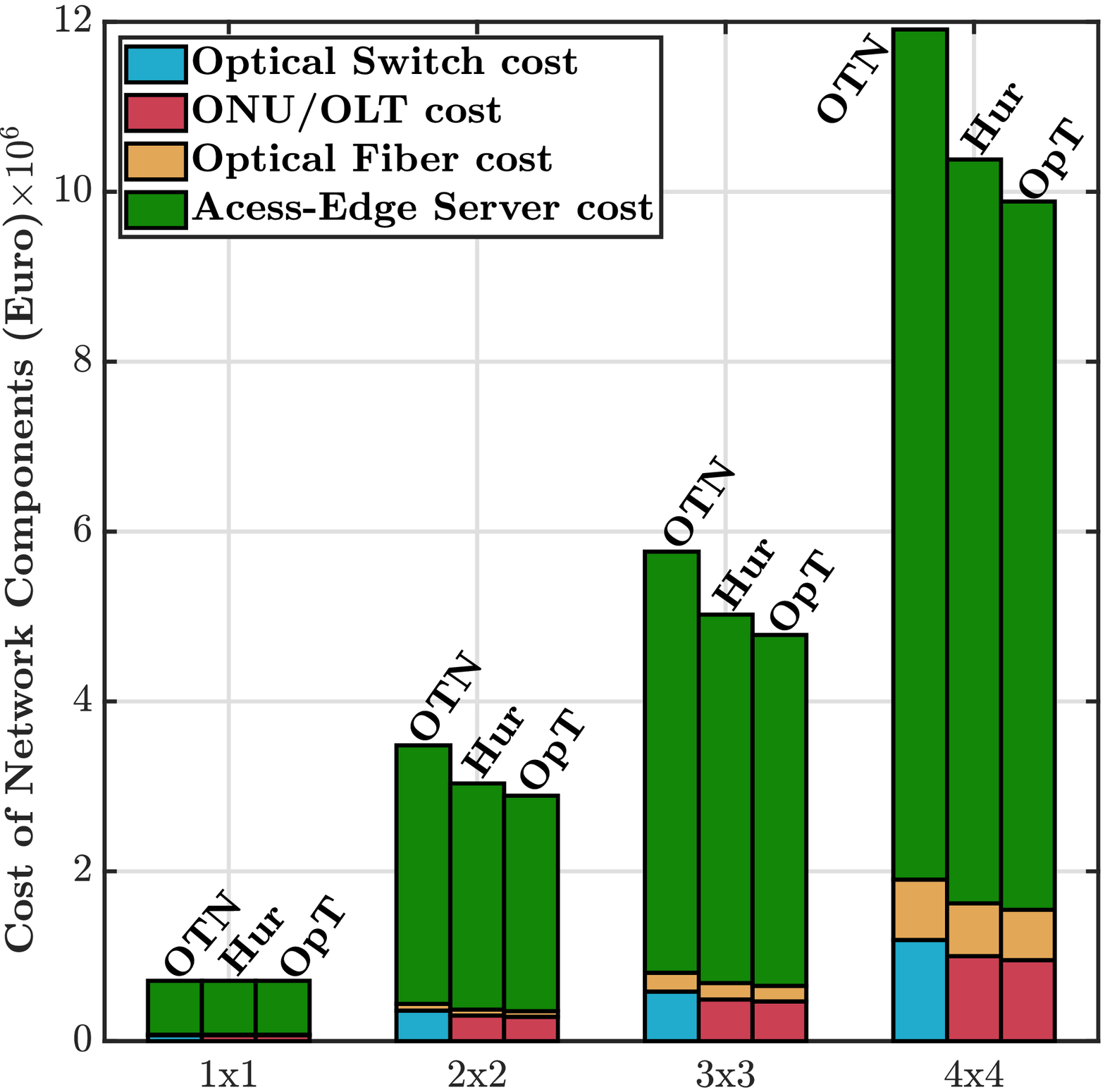}\label{cost1}%
  }
  \subfloat[Urban Area (km $\times$ km)]{%
    \includegraphics[width=0.333\textwidth,height=5.1cm]{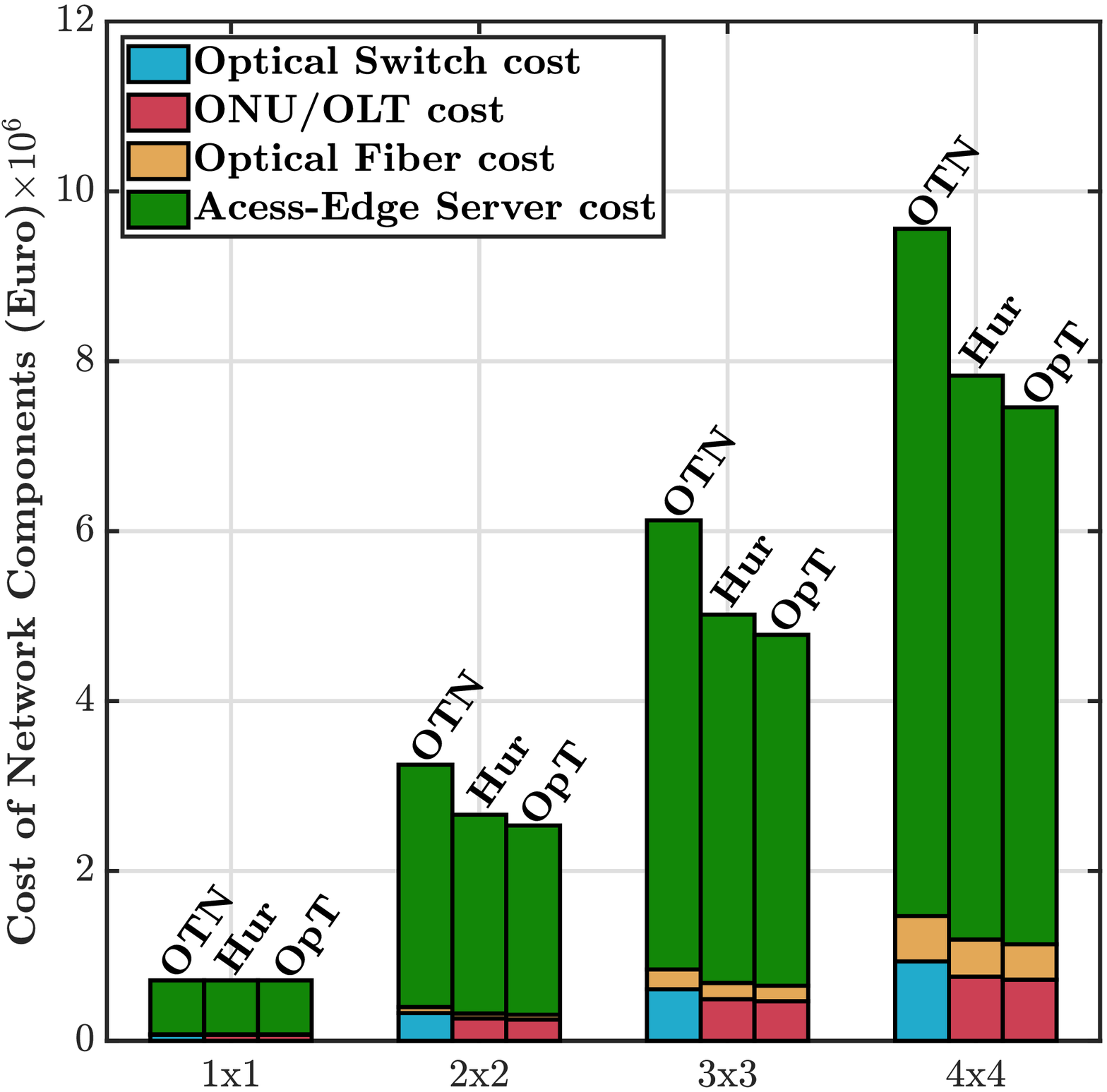}\label{cost2}%
  }
  \subfloat[Rural Area (km $\times$ km)]{%
    \includegraphics[width=0.333\textwidth,height=5.1cm]{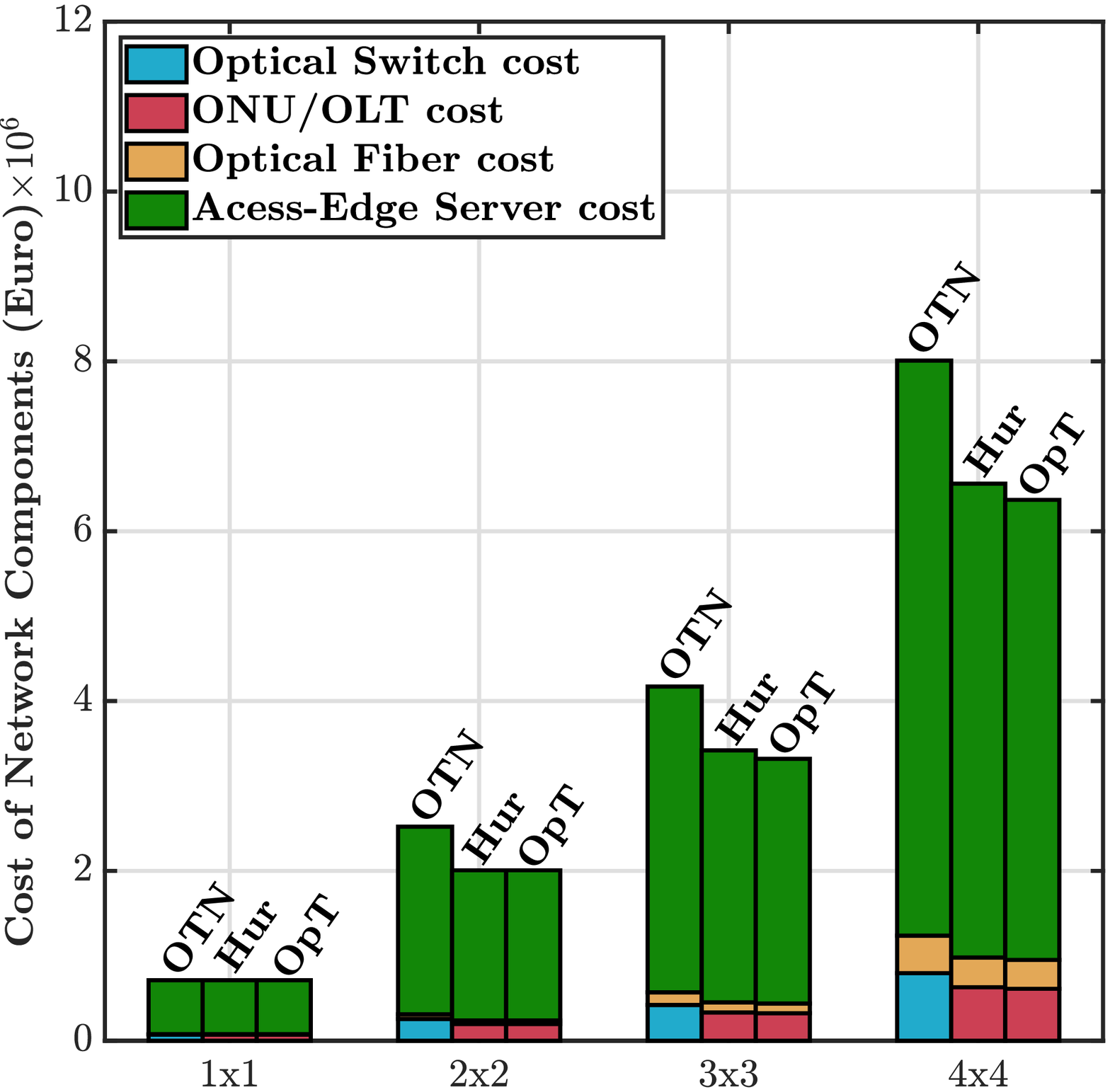}\label{cost3}%
  }

  \caption{Comparison of front/mid-haul interfaces and open access-edge server deployment cost with TWDM-PON-based framework, both optimal (OpT) and heuristic (Hur), and OTN-based framework (OTN) against (a) industrial, (b) urban, and (c) rural areas.}
  \label{cost_comp}
\end{figure*}
\setlength{\textfloatsep}{1pt}
%
Our primary motivation behind proposing a two-stage TWDM-PON-based sliced O-RAN architecture is to provide flexible and cost-efficient RAN deployment options to mobile network operators. Commonly, only Stage-I TWDM-PON front/mid-haul could be the most cost-efficient option, but for certain scenarios, the double-stage TWDM-PON front/mid-haul may prove as a better option. For showing this, we consider two different network configurations and plot the OLT/ONU cost, optical fiber+installation cost, and open access-edge server cost components against industrial, urban, and rural areas. In Fig. \ref{cost_stageI}, only Stage-I TWDM-PON is sufficient because the access-edge servers have sufficient resources ($10^5$ GOPS) to host both the DUs and CUs of the aggregated RUs and hence, including Stage-II TWDM-PONs in the O-RAN architecture is not required. Nonetheless, in Fig. \ref{cost_stageII}, we consider the maximum processing capacities of both the Stage-I and Stage-II servers are $0.5\times10^5$ GOPS such that the servers at Stage-I OLTs have resources only to process the DUs of the aggregated RUs but the CUs are processed by servers at Stage-II OLTs. Thus, our proposed framework is compelled to include the Stage-II TWDM-PON along with Stage-I TWDM-PON in the O-RAN architecture for the CU function processing. In this case, the cost of ONU/OLT and the optical fiber is slightly higher than with only Stage-I, but the overall O-RAN deployment cost reduces as the cost of access-edge servers reduces. From Fig. \ref{cost_stageII}, it is also evident that the cost reduction increases with a bigger area and a higher number of devices. We can see nearly 26\% or lower cost for $4\times 4$ km$^2$ industrial, urban, and rural areas. Note that a much better cost reduction may also be achievable if we use a much higher data rate for the Stage-II TWDM-PON. We consider the maximum 100 Gbps data rate for Stage-I TWDM-PON as earlier but a higher maximum data rate of Stage-II TWDM-PON by aggregating more channels. We assume that the cost of ONUs and OLTs increase linearly with the number of aggregated wavelengths. In Fig. \ref{cost_comp1}, we plot the cost of $4\times 4$ km$^2$ industrial, urban, and rural areas against $N = \lceil \text{(Stage-II datarate)/(Stage-I datarate)} \rceil$. From this plot, we can observe that initially, the overall O-RAN deployment cost decreases steadily as $N$ increases due to a better aggregation of Stage-I OLT-ONU boxes through the Stage-II TWDM-PONs. However, the cost reduction becomes very minimal beyond $N =3$ against the considered scenarios because the cost of OLT/ONUs in Stage-II TWDM-PON increases with $N$.\par

%
To compare the performance of our proposed TWDM-PON-based NGFI framework against OTN-based NGFI frameworks, we consider the OTN-based O-RAN architecture proposed in \cite{oran_opt}. In OTN architecture, each node is connected to optical links through a re-configurable add/drop multiplexer (ROADM) and an electric switch (E-switch). The ROADM can switch traffic on a wavelength basis with negligible latency, while the electric switch performs the optical-electric-optical conversion and electric switching \cite{F_split}. The RUs and servers for DUs (front-haul links) and the servers for DUs and CUs (mid-haul links) are connected by \emph{mesh topology} and the capacity of each optical path is 100 Gbps \cite{oran_opt}. The cost of each ROADM and E-switch is \euro 19200 and a pictorial description of various cost components of both TWDM-PON and OTN are given in Fig. \ref{cost_params}. Now, in Fig. \ref{cost_comp}, we compare the O-RAN deployment costs with both the TWDM-PON and OTN-based NGFI frameworks against industrial, urban, and rural areas. Note that the same RU locations and network configurations are used as input to both the TWDM-PON and OTN-based NGFI frameworks. Both the optimal cost (OpT) and the cost obtained by the heuristic Algorithm \ref{alg2} (Hur) are compared against the cost for the OTN-based framework (OTN). We observe that the RAN deployment cost obtained by the heuristic Algorithm \ref{alg2} is slightly higher than the optimal solution for $\mathcal{P}_2$ in some cases because a higher number of TWDM-PONs are installed. However, the cost of OTN-based frameworks is much higher than the TWDM-PON-based framework against all scenarios. The primary reason behind this is the constraints in establishing optical paths with mesh topology among nodes, which leads to the installation of more nodes in OTN than TWDM-PON. Due to the strict latency requirements of the front-haul interfaces, the traffic can not be aggregated and routed through a large number of hops, especially in sparse rural areas. Furthermore, setting up many optical paths is required with mesh-based OTN architecture, whereas the tree-and-branch architecture of TWDM-PON required a much lower number of fiber links. We could observe the maximum cost saving by TWDM-PON-based framework against industrial area is $\sim 21\%$, against urban area is $\sim 23\%$, and against rural area is $\sim 28\%$.

\section{Conclusion} \label{sec7}
In this paper, we have proposed an efficient framework for the optimal placement of RUs based on long-term network statistics and connecting them to open access-edge servers for hosting the corresponding DU and CU functions over the front/mid-haul interfaces while satisfying the diverse QoS requirements of uRLLC, eMBB, and mMTC applications. We have proposed a two-stage TWDM-PON network architecture that opportunistically allows us to choose either a single-stage or double-stage deployment. We have formulated a two-stage ILP for UE to RU association and installing TWDM-PON-based front/mid-haul interfaces while flexibly deploying open access-edge servers for hosting DUs and CUs. In turn, we have designed Lagrangian relaxation and greedy approach-based heuristics for solving these ILPs in a time-efficient manner. Using this framework, we find the optimal number of RUs required for uRLLC, eMBB, and mMTC slices. We also evaluate the communication latencies of the front/mid-haul interfaces and the virtual BBU (i.e., RU-DU-CU) function processing latency. We have shown that the two-stage TWDM-PON-based architecture can lead to better cost optimization against certain network scenarios than a single-stage architecture. Moreover, we evaluate the cost of O-RAN deployment with the proposed TWDM-PON-based as well as the state-of-the-art OTN-based frameworks to show that the TWDM-PON-based framework is at least 21\% cost-efficient over the OTN-based framework due to better network resource utilization.


%



\section*{Acknowledgment}
This work is financially supported by EU H2020 EDGE/MSCA (grant 713567) and Science Foundation Ireland (SFI) grants 17/CDA/4760 and 13/RC/2077\_P2.

\bibliographystyle{IEEEtran}
\bibliography{IEEEabrv,references}
\vspace{-5ex}
\begin{IEEEbiography}[{\includegraphics[width=1in,height=1.25in,clip,keepaspectratio]{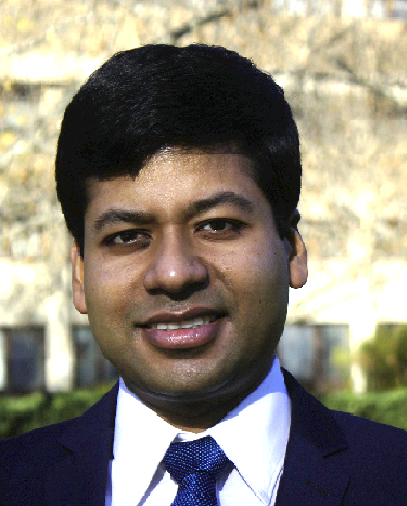}}]{Sourav Mondal}
(GS'16–M'21) received Ph.D. from the Department of Electrical and Electronic Engineering of the University of Melbourne in 2020. He received his M.Tech in Telecommunication Systems Engineering from the Department of Electronics and Electrical Communication Engineering, Indian Institute of Technology Kharagpur, and B.Tech in Electronics and Communication Engineering from Kalyani Govt. Engineering College, affiliated with the West Bengal University of Technology in 2014 and 2012, respectively. He was employed as an Engineer in Qualcomm India Pvt. Ltd. from 2014 to 2016. Currently, he is working as an EDGE/Marie Skłodowska-Curie post-doctoral fellow at CONNECT Centre for Future Networks and Communication in Trinity College Dublin, Ireland.
\end{IEEEbiography} 
\begin{IEEEbiography}[{\includegraphics[width=1in,height=1.25in,clip,keepaspectratio]{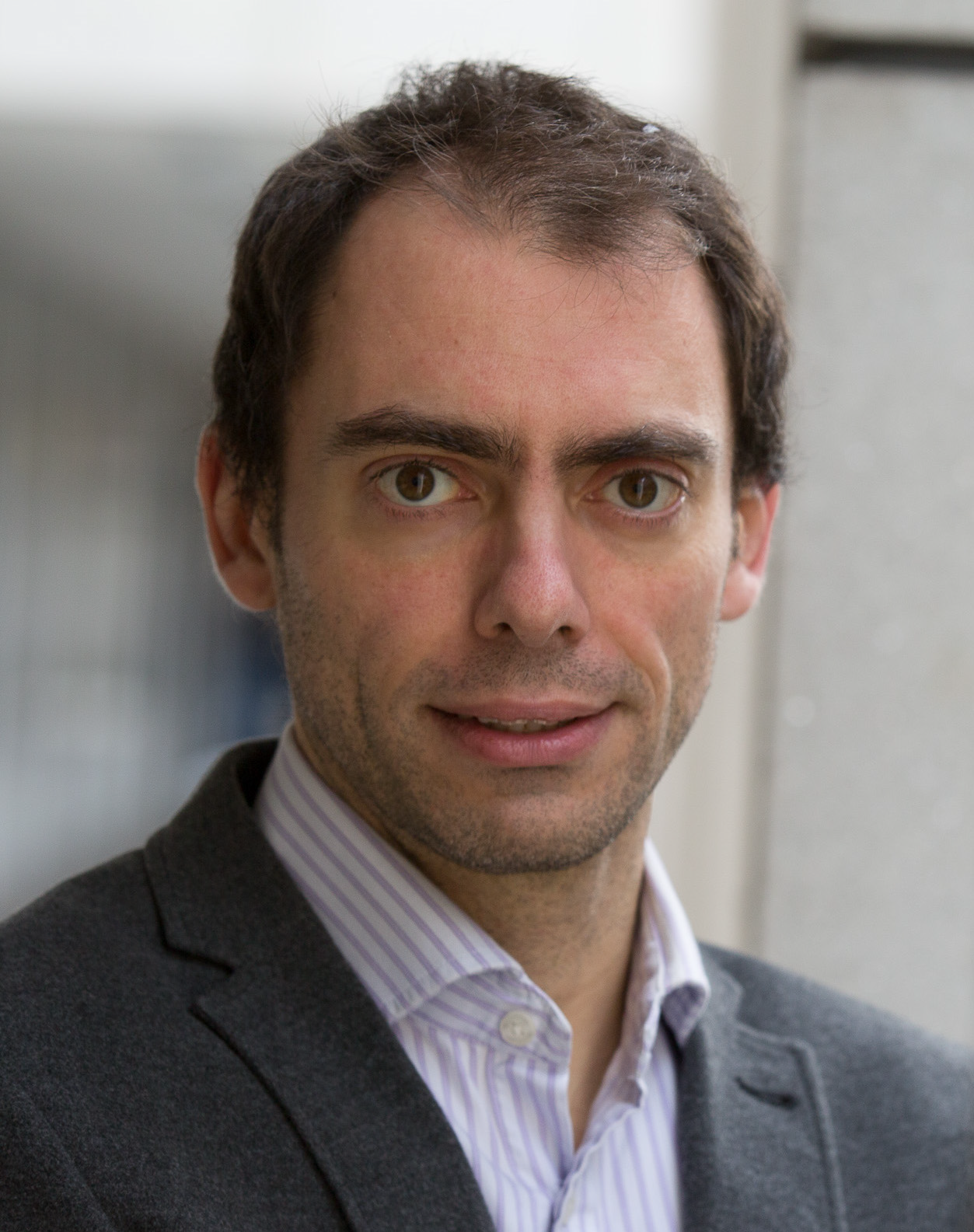}}]{Marco Ruffini} 
received the M.Eng. degree in telecommunications from the Polytechnic University of Marche, Italy, in 2002, and the Ph.D. degree from Trinity College Dublin (TCD) in 2007, where he joined Trinity College Dublin in 2005, after working as a Research Scientist with Philips, Germany. 
He is an Associate Professor and Fellow of Trinity College and he is the Principal Investigator of both the IPIC Photonics Integration Centre and the CONNECT Telecommunications Research Centre. He is currently involved in several Science Foundation Ireland and H2020 projects, including a new research infrastructure to build a beyond 5G testbed in Dublin. Prof. Ruffini leads the Optical Network and Radio Architecture Group, TCD, and has authored over 160 international publications, and over ten patents, and contributed to standards at the broadband forum. He has raised research funding in excess of \euro 8M. His main research is in the area of 5G optical networks, where he carries out pioneering work on the convergence of fixed-mobile and access-metro networks, and on the virtualization of next-generation networks, and has been invited to share his vision through several keynote and talks at the major international conferences across the world. He leads the new SFI-funded Ireland’s Open Networking testbed infrastructure (OpenIreland).
\end{IEEEbiography}

\vfill


\end{document}